\documentclass[11pt,a4paper]{article}

\usepackage{jheppub}
\usepackage{amsmath}
\usepackage{txfonts}
\begin{document}

\newcounter{ENVeqcntsave}
\newenvironment{ENV}

% Use the \preprint command to place your local institutional report
% number in the upper righthand corner of the title page in preprint mode.
% Multiple \preprint commands are allowed.
% Use the 'preprintnumbers' class option to override journal defaults
% to display numbers if necessary
%\preprint{}

%Title of paper
%\title{Exact extended quasiparticle theory of excited-state
% statics and dynamics of materials}
\title{Comment on CFT in AdS and boundary RG flows:
\textbf{\textit{O}(1/\textit{N})} Result}

% repeat the \author .. \affiliation  etc. as needed
% \email, \thanks, \homepage, \altaffiliation all apply to the current
% author. Explanatory text should go in the []'s, actual e-mail
% address or url should go in the {}'s for \email and \homepage.
% Please use the appropriate macro foreach each type of information

% \affiliation command applies to all authors since the last
% \affiliation command. The \affiliation command should follow the
% other information
% \affiliation can be followed by \email, \homepage, \thanks as well.
\author[a]{Kaoru Ohno}
% \altaffiliation[Also at ]{Physics Department, XYZ University.}%Lines break automatically or can be forced with \\
%\affiliation{
%}
\affiliation[a]{
Department of Physics, Yokohama National University,
79-5 Tokiwadai, Hodogaya-ku, Yokohama 240-8501, Japan
}
\author[b]{and Yutaka Okabe}
\affiliation[b]{Department of Physics, Tokyo Metropolitan University,
Hachioji, Tokyo 192-0397, Japan}
%\homepage[]{Your web page}
%\thanks{}
%\altaffiliation{}

\emailAdd{ohno@ynu.ac.jp}
\emailAdd{okabe@phys.se.tmu.ac.jp}

%Collaboration name if desired (requires use of superscriptaddress
%option in \documentclass). \noaffiliation is required (may also be
%used with the \author command).
%\collaboration can be followed by \email, \homepage, \thanks as well.
%\collaboration{}
%\noaffiliation

\date{\today}

\abstract{
In a recent paper [JHEP {\bf 11} (2020) 118], S. Giombi and H. Khanchandani
studied the $1/N$ expansion of the $O(N)$ model in semi-infinite space
within the framework of conformal field theory in anti-de Sitter space.
They presented a series expansion for
the $O(1/N)$ correction to the boundary anomalous dimension
in the case of the ordinary transition.
Although they were unable to sum the series or simplify its form analytically,
they demonstrated numerically that their result matches our earlier,
simple analytic expression given in Prog. Theor. Phys. {\bf 70} (1983) 1226. 
In this paper, we show that their series expansion is in fact exactly 
equivalent to our original expression.
However, since the final formula in eq. (4.57) of their paper,
which is expressed in terms of two different ${}_3F_2$ functions,
cannot produce the correct values,
we derive the correct formulae involving two ${}_3F_2$ functions
in the Appendices.
We comment on the similarity and difference
between their analysis and ours in the cases of the ordinary and special transitions,
and present full details of deriving the anomalous dimension
for all the cases including the extraordinary transition,
which were not written in our earlier paper.
}

\keywords{$O(N)$ model, $1/N$ expansion, semi-infinite space,
boundary anomalous dimension}

%\maketitle must follow title, authors, abstract, \pacs, and \keywords
\maketitle

% body of paper here - Use proper section commands
% References should be done using the \cite, \ref, and \label commands
% Put \label in argument of \section for cross-referencing
%\section{\label{}}
%\subsection{}
%\subsubsection{}

\section{Introduction}
\label{Introduction}

\begin{figure}[t]
\begin{center}
\includegraphics[width=140mm]{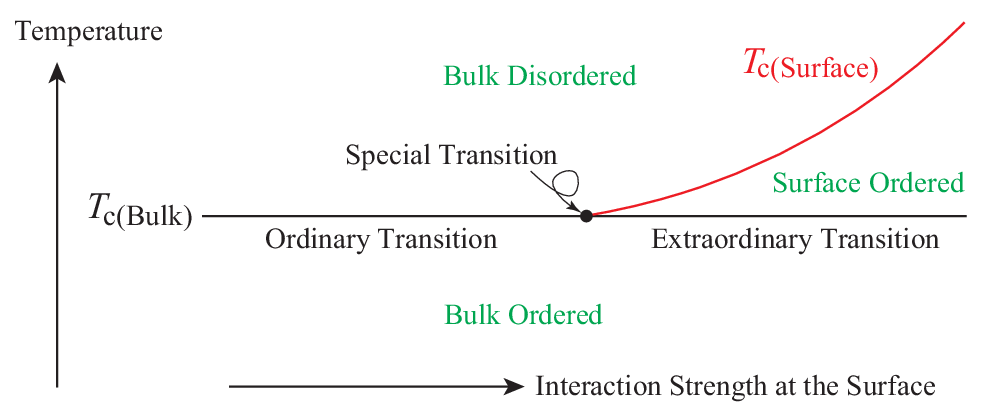}
\caption{Phase diagram of the $O(N)$ model in semi-infinite space.
Depending on the interaction strength at the surface, the phase boundary
at the bulk critical temperature $T_c$ is classified into
the ordinary, special, and extraordinary transitions,
each belonging to distinct universality class.}
\label{Phase}
\end{center}
\end{figure}

The $1/N$ expansion of $O(N)$ model in $d$-dimensional semi-infinte space
[$\mbox{\boldmath $r$}=(\mbox{\boldmath $x$}, y)$, $y>0$]
has recently attracted considerable attention in the context of
conformal field theory in anti-de Sitter (AdS) space \cite{Carmi,Giombi,Ankur}.
There are three distinct universality classes
at the bulk critical temperature $T_c$,
depending on the strength of the exchange interaction at the surface:
the ordinary, special, and extraordinary transitions \cite{BrayMoore},
as schematically illustrated in Fig.~\ref{Phase}.
When the exchange interaction at the surface is anisotropic,
an anisotropic special transition \cite{DiehlEisenriegler} occurs,
which belongs to a further distinct universality class.
Many years ago, we derived a simple analytic expression for
the $O(1/N)$ correction in the $O(N)$ model in semi-infinte space.
In particular, we obtained results for the ordinary transition \cite{JMMM,PLA1,PTP1},
the special transition \cite{PLA2, PTP1}, the anisotropic special transition \cite{PLA3},
and the extraordinary transition \cite{PTP2}.
We showed that two point functions to order $1/N$
are expressed explicitly as $(yy')^{\zeta}\lambda(\upsilon)$,
where $\lambda(\upsilon)$ is only a function of
a conformally invariant variable \cite{Cardy,Mirror,Dynamic}
\begin{align}
\upsilon =
\frac{|\mbox{\boldmath $r$}-\mbox{\boldmath $r$}'|^2}{2yy'} + 1
= \frac{\rho^2+ (y-y')^2}{2yy'} + 1
= \frac{\rho^2 + y^2 + y'^2}{2yy'}
\label{v}
\end{align}
with $\rho=|\mbox{\boldmath $x$}-\mbox{\boldmath $x$}'|$.
This variable $\upsilon$ is convenient as it appears in the argument
of the associated Legendre function; see later Sections.
To perform the $1/N$ expansion, it is always necessary to invert a single bubble.
For this purpose,
we utilized the Fourier-Bessel integration theorem or the Hankel theorem
in mixed space, where $\rho$ is converted to $q$;
see Section \ref{Large N}, and also Appendices \ref{Differential} and
\ref{InverseSpecial}.
We explored two-point functions in real space
by solving differential equations in terms of $\upsilon$
as explained in detail in Section \ref{1/N}
and Appendices \ref{Ordinary} and \ref{Special}.
Later, McAvity and Osborn \cite{McAvity} proposed a quite different approach.
They integrated the two-point functions with respect to $\rho$
to eliminate the $\rho$ dependence completely
by assuming that they depend on a similar conformal invariant variable only.
The results are the functions of only $(y-y')^2/4yy'$,
which were Fourier transformed with respect to $y$ to take inverse.
They also proposed the boundary operator expansion
in terms of the hypergeometric functions of the conformal invariant variable.
Recently, Giombi and Khanchandani \cite{Giombi} derived
a series expansion for the $O(1/N)$ term
of the anomalous dimension (critical exponent)
corresponding to the ordinary transition by extending
the boundary operator expansion by McAvity and Osborn \cite{McAvity}.
Although they were unable to sum the series or simplify its form,
they demonstrated numerically that their result
matches our simple analytic expression derived in refs. \cite{PLA1,PTP1}. 

In Section \ref{Anomalous} of the present paper,
we prove that Giombi and Khanchandani's series expansion
is analytically equivalent to our simple expression.
However, there is an error in the final formula in eq. (4.57) of their paper,
which involves two ${}_3F_2$ functions and a constant term.
Although we could not pinpoint the exact location of the error,
we derive somewhat similar formulae involving two ${}_3F_2$ functions
in the Appendices \ref{Appendix A} and \ref{Appendix B}.
Giombi and Khanchandani also mentioned in their paper \cite{Giombi}
that they could not get to the final result in the case of the special transition
and that the transverse field should not have anomalous dimension
in the case of the extraordinary transition.
We also make some comments on their analysis.
We give detailed derivations of the inverse of a single bubble
and the $1/N$ correction to the correlation function, respectively,
in Sections \ref{Large N} and \ref{1/N};
see also Appendices \ref{Differential} and \ref{InverseSpecial}.
Sections \ref{Large N} and \ref{1/N} contain also
the derivation of the anomalous dimensions for the extraordinary transition.
Full details of calculating the $O(1/N)$ correction
to the anomalous dimension are given in Appendices \ref{Ordinary}
and \ref{Special}, respectively, for the ordinary and special transitions,
which were not written in our earlier papers \cite{PLA1,PLA2,PTP1}.

\section{Anomalous Dimension for the Ordinary Transition}
\label{Anomalous}

According to Metlitski \cite{Metlitski}, the bulk anomalous dimension,
$\Delta_{\phi}=(d-2+\eta)/2$,
is introduced to the correlation function in semi-infinite space as
\begin{align}
\langle\phi(\mbox{\boldmath $r$})\phi(\mbox{\boldmath $r$}')\rangle
&= G_{\rho}(y,y')
%= \frac{f(\upsilon)}{(yy')^{(d-2+\eta)/2}}
= \frac{f(\upsilon)}{(yy')^{\Delta_{\phi}}},
\end{align}
where $f(\upsilon)$ is a function of only the conformal variable $\upsilon$
introduced in eq.~(\ref{v}) and has the asymptotic behaviors like
\begin{align}
&f(\upsilon) \;\;
\begin{cases}
\;\;\; \xrightarrow[\upsilon\rightarrow 1]{}\;
 (\upsilon-1)^{-\Delta_{\phi}}, \\
\;\;\; \xrightarrow[\upsilon\rightarrow\infty]{}\;
 \upsilon^{-\hat{\Delta}_{\phi}},
\end{cases}
\label{f=1/v^D}
\end{align}
where $\hat{\Delta}_{\phi}$ is the boundary anomalous dimension
\begin{align}
\hat{\Delta}_{\phi} = \frac{d - 2 + \eta_{_{/\!/}}}{2}
\label{Delta}
%\\
%\color{Lavender}
%\eta_{_{/\!/}}^O = d-2+2A\left(\frac{1}{N}\right), \;\;\;\;
%\eta_{_{/\!/}}^S &
%\color{Lavender}
%= d-4+2B\left(\frac{1}{N}\right), \;\;\;\;
%\eta_{_{/\!/}}^{E,T} = d+2C\left(\frac{1}{N}\right)
%\;\;\;\; \leftarrow \textrm{ピンクは後で消す}
%\nonumber\\
%\color{Lavender}
%\hat{\Delta}_{\phi}^O = d-2+A\left(\frac{1}{N}\right), \;\;\;\;
%\hat{\Delta}_{\phi}^S &
%\color{Lavender}= d-3+B\left(\frac{1}{N}\right), \;\;\;\;
%\hat{\Delta}_{\phi}^{E,T} = d-1+C\left(\frac{1}{N}\right)
%\;\;\;\; \leftarrow \textrm{ピンクは後で消す}
%\nonumber
\end{align}
related to the surface critical exponent $\eta_{/\!/}$.
In the large $N$ limit, $\hat{\Delta}_{\phi}$
for the ordinary transition is expressed as \cite{BrayMoore,PLA1,PTP1}
\begin{align}
%\color{Lavender}
%\hat{\hat{\Delta}}_{\phi}^O = \frac{\eta_{_{/\!/}}}{2} &
%\color{Lavender}
%= \frac{d}{2} - 1 + \frac{\hat{\gamma}^O}{N} \;\;\;\;
%\leftarrow \textrm{この式は間違い（ピンクは後で消す）}
%\nonumber\\
\hat{\Delta}_{\phi}^O &
= d - 2 + \frac{\hat{\gamma}^O}{N} + O\left(\frac{1}{N^2}\right).
\label{DeltaO}
\end{align}
%\begin{align}
%\hat{\gamma}^O = \frac{(4-d)\Gamma(2d-3}{Nd\Gamma(d-2)\Gamma(d-1)}.
%\color{Lavender} \;\;\;\;
%\leftarrow \textrm{ピンクは後で消す}
%\nonumber
%\end{align}
%which is also written in Giombi and Khanchandani's paper as eq. (4.58).
Giombi and Khanchandani \cite{Giombi}
showed that the boundary operator expansion \cite{McAvity}
contains a tower of operators in the case of the ordinary transition, and
derived a series expansion for $\hat{\gamma}^O$
as [eq. (4.56) in their paper]
\begin{align}
\frac{\Gamma\left(1-\frac{\textstyle d}{\textstyle2}\right)\Gamma(d-2)}{\Gamma\left(\frac{\textstyle d}{\textstyle2}-1\right)}\hat{\gamma}^O
+ \sum_{k=0}^{\infty} \left(\mu_k^O\right)^2
\frac{\Gamma\left(1-\frac{\textstyle d}{\textstyle2}\right)\Gamma(3d-2+4k)}
{\Gamma(d+2k)\Gamma\left(\frac{\textstyle3d}{\textstyle2}-1+2k\right)}
 = 0.
\label{eq4-56}
\end{align}
Here, as written in their eq. (4.53), $\left(\mu_k^O\right)^2$ is given by
\begin{align}
\left(\mu_k^O\right)^2
= \frac{2^{-d-4k+2}\sin\left(\frac{\textstyle\pi d}{\textstyle2}\right)
\Gamma\left(\frac{\textstyle d-1}{\textstyle2}\right)
\Gamma\left(\frac{\textstyle3(d-1)}{\textstyle2}+k\right)
\Gamma\left(\frac{\textstyle d}{\textstyle2}+k\right)\Gamma(d+2k)}
{\pi d(d+2k)(2d+2k-3)\Gamma\left(\frac{\textstyle d}{\textstyle2}-2\right)
\Gamma\left(\frac{\textstyle d}{\textstyle2}\right)\Gamma(k+1)
\Gamma\left(d+k-\frac{\textstyle1}{\textstyle2}\right)
\Gamma\left(\frac{\textstyle3(d-1}{\textstyle2}+2k\right)},
\label{eq4-53}
\end{align}
where we removed $N$ from the denominator to fit the definition
of $\hat{\gamma}^O$ in eq.~(\ref{DeltaO}).
Although Giombi and Khanchandani \cite{Giombi} were unable to sum this series,
they derived eq.~(4.57) in their paper as the final result, 
which involves two hypergeometric functions ${}_3F_2$ and a constant term.
They claimed that their result numerically matches our simple analytic
form \cite{PLA1,PTP1}
\begin{align}
\hat{\gamma}^O = \frac{(4-d)\Gamma(2d-3)}{d\Gamma(d-2)\Gamma(d-1)}
\;\;\;\; \textrm{for $2<d<4$},
\label{Eq_O}
\end{align}
which is explicitly derived in Appendix \ref{Ordinary}.
However, we could neither derive their final formula, eq.~(4.57),
nor numerically confirm this formula,
although we could confirm the numerical equivalence between
their eq.~(\ref{eq4-56}) with eq.~(\ref{eq4-53}) and
our eq.~(\ref{Eq_O}); see below.
Our result, eq.~(\ref{Eq_O}), is consistent with
the $\varepsilon=4-d$ expansion up to $\varepsilon^2$
by Reeve and Guttmann \cite{ReeveGuttmann},
and Diehl and Dietrich \cite{DiehlDietrich},
\begin{align}
%\hat{\Delta}_{\phi} = 2 - \e + \frac{2}{N}
\hat{\gamma}^O = 3\varepsilon - \frac{17}{4}\varepsilon^2
+ {\cal O}\left(\varepsilon^3\right),
\label{4-d}
\end{align}
and also with the $\varepsilon=d-2$ expansion up to $\varepsilon^2$
by Diehl and N\"{u}sser \cite{DiehlNusser},
\begin{align}
%\hat{\Delta}_{\phi} = \e + \frac{2}{N}
\hat{\gamma}^O = \varepsilon - \varepsilon^2
+ {\cal O}\left(\varepsilon^3\right).
\label{d-2}
\end{align}
\begin{figure}[b]
\begin{center}
\includegraphics[width=150mm]{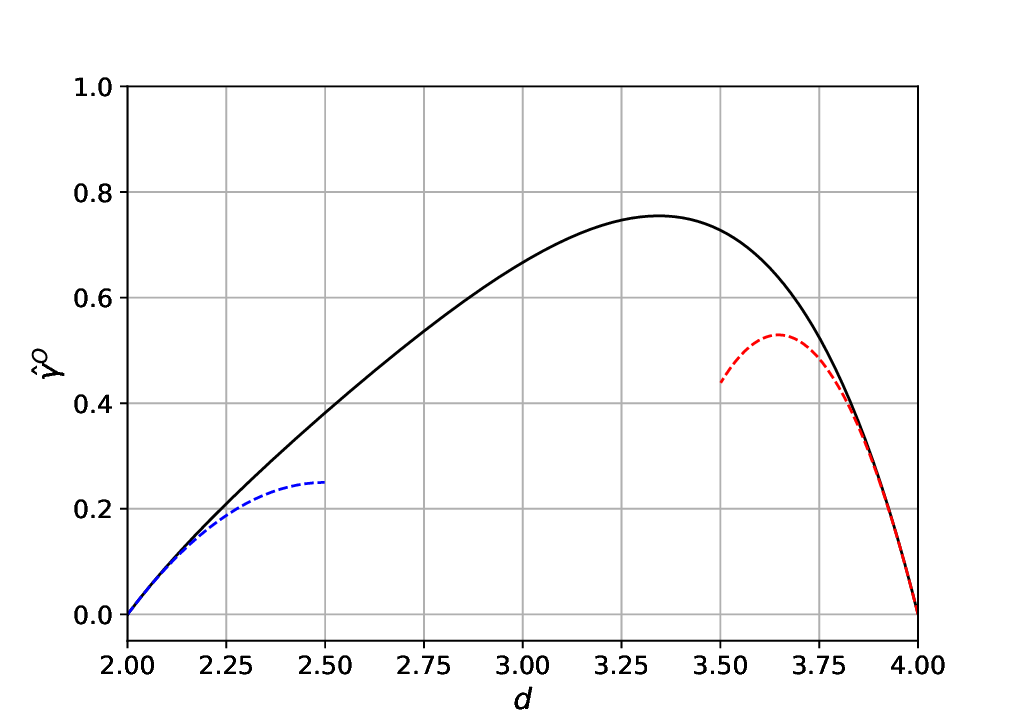}
\caption{Plots of $\hat{\gamma}^O$ versus the space dimension $d$
in the case of the ordinary transition.
Black solid curve is eq.~(\ref{Eq_O}) of the $1/N$ expansion \cite{PLA1,PTP1},
red dashed curve near $d=4$ is eq.~(\ref{4-d}) of the $\epsilon=4-d$
expansion \cite{ReeveGuttmann,DiehlDietrich},
and blue dashed curve near $d=2$ is eq.~(\ref{d-2}) of the $\epsilon=d-2$
expansion \cite{DiehlNusser}.}
\label{gamma_O}
\end{center}
\end{figure}
Equations (\ref{Eq_O}), (\ref{4-d}), and (\ref{d-2}) are plotted versus $d$
in Fig.~\ref{gamma_O}. Our result (\ref{Eq_O}) is smooth around $d=3$
and matches eqs.~(\ref{4-d}) and (\ref{d-2}) near $d=4$ and $d=2$,
respectively; for example, the slopes near $d=4$ and $d=2$ are
exactly equal to 6 and 2, respectively, and all three curves are
concave down.
It is also possible to compare our result, eq.~(\ref{Eq_O}),
with the $\epsilon$ expansion and the high-temperature expansion
(for arbitrary $N$) \cite{HTE}
as a function of $1/N$ as plotted in Fig. 6(a) of ref.~\cite{HTE}.

By using $\Gamma(x)\Gamma(1-x)=\pi/\sin(\pi x)$ at $x=d/2$,
eq. (\ref{eq4-56}) with eq. (\ref{eq4-53}) can be rewritten as
\begin{align}
\hat{\gamma}^O = \frac{(4-d)}{d\Gamma(d-2)}
\frac{2^{-d+1}\Gamma\left(\frac{\textstyle d-1}{\textstyle2}\right)}
{\Gamma\left(\frac{\textstyle d}{\textstyle2}\right)
\Gamma\left(\frac{\textstyle d}{\textstyle2}\right)
\Gamma\left(1-\frac{\textstyle d}{\textstyle2}\right)}\; R
\label{gammaO1}
\end{align}
with
\begin{align}
R =  \sum_{k=0}^{\infty}
\frac{2^{-4k}\Gamma\left(\frac{\textstyle3(d-1)}{\textstyle2}+k\right)
\Gamma\left(\frac{\textstyle d}{\textstyle2}+k\right)\Gamma(3d-2+4k)}
{(d+2k)(2d+2k-3)\Gamma\left(d+k-\frac{\textstyle1}{\textstyle2}\right)
\Gamma\left(\frac{\textstyle3(d-1)}{\textstyle2}+2k\right)
\Gamma\left(\frac{\textstyle3d}{\textstyle2}-1+2k\right)}\frac{1}{k!}.
\label{R1}
\end{align}
Using
\begin{align}
\Gamma(2x)
 = \frac{2^{2x}}{2\sqrt{\pi}}\Gamma(x)\Gamma\left(x+\frac{1}{2}\right),
\label{Gam(2x)}
\end{align}
we can rewrite $\Gamma(3d-2+4k)$ in eq.~(\ref{R1}) as
\begin{align}
\Gamma(3d-2+4k)
&= \frac{2^{3d-2+4k}}{2\sqrt{\pi}}\Gamma\left(\frac{3d}{2}-1+2k\right)
\Gamma\left(\frac{3d}{2}-\frac{1}{2}+2k\right)
\nonumber\\
&=  \frac{2^{3d-2+4k}}{2\sqrt{\pi}}\Gamma\left(\frac{3d}{2}-1+2k\right)
\left(\frac{3(d-1)}{2}+2k\right)\Gamma\left(\frac{3(d-1)}{2}+2k\right).
\label{Gam(3d-2+4k)}
\end{align}
Then, we have
\begin{align}
R = \frac{2^{3d-3}}{\sqrt{\pi}} \sum_{k=0}^{\infty}
\frac{\left(\frac{\textstyle3(d-1)}{\textstyle2}+2k\right)
\Gamma\left(\frac{\textstyle3(d-1)}{\textstyle2}+k\right)
\Gamma\left(\frac{\textstyle d}{\textstyle2}+k\right)}
{(d+2k)(2d-3+2k)\Gamma\left(d-\frac{\textstyle1}{\textstyle2}+k\right)}\frac{1}{k!},
\label{R2}
\end{align}
which can be rewritten as
\begin{subequations}
\begin{align}
R &= \frac{2^{3d-4}}{\sqrt{\pi}} \sum_{k=0}^{\infty}
\frac{\left(\frac{\textstyle3(d-1)}{\textstyle4}+k\right)
\Gamma\left(\frac{\textstyle3(d-1)}{\textstyle2}+k\right)
\Gamma\left(\frac{\textstyle d}{\textstyle2}+k\right)}
{\left(\frac{\textstyle d}{\textstyle2}+k\right)\left(d-\frac{\textstyle3}{\textstyle2}+k\right)
\Gamma\left(d-\frac{\textstyle1}{\textstyle2}+k\right)}\frac{1}{k!}
\nonumber\\
&= \frac{2^{3d-4}}{\sqrt{\pi}} \sum_{k=0}^{\infty}
\frac{\Gamma\left(\frac{\textstyle3d+1}{\textstyle4}+k\right)
\Gamma\left(\frac{\textstyle d}{\textstyle2}+k\right)
\Gamma\left(d-\frac{\textstyle3}{\textstyle2}+k\right)
\Gamma\left(\frac{\textstyle3(d-1)}{\textstyle2}+k\right)
\Gamma\left(\frac{\textstyle d}{\textstyle2}+k\right)}
{\Gamma\left(\frac{\textstyle d}{\textstyle2}+1+k\right)
\Gamma\left(d-\frac{\textstyle1}{\textstyle2}+k\right)
\Gamma\left(\frac{\textstyle3(d-1)}{\textstyle4}+k\right)
\Gamma\left(d-\frac{\textstyle1}{\textstyle2}+k\right)}\frac{1}{k!}
\nonumber\\
&= \frac{2^{3d-6}}{\sqrt{\pi}}
\frac{3(d-1)\Gamma\left(\frac{\textstyle d}{\textstyle2}\right)
\Gamma\left(d-\frac{\textstyle3}{\textstyle2}\right)
\Gamma\left(\frac{\textstyle3(d-1)}{\textstyle2}\right)
\Gamma\left(\frac{\textstyle d}{\textstyle2}\right)}
{\Gamma\left(\frac{\textstyle d}{\textstyle2}+1\right)
\Gamma\left(d-\frac{\textstyle1}{\textstyle2}\right)
\Gamma\left(d-\frac{\textstyle1}{\textstyle2}\right)}\; A
\label{R3}
\end{align}
with the generalized hypergeometric function (GHF)
\begin{align}
A= {}_5F_4\left(\frac{3d+1}{4},\,\frac{d}{2},\,d-\frac{3}{2},\frac{3(d-1)}{2},\,\frac{d}{2};\,
\frac{d}{2}+1,d-\frac{1}{2},\frac{3(d-1)}{4},d-\frac{1}{2};\,1\right).
\label{A_O}
\end{align}
\label{R_A_O}
\end{subequations}
\begin{table}[t]
\begin{center}
\caption{Numerical values of eq.~(\ref{Eq_O}),
eq.~(\ref{gammaO1}) with (\ref{R_A_O}),
and eq. (4.57) of ref.~\cite{Giombi}
calculated with python (mpmath).}
%\begin{threeparttable}
{\tabcolsep = 2mm
\begin{tabular}{crrr}\hline\hline
$d$ & eq.~(\ref{Eq_O}) \hspace{10mm} &
 eq.~(\ref{gammaO1}) with (\ref{R_A_O})
& eq. (4.57) of ref.~\cite{Giombi} \\ \hline\hline
3.9999 & 0.000299957502873 & 0.000299957502871 & 0.000299967345168 \\ \hline
%3.9000 & 0.260247102635293 & 0.260247102635292 & 0.26880023357305 \\ \hline
3.8500 & 0.363445472579003 & 0.363445472579002 & 0.381483580425512 \\ \hline
%3.8000 & 0.451039326784964 & 0.451039326784962 & 0.481212111999457 \\ \hline
3.7000 & 0.585543391341222 & 0.585543391341222 & 0.646398317010116 \\ \hline
%3.6000 & 0.674698713429900 & 0.674698713429900 & 0.773570393059822 \\ \hline
3.5000 & 0.727565454134379 & 0.727565454134378 & 0.872343408980861 \\ \hline
%3.4000 & 0.751619359170797 & 0.751619359170797 & 0.95414756802933 \\ \hline
3.3000 & 0.753000980328117 & 0.753000980328116 & $1.03669228023846\;\;$ \\ \hline
%3.2000 & 0.736725672371363 & 0.736725672371363 & 1.16182421756964 \\ \hline
3.1000 & 0.706859685601035 & 0.706859685601035 & $1.52022234075515\;\;$ \\ \hline
3.0001 & 0.666711107262146 & 0.666711107262146 & $748.480756311349\;\;\;\;\;\,$ \\ \hline
2.9999 & 0.666622218374012 & 0.666622218374012 & $-746.992471963552\;\;\;\;\;\,$ \\ \hline
2.9000 & 0.618727871474671 & 0.618727871474671 & $-0.042272289195445$ \\ \hline
%2.8000 & 0.565038277510834 & 0.565038277510834 & 0.284619075115068 \\ \hline
2.7000 & 0.507079372079706 & 0.507079372079706 & 0.355547192130489 \\ \hline
%2.6000 & 0.445867331602277 & 0.445867331602277 & 0.358606001360936 \\ \hline
2.5000 & 0.381971863420549 & 0.381971863420549 & 0.331928816283681 \\ \hline
%2.4000 & 0.315494545824779 & 0.315494545824779 & 0.288282780052966 \\ \hline
2.3000 & 0.245982269000819 & 0.245982269000820 & 0.232782572553323 \\ \hline
%2.2000 & 0.172221585749150 & 0.172221585749150 & 0.167137079244346 \\ \hline
2.1500 & 0.133058741866088 & 0.133058741866088 & 0.130398919874215 \\ \hline
%2.1000 & 0.091785792508600 & 0.091785792508600 & 0.090689182952722 \\ \hline
2.0001 & 0.000099990002145 & 0.000099990002145 & 0.000099989072924 \\
\hline\hline
%--------------------------------------------------------------------------------
\end{tabular}
}
%\end{threeparttable}
\label{Comparison}
\end{center}
\end{table}
%%%%%%%%%%%%%%%%%%%%%%%%%%%
In Table~\ref{Comparison},
we numerically confirmed that
eq.~(\ref{gammaO1}) with eqs. (\ref{R3}) and (\ref{A_O})
matches our simple form, eq. (\ref{Eq_O}), up to the 14 decimal places
after the decimal point for all $2<d<4$ by using python (mpmath),
where mpmath is the mathematical library to compute the hypergeometric function.
From this table, it is clear that
the slope near $d=4$ (value at $d=3.9999$) is three times larger than
the slope near $d=2$ (value at $d=2.0001$).
Although Giombi and Khanchandani \cite{Giombi} presented
a different formula in their eq. (4.57) as the final result,
which involves two ${}_3F_2$ functions and a constant term,
this formula likely contains some misprints.
In fact, it produces rather unexpected values, as observed in Table~\ref{Comparison}.
In particular, it is singular at $d=3$ in contrast to the smooth behavior
in Fig.~\ref{gamma_O}.
This singularity in their expression obviously comes from $1/\sin(\pi d)$ in the factor
of the hypergeometric functions and $\Gamma(3/2-d/2)$ in the constant term,
which do not cancel at $d=3$.
Although we could not pinpoint the exact location of the error,
somewhat similar formulae involving two ${}_3F_2$ functions
can be derived from eq.~(\ref{R2}).
At least two such expressions are explicitly derived
in Appendices \ref{Appendix A} and \ref{Appendix B}.

Now, we notice the mathematical formula,
eq. (1) of Section 4.4 (p.27) of ref. \cite{HGS},
\begin{align}
&{}_5F_4\left(\alpha,1+\frac{\alpha}{2},\,\gamma,\,\delta,\,\varepsilon;\,
\frac{\alpha}{2},1+\alpha-\gamma,1+\alpha-\delta,1+\alpha-\varepsilon;\,1\right)
\nonumber\\
= &\frac{\Gamma(1+\alpha-\gamma)\Gamma(1+\alpha-\delta)
\Gamma(1+\alpha-\varepsilon)\Gamma(1+\alpha-\gamma-\delta-\varepsilon)}
{\Gamma(1+\alpha)\Gamma(1+\alpha-\delta-\varepsilon)
\Gamma(1+\alpha-\gamma-\varepsilon)\Gamma(1+\alpha-\gamma-\delta)}.
\label{5F4}
\end{align}
Putting $\alpha=3(d-1)/2$, $1+\alpha/2=(3d+1)/4$, $\gamma=\delta=d/2$,
$\varepsilon=d-3/2$,
we have $1+\alpha=3d/2-1/2$, $\alpha/2=3(d-1)/4$,
$1+\alpha-\gamma=1+\alpha-\delta=d-1/2$,
and $1+\alpha-\varepsilon=d/2+1$, which perfectly matches the GHF in eq.~(\ref{A_O}).
Therefore, eq.~(\ref{A_O}) can be simplified as
\begin{align}
A = \frac{\Gamma\left(d-\frac{\textstyle1}{\textstyle2}\right)
\Gamma\left(d-\frac{\textstyle1}{\textstyle2}\right)
\Gamma\left(\frac{\textstyle d}{\textstyle2}+1\right)
\Gamma\left(1-\frac{\textstyle d}{\textstyle2}\right)}
{\Gamma\left(\frac{\textstyle3d}{\textstyle2}\right)
\Gamma\left(\frac{\textstyle d-1}{\textstyle2}\right)}.
\label{A2}
\end{align}
Inserting this into eq.~(\ref{R3}),
we have
\begin{align}
R = \frac{2^{3d-6}}{\sqrt{\pi}}3(d-1)\Gamma\left(\frac{d}{2}\right)
\Gamma\left(d-\frac{3}{2}\right)
\Gamma\left(\frac{3(d-1)}{2}\right)\Gamma\left(\frac{d}{2}\right)
\frac{\Gamma\left(1-\frac{\textstyle d}{\textstyle2}\right)}
{\Gamma\left(\frac{\textstyle3d}{\textstyle2}-\frac{\textstyle1}{\textstyle2}\right)
\Gamma\left(\frac{\textstyle d}{\textstyle2}-\frac{\textstyle1}{\textstyle2}\right)}.
\label{R4}
\end{align}
Thus, from eq.~(\ref{gammaO1}), we have
\begin{align}
\hat{\gamma}^O = \frac{(4-d)2^{-d+1}}{d\Gamma(d-2)}
\frac{2^{3d-5}}{\sqrt{\pi}}\Gamma\left(d-\frac{3}{2}\right).
\label{gammaO2}
\end{align}
Using again eq.~(\ref{Gam(2x)}) with $x=d-3/2$, we finally obtain
eq.~(\ref{Eq_O}).

\section{Large \textit{N} Limit and Bubble Summation}
\label{Large N}

In the large $N$ limit, the correlation function can be expressed as
\begin{align}
G^{(0)}_{\rho}(y,y') = C_f (yy')^{1-d/2} g^{(0)}(\upsilon),
\label{G^0}
\end{align}
%$G^{(0)}_{\rho}(y,y')=(yy')^{1-d/2}g^{(0)}(\upsilon)$
where the coefficient $C_f$ is defined in eq.~(\ref{C_f}) below.
Equation~(\ref{G^0}) satisfies
\begin{align}
{\cal L}\; G^{(0)}_{\rho}(y,y') =
\delta({\mbox{\boldmath $r$}} - {\mbox{\boldmath $r$}}')
\label{LG=d}
\end{align}
with the differential operator \cite{BrayMoore,PTP1,PTP2}
\begin{align}
{\cal L} =\; -\, \nabla_{\rho}^2 - \frac{\partial^2}{\partial y^2}
 + \frac{\mu^2-1/4}{y^2}.
\label{L}
\end{align}
In addition, in the case of the extraordinary transition,
the magnetization profile satisfies \cite{BrayMoore,PTP2}
\begin{align}
\left[\, -\; \frac{\partial^2}{\partial y^2} + \frac{\mu^2-1/4}{y^2}\, \right] M^0(y) = 0;
\label{LM^0=0}
\end{align}
Substitution of the expected dimensional dependence,
$M^0(y)\propto 1/y^{(d-2)/2}$, into eq.~(\ref{LM^0=0})
yields $\mu=(d-1)/2$. In the other cases, $M^0(y)=0$.
In mixed space, where $\rho$ is Fourier transformed to $q$,
the correlation function
(for the $N-1$ transverse fields in the case of the extraordinary transition)
is given by \cite{BrayMoore,PLA1,PLA2,PTP1,PTP2}
\begin{align}
G_q^{(0)}(y,y') &=
\sqrt{yy'} \int_0^{\infty} dt\,t\,J_{\mu}(ty)J_{\mu}(ty')\,(t^2+q^2)^{-1}
\nonumber\\
&=\; \sqrt{yy'}\,
\begin{cases}
\; I_{\mu}(qy)K_{\mu}(qy') & \text{for $y<y'$,} \\
\; K_{\mu}(qy)I_{\mu}(qy') & \text{for $y>y'$},
\end{cases}
\label{JJ}
\end{align}
which coincides with the bulk correlation function
$\exp(-q|y-y'|)/2q$ in the limit $y,y'\rightarrow\infty$.
%
%Bulk correlation function is
%\begin{align}
%G^{(0)}_q(k) = \frac{1}{t + k^2 + q^2}
%\end{align}
%Fourier transform in the $y$ direction perpendicular to the surface,
%\begin{align}
%G^{(0)}_q(y-y') &= \int_{-\infty}^{\infty} \frac{dq}{2\pi} \frac{e^{iq(y-y')}}{t + k^2 + q^2}
%nonumber\\
%&= \int_{-\infty}^{\infty} \frac{dk}{2\pi} \frac{e^{ik(y-y')}}{(k-i\kappa)(k+i\kappa)}
%\nonumber\\
%&= \frac{2\pi i}{2\pi} \frac{1}{2i\kappa}e^{-\kappa|y-y'|}
%=  \frac{1}{2\kappa}e^{-\kappa|y-y'|},
%\end{align}
%where we put $\kappa=\sqrt{t+q^2}$.
%
Moreover, the saddle point condition \cite{PTP1,PTP2}
\begin{align}
\lim_{\rho\rightarrow 0}
\left[ G^{(0)}_{\rho}(y,y) - \lim_{y\rightarrow\infty}G^{(0)}_{\rho}(y,y) \right]
+ \left[ M^0(y) \right]^2 = 0
\label{saddle}
\end{align}
should be satisfied.
The related integral is evaluated in the Appendix of Bray and Moore's
paper \cite{BrayMoore}.
Equation~(\ref{saddle}) then becomes
\begin{align}
%&= K_{d-1}\int_0^{\infty}dq\,q^{d-2}\left[ y I_{\mu}(qy)K_{\mu}(qy) - \frac{1}{2q} \right]
%+ \left[ M^0(y) \right]^2
%\nonumber\\
%&= 
\frac{K_{d-1}}{y^{d-2}} \frac{\Gamma(2-d)
\Gamma\left(\frac{\textstyle d-1}{\textstyle2}+\mu\right)
\Gamma\left(\frac{\textstyle d-1}{\textstyle2}\right)}
{2^{3-d}\Gamma\left(\frac{\textstyle3-d}{\textstyle2}\right)
\Gamma\left(\frac{\textstyle3-d}{\textstyle2}+\mu\right)}
+ \left[ M^0(y) \right]^2 = 0,
\label{saddle'}
\end{align}
where
\begin{align}
 K_{d-1} =\left[\,2^{d-2}\pi^{(d-1)/2}
 \Gamma\left(\frac{\textstyle d-1}{\textstyle2}\right)\,\right]^{-1}
\label{K_d-1}
\end{align}
is the surface area of unit ($d-1$)-dimensional sphere divided by $(2\pi)^{d-1}$.
From this condition, the parameter $\mu$ is identified to be
$\mu=(d-3)/2$ for the ordinary (or anisotropic special) transition
and $\mu=(d-5)/2$ for the special transition,
while $\left[ M^0(y) \right]^2=A/y^{d-2}$ is identified for the extraordinary transition;
see eq.~(\ref{A}) below.
The condition (\ref{saddle}) is necessary also to retrieve
the potential $(\mu^2-1/4)/y^2$ in eqs.~(\ref{L}) and (\ref{LM^0=0}) self-consistently;
see ref.~\cite{BrayMoore} for detail.

Equation~(\ref{JJ}) can be inversely Fourier transformed into real space
using the integral formulae (6.578.11) and (6.578.16) in ref.~\cite{Gradshteyn};
see also eqs.~(\ref{G^0_int}) and (\ref{Fourier}) in Appendix \ref{Differential}.
Here, it is convenient to define the coefficient $C_f$ in eq.~(\ref{G^0}) as
\begin{align}
C_f = \frac{1}{2}K_{d-1}\Gamma(d-2)
%= \frac{\Gamma(d-2)}
%{2^{d-1}\pi^{(d-1)/2}\Gamma\left(\frac{\textstyle d-1}{\textstyle2}\right)}
= \frac{\Gamma\left(\frac{\textstyle d}{\textstyle2}-1\right)}
{4\pi^{\,d/2}}.
\label{C_f}
\end{align}
Then, the explicit forms of $g^{(0)}(\upsilon)$ are given as follows:
\begin{subequations}
\begin{align}
g^{(0)}(\upsilon) &=
\frac{\Gamma\left(2-\frac{\textstyle d}{\textstyle2}\right)}{2^{d/2+1}}
\frac{P_{-\mu-1/2}^{\,d/2-1}(\upsilon)}{(\upsilon^2-1)^{(d-2)/4}}
= \frac{1}{(\upsilon^2 - 1)^{d/2-1}},
 \;\;\;\; \mu=(d-3)/2
\label{g_O}
\end{align}
for the ordinary (or anisotropic special) transition and
\begin{align}
g^{(0)}(\upsilon) &=
\frac{\Gamma\left(2-\frac{\textstyle d}{\textstyle2}\right)}{2^{d/2+1}}
\frac{P_{-\mu-1/2}^{\,d/2-1}(\upsilon)}{(\upsilon^2-1)^{(d-2)/4}}
= \frac{\upsilon}{(\upsilon^2 - 1)^{d/2-1}},
 \;\;\;\; \mu=(d-5)/2
\label{g_S}
\end{align}
for the special transition,
where $P_{-\mu-1/2}^{\,d/2-1}(\upsilon)=P_{\mu-1/2}^{\,d/2-1}(\upsilon)$
is the associated Legendre function of the first kind.
For the transverse correlation function in the case of the extraordinary transition,
\begin{align}
g^{(0)}(\upsilon) &=
% \frac{\Gamma(d-1)}{2(d-1)} K_{d-1}
\frac{2^{(d-1)/2}}{\Gamma(d-2)}
\Gamma\left(\frac{d-1}{2}\right)
\frac{e^{-i\pi(d-2)/2}}{\sqrt{2\pi}}
\frac{Q_{\mu-1/2}^{d/2-1}(\upsilon)}{(\upsilon^2-1)^{(d-2)/4}}
%\nonumber\\
%&= 2^{(d-3)/2}\left(\frac{d-2}{d-1}\right)\frac{2(d-1)}{\Gamma(d-1)}
%\Gamma\left(\frac{d-1}{2}\right)
%\frac{e^{-\pi i(d/2-1)}}{\sqrt{2\pi}}
%\frac{Q_{d/2-1}^{d/2-1}(\upsilon)}{(\upsilon^2-1)^{(d-2)/4}}
%\nonumber\\
%&= \left(\frac{d-2}{d-1}\right)
%\frac{1}{\upsilon^{d-1}}
%F\left(\frac{d-1}{2},\,\frac{d}{2},\,\frac{d+1}{2};\,\frac{1}{\upsilon^2}\right)
%\nonumber\\
%&\rightarrow \frac{1}{2}K_{d-1}\Gamma(d-2)
%\frac{2^{(d-1)/2}}{\Gamma(d-2)}
%\Gamma\left(\frac{d-1}{2}\right)
%\frac{e^{-\pi i(d/2-1)}}{\sqrt{2\pi}}
%\frac{Q_{d/2-1}^{d/2-1}(\upsilon)}{(\upsilon^2-1)^{(d-2)/4}}
%\nonumber\\
%&= 2^{(d-3)/2}\Gamma\left(\frac{d-1}{2}\right)K_{d-1}
%\frac{e^{-\pi i(d/2-1)}}{\sqrt{2\pi}}
%\frac{Q_{d/2-1}^{d/2-1}(\upsilon)}{(\upsilon^2-1)^{(d-2)/4}},
\nonumber\\
&= \frac{d-2}{2\mu}
\frac{1}{\upsilon^{2\mu}}
F\left(\mu,\,\frac{1}{2}+\mu,\,1+\mu;\,\frac{1}{\upsilon^2}\right),
 \;\;\;\; \mu=(d-1)/2
\label{g_E}
\end{align}
\end{subequations}
where $Q_{\mu-1/2}^{d/2-1}(\upsilon)$ is
the associated Legendre function of the second kind.
Comparing the asymptotic behavior of these functions
for $\upsilon\rightarrow\infty$ with eq.~(\ref{f=1/v^D}) yields
$\hat{\Delta}_{\phi}^O$ of eq.~(\ref{DeltaO}) for the ordinary transition,
\begin{align}
\hat{\Delta}_{\phi}^S = d - 3 + \frac{\hat{\gamma}^S}{N} + O\left(\frac{1}{N^2}\right)
\label{DeltaS}
\end{align}
for the special transition, and
\begin{align}
\hat{\Delta}_{\phi}^{T} = d - 1 + \frac{\hat{\gamma}^T}{N} + O\left(\frac{1}{N^2}\right).
\label{DeltaT}
\end{align}
for the transverse fields of the extraordinary transition.
From eqs.~(\ref{g_E}) and (\ref{saddle}), we have \cite{PTP2}
\begin{align}
A %&
= y^{d-2} \left[ M^0(y) \right]^2
%\nonumber\\
%&= -\, K_{d-1}\, \frac{\Gamma(2-d)\Gamma(d-1)
%\Gamma\left(\frac{\textstyle d-1}{\textstyle2}\right)}
%{2^{3-d}\Gamma\left(\frac{\textstyle3-d}{\textstyle2}\right)}
%\nonumber\\
%&= -\, K_{d-1}\,
%\frac{2^{2-d}\Gamma\left(\frac{\textstyle2-d}{\textstyle2}\right)
%\Gamma(d-1)
%\Gamma\left(\frac{\textstyle d-1}{\textstyle2}\right)}
%{2\!\sqrt{\pi}2^{3-d}}
%\nonumber\\
%&= -\, K_{d-1}\,
%\frac{\Gamma(2\mu)\Gamma(\mu)\Gamma\left(\frac{\textstyle1}{\textstyle2}-\mu\right)}
%{4\!\sqrt{\pi}}
%\nonumber\\
%&
%= -\, C_f\, g^{(0)}(1)
%= -\, C_f \left(\frac{d-2}{d-1}\right)
%\frac{\Gamma\left(\frac{\textstyle d+1}{\textstyle2}\right)
%\Gamma\left(\frac{\textstyle2-d}{\textstyle2}\right)}{\sqrt{\pi}},
%= -\, C_f\, \left(\frac{d-2}{d-1}\right)
%\frac{\Gamma(1+\mu)\Gamma\left(\frac{\textstyle1}{\textstyle2}-\mu\right)}
%{\sqrt{\pi}},
= -\, C_f \frac{d-2}{2\mu}
\frac{\Gamma(1+\mu)\left(\frac{\textstyle1}{\textstyle2}-\mu\right)}{\sqrt{\pi}},
\label{A}
\end{align}
which can be derived also from eq.~(\ref{saddle'}).

As explained in Appendix \ref{Differential}, the function $g^{(0)}(\upsilon)$
in eqs.~({\ref{g_O})-(\ref{g_E}) satisfies the differential equation
\begin{align}
{\cal D}\, g^{(0)}(\upsilon) = 0
\label{Dg^(0)=0}
\end{align}
with
\begin{subequations}
\begin{align}
{\cal D} = (\upsilon^2-1) \frac{d^2}{d\upsilon^2}
+ (3+2\mu)\, \upsilon \frac{d}{d\upsilon} + (1+2\mu), \;\;\;\;
\mu=(d-3)/2
\label{D_O}
\end{align}
for the ordinary (or anisotropic special) transition,
\begin{align}
{\cal D} = 
(\upsilon^2-1)\frac{d^2}{d\upsilon^2}+(5+2\mu)\upsilon\frac{d}{d\upsilon}+4(1+\mu),
 \;\;\;\; \mu=(d-5)/2
\label{D_S}
\end{align}
for the special transition, and
\begin{align}
{\cal D} = (\upsilon^2-1)\frac{d^2}{d\upsilon^2}+ (1+2\mu)\upsilon\frac{d}{d\upsilon},
 \;\;\;\; \mu=(d-1)/2
\label{D_E}
\end{align}
for the extraordinary transition.
\end{subequations}

A single bubble in Fig. \ref{Bubble}(a) is expressed as $\left[G^{(0)}_{\rho}(y,y')\right]^2$.
Including the extraordinary transition, it is more generally expressed as
\begin{align}
\Pi_{\rho}(y,y') = \left[G^{(0)}_{\rho}(y,y')\right]^2 + 2M^0(y)M^0(y')G^{(0)}_{\rho}(y,y').
\label{Pi_rho}
\end{align}
It is given by $(yy')^{2-d}(\upsilon^2 - 1)^{2-d}$ and
$(yy')^{2-d}\left[(\upsilon^2 - 1)^{2-d}+(\upsilon^2 - 1)^{3-d}\right]$,
respectively, for the ordinary (or anisotropic special) and the special
transitions, which can be reproduced by the integral formula
(6.578.15) of ref. \cite{Gradshteyn},
\begin{figure}[h]
\begin{center}
\includegraphics[width=105mm]{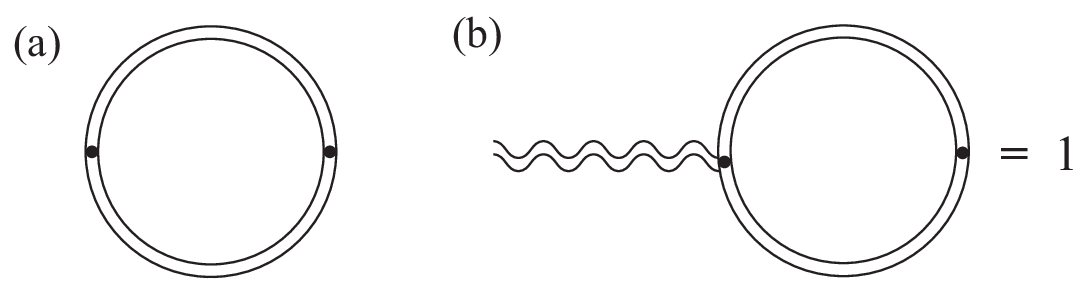}
\caption{A single bubble (a) and the relation to its inverse
(i.e., the bubble summation represented by a double wavy line) (b),
which are relevant in the large $N$ expansion.
Here, a double solid line represents the correlation function in the large $N$ limit.}
\label{Bubble}
\end{center}
\end{figure}
%\begin{align}
%\int_0^{\infty} dt\,t^{1+\alpha}J_{\alpha}(ty)J_{\alpha}(ty')K_{\alpha}(t\rho)
%=\frac{2^{3\alpha}}{\sqrt{\pi}}\Gamma\left(\alpha+\frac{1}{2}\right)
%\frac{(yy'\rho)^{\alpha}}
%{\left[(y^2+y'^2+\rho^2)^2-(2yy')^2\right]^{\alpha+1/2}}
%\label{JJK}
%\end{align}
\begin{align}
\frac{\sqrt{yy'}}{\rho^{\alpha}}
\int_0^{\infty} dt\,t^{1+\alpha}J_{\alpha}(ty)J_{\alpha}(ty')K_{\alpha}(t\rho)
=\Gamma\left(\alpha+\frac{1}{2}\right)
\frac{2^{\alpha-1}}{\sqrt{\pi}\left[yy'(\upsilon^2-1)\right]^{\alpha+1/2}},
\label{JJK}
\end{align}
by putting $\alpha=d-5/2$ and $\alpha=d-7/2$.
The factor $t^{\alpha}K_{\alpha}(t\rho)$ in eq.~(\ref{JJK}) can be
reproduced by the ($d-1$)-dimensional Fourier transform
[see Appendix \ref{Differential}]
of $(t^2+q^2)^{\alpha-(d-1)/2}$ using the integral formula (6.565.4) of ref.~\cite{Gradshteyn} as
%{\color{Bittersweet} $-\mu-1=\alpha-(d-1)/2$, $-\mu=\alpha-(d-3)/2$,
%$\nu-\mu=\alpha-(d-3)/2-((d-3)/2=\alpha$}
%森口他、数学公式集III, p.196 の1段目の公式
%\begin{align}
%\int_0^{\infty} dx\,x^{\nu+1}J_{\nu}(ax)(x^2+y^2)^{-\mu-1}
%= \frac{a^{\mu}y^{\nu-\mu}K_{\nu-\mu}(ay)}{2^{\mu}\Gamma(\mu+1)}
%\end{align}
%
%\begin{align}
%\int_0^{\infty} dq\,q^{(d-1)/2}J_{(d-3)/2}(q\rho)(t^2+q^2)^{-\beta-1}
%= \frac{\rho^{\beta}t^{(d-3)/2-\beta}K_{(d-3)/2-\beta}(t\rho)}{2^{\beta}\Gamma(\beta+1)}
%\end{align}
%$(d-3)/2-\beta=\alpha\;\;\rightarrow\;\;\beta=(d-3)/2-\alpha$
%\begin{align}
%\int_0^{\infty} dq\,q^{(d-1)/2}J_{(d-3)/2}(q\rho)(t^2+q^2)^{\alpha-(d-1)/2}
%= \frac{\rho^{(d-3)/2-\alpha}t^{\alpha}K_{\alpha}(t\rho)}
%{2^{(d-3)/2-\alpha}\Gamma\left(\frac{\textstyle d-3}{\textstyle2}-\alpha+1\right)}
%\end{align}
%
\begin{align}
\frac{C_F}{\rho^{(d-3)/2}}
\int_0^{\infty} dq\,q^{(d-1)/2}J_{(d-3)/2}(q\rho)(t^2+q^2)^{\alpha-(d-1)/2}
=  \frac{C_F}{\rho^{\alpha}}
\frac{t^{\alpha}K_{\alpha}(t\rho)}
{2^{(d-3)/2-\alpha}\Gamma\left(\frac{\textstyle d-1}{\textstyle2}-\alpha\right)},
\label{K_alpha}
\end{align}
where we put the coefficient for the ($d-1$)-dimensional transform as
\begin{align}
C_F = 2^{(d-3)/2}\Gamma\left(\frac{d-1}{2}\right)K_{d-1}.
\label{C_F}
\end{align}
Therefore, $(yy')^{2-d}(\upsilon^2 - 1)^{2-d}$ and $(yy')^{2-d}(\upsilon^2 - 1)^{3-d}$
correspond to $(t^2+q^2)^{d/2-2}$ and $(t^2+q^2)^{d/2-3}$
in the Fourier--Bessel integral representation in mixed space ($\rho\rightarrow q$).
In the case of the ordinary (or anisotropic special)
transition [$\mu=(d-3)/2$],
the single bubble in mixed space is given by \cite{PLA1,PTP1}
\begin{align}
\Pi_q(y,y') = C_{\Pi}\sqrt{yy'}
\int_0^{\infty} dt\,t\,J_{1/2+2\mu}(ty)J_{1/2+2\mu}(ty')(t^2+q^2)^{-1/2+\mu}
\label{Pi}
\end{align}
with
\begin{align}
C_{\Pi} &= \Gamma\left(\frac{\textstyle d}{\textstyle2}-1\right)
\Gamma\left(2-\frac{\textstyle d}{\textstyle2}\right)
 \frac{K_{d-1}}{2^{d-1}}
\; = \frac{\Gamma\left(\frac{\textstyle d}{\textstyle2}-1\right)
\Gamma\left(2-\frac{\textstyle d}{\textstyle2}\right)}
 {2^{d-2}\Gamma(d-2)}\,C_f. \;\;
%C_f = \frac{1}{2}K_{d-1}\Gamma(d-2),} \;\;\;\;
% {\color{Lavender}
% K_{d-1} = \frac{1}{2^{d-2}\pi^{(d-1)/2}
% \Gamma\left(\frac{\textstyle(d-1)}{\textstyle2}\right)},
\label{C_Pi}
\end{align}
In the case of the special transition [$\mu=(d-5)/2$],
it is given by \cite{PLA2,PTP1}
\begin{align}
\Pi_q(y,y') &= C_{\Pi}\sqrt{yy'} \left[
 \int_0^{\infty} dt\,t\,J_{5/2+2\mu}(ty)J_{5/2+2\mu}(ty')(t^2+q^2)^{1/2+\mu}
\right.
\nonumber\\
&\left. - \; \frac{4(1+\mu)(1+2\mu)}{yy'}
\int_0^{\infty} dt\,t\,J_{3/2+2\mu}(ty)J_{3/2+2\mu}(ty')(t^2+q^2)^{-1/2+\mu}
\right]
\label{Pi_special}
\end{align}
with the same $C_{\Pi}$ defined in eq.~(\ref{C_Pi}).

To perform the $1/N$ expansion, we have to invert this single bubble
according to the relation described by Fig.~\ref{Bubble}(b).
To explicitly write down the relation between a bubble and its inverse
of Fig.~\ref{Bubble}(b) in a convenient formula,
we used the Fourier--Bessel integration theorem,
which is known as a Hankel transform,
\begin{align}
\int_0^{\infty} \Pi_q(y,y'')\, \Pi^{-1}_q(y'',y')\, dy'' = 
\sqrt{yy'}\int_0^{\infty} dt\,t\,J_{\alpha}(ty)J_{\alpha}(ty') = \delta(y-y').
\label{Bessel}
\end{align}
Then, its inverse is readily obtained as \cite{PLA1,PTP1}
\begin{align}
\Pi^{-1}_q(y,y') = C^{-1}_{\Pi}\!\sqrt{yy'}
\int_0^{\infty} dt\,t\,J_{1/2+2\mu}(ty)J_{1/2+2\mu}(ty')(t^2+q^2)^{1/2-\mu}
\label{Pi^-1}
\end{align}
in the case of the ordinary transition.
It is possible to transform $\Pi^{-1}_q(y,y')$ back into real space
in a form
\begin{align}
\Pi^{-1}_\rho(y,y')=(yy')^{-2}\nu(\upsilon),
\label{Pi-1=nu}
\end{align}
using eq.~(\ref{K_alpha}) and the integral formula (6.578.7) of ref.~\cite{Gradshteyn},
where $\nu(\upsilon)$ is a function of $\upsilon$ defined in eq.~(\ref{v}).
The result is \cite{PLA1,PTP1}
\begin{align}
\nu(\upsilon)
%&= - \; \frac{2^{2+2\mu}(1-2\mu)\Gamma(\mu)\Gamma(1+2\mu)}
%{\sqrt{\pi}\Gamma(2\mu)\left[\Gamma\left(\frac{\textstyle1}{\textstyle2}+\mu\right)
%\right]^2\Gamma\left(\frac{\textstyle1}{\textstyle2}-\mu\right)}\;
%この結果は係数が2余計：ノート[II],PP.2-3参照、2/nの2を含んでいる
%\nonumber\\
%&
= \frac{C_X}{C_f}\,
 \frac{Q_{2\mu}^2(\upsilon)}{\upsilon^2-1}
%
%\nonumber\\
%\tilde{h}^{(a)}(\upsilon) 
%&= A\,\frac{Q_{2\mu}^2(\upsilon)}{(\upsilon^2-1)}
= %A
\frac{C_X}{C_f}\,
 \frac{\sqrt{\pi}\Gamma(3+2\mu)}{2^{1+2\mu}
\Gamma\left(\frac{\textstyle3}{\textstyle2}+2\mu\right)}\frac{1}{\upsilon^{3+2\mu}}
F\left(2+\mu,\,\frac{3}{2}+\mu,\,\frac{3}{2}+2\mu;\,\frac{1}{\upsilon^2}\right),
\label{nu_O}
\end{align}
where the coefficient $C_X$ can be calculated from
the coefficient in eq.~(\ref{K_alpha}), i.e.,
$1/2^{(d-3)/2-\alpha}\Gamma[(d-1)/2-\alpha]$ with $\alpha=3/2$, as
\begin{align}
C_X = C_f\,\frac{C_F}{C_{\Pi}}\,
\frac{1}{2^{\mu-3/2}
\Gamma\left(\mu-\frac{\textstyle1}{\textstyle2}\right)}
\frac{e^{-2\pi i}}{\sqrt{2\pi}}
= -\; C_f\, \frac{2^{2+2\mu}(1-2\mu)\Gamma(1+\mu)}
{\sqrt{\pi}\left[\Gamma\left(\frac{\textstyle1}{\textstyle2}+\mu\right)\right]^2
\Gamma\left(\frac{\textstyle1}{\textstyle2}-\mu\right)}.
%
%= -\, C_f \frac{2^{2+2\mu}(1-2\mu)\Gamma(\mu)\Gamma(1+2\mu)}
%{\sqrt{\pi}\Gamma(2\mu)
%\left[\Gamma\left(\frac{\textstyle1}{\textstyle2}+\mu\right)\right]^2
%\Gamma\left(\frac{\textstyle1}{\textstyle2}-\mu\right)}.
%この結果は係数が2余計：ノート[II],PP.2-3参照、2/nの2を含んでいる
\label{C_X}
\end{align}
It is easy to confirm that this formula is identical to eq.~(4$\cdot$7)
in ref.~\cite{PTP1} except for an extra factor of 2 in ref.~\cite{PTP1},
which will be taken into account as a factor of $-2/N$
in eq.~(\ref{f(1)^a}) in Appendix \ref{Ordinary}.
Equation (\ref{nu_O}) was also derived by McAvity and Osborn \cite{McAvity},
who used a method mentioned briefly in Section \ref{Introduction};
see (4.32) and (B.6) in their paper. This result is also written in eq.~(4.7)
of Giombi and Khanchandani's paper \cite{Giombi}.

In the case of the special transition,
two Fourier--Bessel integrations appear
in a single bubble in mixed space as in eq.~(\ref{Pi_special}).
In such a case, to use some mathematical trick may solve the problem.
Instead of directly inversing the single bubble,
we evaluated the product of $\Pi_q(y,y')$ and the function
\begin{align}
\Xi_q(y,y') &= - \; C^{-1}_{\Pi}(yy')^{3/2}
\int_0^{\infty} dt\,t\,J_{3/2+2\mu}(ty)J_{3/2+2\mu}(ty')(t^2+q^2)^{1/2-\mu}
\label{Xi_special}
\end{align}
as
\begin{align}
&\int_0^{\infty} \Pi_q(y,y'')\, \Xi_q(y'',y')\, dy'' = -\; 4(1+\mu)(1+2\mu)\delta(y-y')
- \sqrt{yy'}\,y' \int_0^{\infty} dt\,t\,J_{5/2+2\mu}(ty)
\nonumber\\
&\times (t^2+q^2)^{1/2+\mu}
\int_0^{\infty} dt'\,t'\,J_{3/2+2\mu}(t'y')(t'^2+q^2)^{1/2-\mu}
\int_0^{\infty} dy'' y''^2 J_{5/2+2\mu}(ty'') J_{3/2+2\mu}(t'y'').
\label{PiXi}
\end{align}
Since the last integral with respect to $y''$ can be expressed as
\begin{align}
\int_0^{\infty} dy'' y''^2 J_{5/2+2\mu}(ty'') J_{3/2+2\mu}(t'y'')
= \frac{1}{t^{1/2}t'^{5/2+2\mu}}\frac{\partial}{\partial t'}t'^{2+2\mu}\delta(t-t'),
\label{J5/2J3/2=dd}
\end{align}
which can easily be derived by differentiating the Hankel formula, eq.~(\ref{Bessel}),
a partial integration of eq.~(\ref{PiXi}) gives
\begin{align}
&\int_0^{\infty} dt'\,t'\,J_{3/2+2\mu}(t'y')(t'^2+q^2)^{1/2-\mu}
\int_0^{\infty} dy'' y''^2 J_{5/2+2\mu}(ty'') J_{3/2+2\mu}(t'y'')
\nonumber\\
&=\; - \frac{1}{t^{1/2}} \int_0^{\infty} dt'\,t'^{2+2\mu}\,\delta(t-t')\,
 \frac{\partial}{\partial t'}
\frac{1}{t'^{3/2+2\mu}} J_{3/2+2\mu}(t'y') (t'^2+q^2)^{1/2-\mu}
\nonumber\\
&= y' J_{5/2+2\mu}(ty')(t^2+q^2)^{1/2-\mu}
- (1-2\mu)\, t\, J_{3/2+2\mu}(ty') (t^2+q^2)^{-1/2-\mu}.
\end{align}
Thus, eq.~(\ref{PiXi}) becomes
\begin{align}
\int_0^{\infty} \Pi_q(y,y'')\, \Xi_q(y'',y')\, dy''
 &=
  -\; 4(1+\mu)(1+2\mu)\delta(y-y')
\nonumber\\
&
\;\;\;\; + \sqrt{yy'}\,y'^2 \int_0^{\infty} dt\,t\,J_{5/2+2\mu}(ty)J_{5/2+2\mu}(ty')(t^2+q^2)
\nonumber\\
&
\;\;\;\; -\; (1-2\mu)\sqrt{yy'}\,y' \int_0^{\infty} dt\,t^2\,J_{5/2+2\mu}(ty)J_{3/2+2\mu}(ty')
\nonumber\\
& = \Biggl[ -\; 4(1+\mu)(1+2\mu)
\nonumber\\
&
\;\;\;\;\;\, - \sqrt{yy'}\,y'^2 \left( -y'^{1/2+2\mu}\frac{\partial}{\partial y'} y'^{-4-4\mu}
\frac{\partial}{\partial y'} y'^{5/2+2\mu} + q^2 \right) \frac{1}{\sqrt{yy'}}
\nonumber\\
&
\;\;\;\;\;\,
-\; (1-2\mu)\sqrt{yy'}\,y'
 \frac{1}{y^{1/2}y'^{5/2+2\mu}} \frac{\partial}{\partial y'} y'^{2+2\mu}
\; \Biggr]\; \delta(y-y')
\nonumber\\
&= y^2\left[\frac{\partial^2}{\partial y^2} + (1-2\mu)\frac{1}{y}\frac{\partial}{\partial y} -q^2\right] \delta(y-y'),
\label{PiXi'}
\end{align}
where we used again eq.~(\ref{J5/2J3/2=dd}) with different variables, and the relation
\begin{align}
t^2 J_{5/2+2\mu}(ty') &= -\,ty'^{3/2+2\mu}\frac{\partial}{\partial y'} y'^{-3/2-2\mu}
J_{3/2+2\mu}(ty')
\nonumber\\
&= -\,y'^{3/2+2\mu}\frac{\partial}{\partial y'}y'^{-3/2-2\mu}
y'^{-5/2-2\mu}\frac{\partial}{\partial y'}y'^{5/2+2\mu}J_{5/2+2\mu}(ty')
\end{align}
along with the Hankel formula, eq.~(\ref{Bessel}).
Comparing eq.~(\ref{PiXi'}) with eq.~(\ref{Bessel}), we derived \cite{PLA2,PTP1}
\begin{align}
y^2\left[\frac{\partial^2}{\partial y^2} + (1-2\mu)\frac{1}{y}\frac{\partial}{\partial y} -q^2\right] \Pi^{-1}_q(y,y') = \Xi_q(y,y').
\label{Pi^-1_special}
\end{align}
Since $\Pi^{-1}_{\rho}(y,y')$ has a form of eq.~(\ref{Pi-1=nu}),
$\Xi_{\rho}(y,y')$ has a similar form,
\begin{align}
\Xi_{\rho}(y,y') = (yy')^{-2}\xi(\upsilon).
\label{Xi=xi}
\end{align}
Then, the differential equation~(\ref{Pi^-1_special}) is transformed
into real space as
\begin{align}
{\cal D}\,\nu(\upsilon) = \xi(\upsilon),
\label{Dnu=xi}
\end{align}
where $\cal D$ is the same as eq.~(\ref{D_S}) for $g^{(0)}(\upsilon)$.
Indeed, as described in Appendix \ref{Differential},
the coefficient $\alpha=1-2\mu=6-d$ on $(1/y)(\partial/\partial y)$
in eq.~(\ref{Pi^-1_special})
and the dimension $\zeta=-2$ of $\Pi^{-1}_{\rho}(y,y')$ in eq.~(\ref{Pi-1=nu})
satisfies the necessary condition,
eq.~(\ref{condition});
and eq.~(\ref{D}) with $\beta=0$ becomes eq.~(\ref{D_S}).
Therefore, the two independent homogeneous solutions of eq.~(\ref{Dnu=xi})
are given by eq.~(\ref{g_S}) of $g^{(0)}(\upsilon)$
and eq.~(\ref{w_S}) of $w^{(0)}(\upsilon)$.
Equation (\ref{Xi=xi}), i.e., Fourier transform of eq.~(\ref{Xi_special}),
is calculated by using eq.~(\ref{K_alpha}) and the integral formula (6.578.7)
of ref.~\cite{Gradshteyn} as
\begin{align}
\xi(\upsilon) &= -\,\frac{C_F}{C_{\Pi}}\,
\frac{1}{2^{\mu-3/2}
\Gamma\left(-\frac{\textstyle1}{\textstyle2}+\mu\right)}
\frac{e^{-3\pi i}}{\sqrt{2\pi}}\frac{Q_{1+2\mu}^3(\upsilon)}{(\upsilon^2-1)^{3/2}}
%\nonumber\\
%&= -\,\frac{C_F}{C_{\Pi}}\,
%\frac{1}{2^{\mu-3/2}
%\Gamma\left(-\frac{\textstyle1}{\textstyle2}+\mu\right)}
%\frac{\Gamma(5+2\mu)}{2^{5/2+2\mu}
%\Gamma\left(\frac{\textstyle5}{\textstyle2}+2\mu\right)}
%\frac{1}{\upsilon^{5+2\mu}}
%F\left(3+\mu,\,\frac{5}{2}+\mu,\,\frac{5}{2}+2\mu;\,\frac{1}{\upsilon^2}\right)
\nonumber\\
&= \frac{C_{\Xi}}{C_f}\,\frac{1}{\upsilon^{5+2\mu}}
F\left(3+\mu,\,\frac{5}{2}+\mu,\,\frac{5}{2}+2\mu;\,\frac{1}{\upsilon^2}\right)
%\nonumber\\
%&= \frac{C_{\Xi}}{C_f}\,\frac{1}{\upsilon^{5+2\mu}}
%\left(1-\frac{1}{\upsilon^2}\right)^{-3-\mu}
%F\left(3+\mu,\,\mu,\,\frac{5}{2}+2\mu;\,\frac{1}{1-\upsilon^2}\right)
%\nonumber\\
%&= \frac{C_{\Xi}}{C_f}\,
%\frac{\upsilon}{(\upsilon^2-1)^{3+\mu}}
%F\left(3+\mu,\,\mu,\,\frac{5}{2}+2\mu;\,\frac{1}{1-\upsilon^2}\right)
\nonumber\\
&= \frac{C_{\Xi}}{C_f}\,\upsilon
\frac{1}{(\upsilon^2-1)^{3+\mu}}
F\left(3+\mu,\,\mu,\,\frac{5}{2}+2\mu;\,-\frac{1}{\upsilon^2-1}\right),
\label{xi}
\end{align}
where we put \cite{PTP1}
\begin{align}
C_{\Xi} &= -\, C_f\,\frac{C_F}{C_{\Pi}}\,\frac{2^{-1-3\mu}\Gamma(5+2\mu)}
{\Gamma\left(-\frac{\textstyle1}{\textstyle2}+\mu\right)
\Gamma\left(\frac{\textstyle5}{\textstyle2}+2\mu\right)}
%
%コメント始め
%\nonumber\\
%&
%{\color{NavyBlue}
%= -\, C_f\,\frac{2^{-1-3\mu}\Gamma(5+2\mu)}
%{\Gamma\left(-\frac{\textstyle1}{\textstyle2}+\mu\right)
%\Gamma\left(\frac{\textstyle5}{\textstyle2}+2\mu\right)}
%\frac{2^{2\mu+4}2^{\mu+1}\Gamma(\mu+2)}
%{\Gamma\left(\frac{\textstyle3}{\textstyle2}+\mu\right)
%\Gamma\left(-\frac{\textstyle1}{\textstyle2}-\mu\right)}
%}
%
%\nonumber\\
%&
%{\color{NavyBlue}
%= -\, C_f\,\frac{2^{4}\Gamma(5+2\mu)\Gamma(\mu+2)}
%{\Gamma\left(-\frac{\textstyle1}{\textstyle2}+\mu\right)
%\Gamma\left(\frac{\textstyle5}{\textstyle2}+2\mu\right)
%\Gamma\left(\frac{\textstyle3}{\textstyle2}+\mu\right)
%\Gamma\left(-\frac{\textstyle1}{\textstyle2}-\mu\right)}
%}
%コメント終わり
%
\nonumber\\
&= -\, C_f\,
\left[\frac{2^{1+\mu}\Gamma(1+\mu)}{\Gamma(2+2\mu)}\right]^2
\frac{2^{2}\Gamma(5+2\mu)
\Gamma(3+2\mu)}
{\sqrt{\pi}\Gamma\left(-\frac{\textstyle1}{\textstyle2}+\mu\right)
\Gamma\left(\frac{\textstyle5}{\textstyle2}+2\mu\right)
\Gamma\left(-\frac{\textstyle1}{\textstyle2}-\mu\right)}.
\label{C_Xi}
\end{align}
%
%
%
%コメント始め
%{\color{NavyBlue}
%\begin{align}
%\frac{C_F}{C_{\Pi}} = \frac{2^{d-1}2^{(d-3)/2}\Gamma\left(\frac{d-1}{2}\right)}
%{\Gamma\left(\frac{\textstyle d}{\textstyle2}-1\right)
%\Gamma\left(2-\frac{\textstyle d}{\textstyle2}\right)}
%= \frac{2^{2\mu+4}2^{\mu+1}\Gamma(\mu+2)}
%{\Gamma\left(\frac{\textstyle3}{\textstyle2}+\mu\right)
%\Gamma\left(-\frac{\textstyle1}{\textstyle2}-\mu\right)}
%\nonumber
%\end{align}
%
%\begin{align}
%C_{\Pi} = \Gamma\left(\frac{\textstyle d}{\textstyle2}-1\right)
%\Gamma\left(2-\frac{\textstyle d}{\textstyle2}\right)
%\frac{K_{d-1}}{2^{d-1}}, \;\;\;\;
%C_F = 2^{(d-3)/2}\Gamma\left(\frac{d-1}{2}\right)K_{d-1}.
%\nonumber
%\end{align}
%}
%コメント終わり
%
%
%
Then, we solved the differential equation~(\ref{Dnu=xi})
without ambiguity as explained in detail in Appendix \ref{InverseSpecial}.
The result is \cite{PTP2}
\begin{align}
\nu^{\rm sp}(\upsilon) &= 
\frac{C_{\Xi}}{C_f}\left[\frac{1}{3(3+2\mu)}
\frac{\upsilon}{(\upsilon^2-1)^{3+\mu}}\,
{}_3F_2\left(3+\mu,1+\mu,\frac{3}{2};\,\frac{5}{2}+2\mu,\frac{5}{2};\,
\frac{1}{1-\upsilon^2}\right) \right.
\nonumber\\
&\hskip10mm \left. +\, C_H\, \frac{1}{\upsilon^2}
F\left(\frac{3}{2},\,1,\,1-\mu;\,\frac{1}{\upsilon^2}\right) \right],
\label{nu_S}
\end{align}
where $C_H$ is given by
\begin{align}
C_H =-\, \frac{1}{2(3+2\mu)}
\frac{2^{-2}\pi\Gamma\left(\frac{\textstyle5}{\textstyle2}+2\mu\right)
\Gamma\left(-\frac{\textstyle1}{\textstyle2}+\mu\right)}
{\Gamma(3+\mu)\Gamma(1+\mu)\Gamma(1-\mu)\Gamma(1+2\mu)}.
\label{C_H}
\end{align}
McAvity and Osborn \cite{McAvity} used a method explained briefly
in Section \ref{Introduction} to invert the sum of the two terms,
$(\upsilon^2 - 1)^{2-d}+(\upsilon^2 - 1)^{3-d}$,
and proposed two approaches.
One is to remove one term in the sum by invoking a differential equation.
Their eq.~(B.13) is essentially identical to eq.~(\ref{Dnu=xi})
with $\cal D$ of eq.~(\ref{D_S}) and $\xi(\upsilon)$ of eq.~(\ref{xi}).
The other is to obtain $\nu(\upsilon)$ more directly by evaluating
the inverse Fourier transform using contour integration
as written in the Appendix B of their paper.
Their result, eq.~(4.40) in their paper, which is also written
in eq.~(4.10) of Giombi and Khanchandani's paper \cite{Giombi},
is identical to our eq.~(\ref{nu_S}).

In the case of the extraordinary transition,
$g^{(0)}(\upsilon)$ is given by eq.~(\ref{g_E}),
which satisfies the differential equation
(\ref{Dg^(0)=0}) with $\cal D$ of eq.~(\ref{D_E}).
If we write eq.~(\ref{Pi_rho}) as 
$\Pi_{\rho}(y,y')=(yy')^{1-2\mu}\chi(\upsilon)$,
the function $\chi(\upsilon)$ is given by
\begin{align}
\chi(\upsilon) = C_f^2\left[g^{(0)}(\upsilon)\right]^2 + 2AC_f g^{(0)}(\upsilon),
\end{align}
where $A$ is given by eq.~(\ref{A}).
Operating ${\cal D}$ onto $\chi(\upsilon)$, we obtained
\begin{align}
{\cal D}\, \chi(\upsilon)
&= 2C_f^2(\upsilon^2-1)\left[\frac{d}{d\upsilon}g^{(0)}(\upsilon)\right]^2
\nonumber\\
&= 2\left[\frac{\Gamma(d-2)}{2}K_{d-1}\right]^2(\upsilon^2-1)
\left[ -\,\frac{d-2}{\mu} \frac{1}{\upsilon^3}
\frac{d}{d(\upsilon^{-2})} \bigl(\upsilon^{-2}\bigr)^{^{\scriptstyle\mu}}
F\left(\mu,\,\frac{1}{2}+\mu,\,1+\mu;\,\frac{1}{\upsilon^2}\right) \right]^2
\nonumber\\
&= 2\left[\frac{\Gamma(d-1)}{2}K_{d-1}\right]^2(\upsilon^2-1)
\left[ -\frac{1}{\upsilon^{1+2\mu}}
F\left(\mu,\,\frac{1}{2}+\mu,\,\mu;\,\frac{1}{\upsilon^2}\right) \right]^2
\nonumber\\
&= 2\left[\frac{\Gamma(d-1)}{2}K_{d-1}\right]^2
\frac{1}{(\upsilon^2 - 1)^{2\mu}} \equiv b(\upsilon).
\end{align}
The function $B_{\rho}(y,y') = (yy')^{1-2\mu}\,b(\upsilon)$ can be transformed
into mixed space, yielding a single Fourier--Bessel integral
\begin{align}
B_q(y,y') = -\, C_{\Pi}\,(yy')^{3/2}
\int_0^{\infty} dt\,t\,J_{-1/2+2\mu}(ty)J_{-1/2+2\mu}(ty')(t^2+q^2)^{-1/2+\mu},
\label{B_q}
\end{align}
where $C_{\Pi}$ is the same as that defined in eq.~(\ref{C_Pi}).
Which differential operator $\cal E$ in mixed space corresponds to ${\cal D}$?
The general form of $\cal E$ is given in eq.~(\ref{E})
in Appendix \ref{Differential}.
It must have $\beta=\zeta(1-d-\zeta)=d-2=2\mu-1$,
because the last constant term of $\cal D$ in eq.~(\ref{D}) is zero,
and $\zeta=1-2\mu=2-d$ in this case.
Thus, the desired $\cal E$ is expressed in a form
\begin{align}
{\cal E} = y^2 \left[\, \nabla_{\rho}^2 + \frac{\partial^2}{\partial y^2}
 - (1 - 2\mu)\frac{\partial}{\partial y}
 + (1 - 2\mu)\frac{1}{y^2}\, \right]
\end{align}
in front of the bubble function
$\Pi_{\rho}(y,y')$ as ${\cal E}\, \Pi_{\rho}(y,y') = B_{\rho}(y, y')$.
The operator $\widetilde{\cal E}$ in mixed space is given by
\begin{align}
\widetilde{\cal E} = y^2 \left[ -\, q^2 + \frac{\partial^2}{\partial y^2}
- (1-2\mu)\frac{1}{y}\frac{\partial}{\partial y} + (1-2\mu)\frac{1}{y^2} \right],
\label{E'}
\end{align}
which enables us to write
\begin{align}
 \widetilde{\cal E}\;\Pi_q(y,y') = B_q(y,y').
 \label{EPi=B}
 \end{align}
 Therefore, we were led to
\begin{align}
\int_0^{\infty} B_q(y,y'')\, \Pi^{-1}_q(y'',y')\, dy''
 = \widetilde{\cal E}\; \delta(y-y').
\label{BPi=Ed}
\end{align}
%{\color{Lavender}\sout{which is identical to eq.~(\ref{BPi=Ed}).}}
To solve this equation, we assumed $\Pi^{-1}_q(y,y')$ in a form
\begin{align}
\Pi^{-1}_q(y,y') = a_1 X_q(y,y') + a_2 Y_q(y,y')
\end{align}
with
\begin{subequations}
\begin{align}
X_q(y,y') &= \frac{1}{\sqrt{yy'}}
\int_0^{\infty} dt\,t\,J_{-1/2+2\mu}(ty)J_{-1/2+2\mu}(ty')(t^2+q^2)^{1/2-\mu},
\\
Y_q(y,y') &= \sqrt{yy'}
\int_0^{\infty} dt\,t\,J_{1/2+2\mu}(ty)J_{1/2+2\mu}(ty')(t^2+q^2)^{3/2-\mu}.
\end{align}
\end{subequations}
Then, from
\begin{subequations}
\begin{align}
\int_0^{\infty} B_q(y,y'')\, X_q(y'',y')\, dy'' &= -\,C_{\Pi}\, \delta(y-y'),
\\
\int_0^{\infty} B_q(y,y'')\, Y_q(y'',y')\, dy'' &= \,C_{\Pi}\, 
y^2\left[-q^2+\frac{\partial^2}{\partial y^2} + (2\mu-1)\frac{1}{y}\frac{\partial}{\partial y}
- \frac{4\mu}{y^2} \right]\, \delta(y-y').
\end{align}
\end{subequations}
we found \cite{PTP2}
\begin{align}
\Pi^{-1}_q(y,y'') &= C^{-1}_{\Pi}\!\sqrt{yy'}
\int_0^{\infty} dt\,t\,J_{1/2+2\mu}(ty)J_{1/2+2\mu}(ty')(t^2+q^2)^{3/2-\mu}
\nonumber\\
& -\; (1+2\mu)\,C^{-1}_{\Pi}\frac{1}{\sqrt{yy'}}
\int_0^{\infty} dt\,t\,J_{-1/2+2\mu}(ty)J_{-1/2+2\mu}(ty')(t^2+q^2)^{1/2-\mu}.
\label{Pi^-1_E}
\end{align}
Compared with the special transition,
this case exhibits a reversed relationship between the expressions of
$\Pi_q(y,y')$ and $\Pi^{-1}_q(y,y')$.
Specifically,
$\Pi_q(y,y')$ is replaced by its $\widetilde{\cal E}$-operated form,
$B_q(y,y')$ in eq.~(\ref{EPi=B}),
whereas $\Pi^{-1}_q(y,y')$ is expressed by two Fourier--Bessel integrals,
as shown in eq.~(\ref{Pi^-1_E}).
The order is vice versa in the special transition:
$\Pi_q(y,y')$ is expressed by two Fourier--Bessel integrals,
as in eq.~(\ref{Pi_special}), whereas $\Pi^{-1}_q(y,y')$ is replaced
by $\Xi_q(y,y')$, as in eq.~(\ref{Pi^-1_special}).

The longitudinal correlation function in the large $N$ limit reads
\begin{align}
G^{L\,(0)}_q(y,y') = G^{(0)}_q(y,y') - 2 \int_0^{\infty}dy_1\int_0^{\infty}dy_2\,
G^{(0)}_q(y,y_1)M^0(y_1)\Pi^{-1}_q(y_1,y_2)M^0(y_2)G^{(0)}_q(y_2,y').
\end{align}
Let ${\cal L}'$ be the differential operator,
where $y$ is replaced with $y'$ in $\cal L$ of eq.~(\ref{L}).
Then, $G^{L\,(0)}_{\rho}(y,y')=(yy')^{1-d/2}f(\upsilon)$ satisfies
\begin{align}
{\cal L}\,{\cal L}'\,G^{L\,(0)}_{\rho}(y,y') = -\,2 M^0(y)\Pi^{-1}_{\rho}(y,y')M^0(y'),
\end{align}
which can be rewritten as
\begin{align}
{\cal D}^2 f(\upsilon) = -\,2A \nu(\upsilon),
\label{D2f=-2Anu}
\end{align}
where ${\cal D}$ and $\nu(\upsilon)$ are defined in eq.~(\ref{D_E}) and
eq.~(\ref{Pi-1=nu}), respectively, $A$ is given by eq.~(\ref{A}).
The explicit form of $\nu(\upsilon)$ can be
obtained by transforming eq.~(\ref{Pi^-1_E})
back into real space using eq.~(\ref{K_alpha})
and the integral formula (6.578.7) of
ref.~\cite{Gradshteyn} as \cite{PTP2}
\begin{align}
\nu(\upsilon) &= \frac{C_F}{C_{\Pi}}
\frac{1}{\sqrt{2\pi}2^{\mu-5/2}\Gamma\left(\mu-\frac{\textstyle3}{\textstyle2}\right)}
\left[ \frac{Q^2_{2\mu}(\upsilon)}{\upsilon^2-1}
 + \frac{1+2\mu}{2\mu-3}\frac{Q^1_{-1+2\mu}(\upsilon)}{(\upsilon^2-1)^{1/2}} \right]
 \nonumber\\
&= \frac{C_F}{C_{\Pi}}
\frac{\Gamma(3+2\mu)}{2^{3\mu-1}
\Gamma\left(-\frac{\textstyle3}{\textstyle2}+\mu\right)
\Gamma\left(\frac{\textstyle3}{\textstyle2}+2\mu\right)}
\left[ \frac{1}{\upsilon^{3+2\mu}}
F\left(2+\mu,\frac{3}{2}+\mu,\frac{3}{2}+2\mu;\frac{1}{\upsilon^2}\right) \right.
\nonumber\\
&\hskip27.5mm \left. -\, \frac{\frac{\textstyle1}{\textstyle2}+2\mu}{(2\mu-3)(1+\mu)}
\frac{1}{\upsilon^{1+2\mu}}
F\left(1+\mu,\frac{1}{2}+\mu,\frac{1}{2}+2\mu;\frac{1}{\upsilon^2}\right) \right].
\label{nu_E}
\end{align}
Since this $\nu(\upsilon)$ has an asymptotic behavior
$\nu(\upsilon)\rightarrow\upsilon^{-1-2\mu}\propto \upsilon^{-d}$
for $\upsilon\rightarrow\infty$,
the solution $f(\upsilon)$ of eq.~(\ref{D2f=-2Anu})
has the same asymptotic behavior. Thus, we found \cite{PTP2}
\begin{align}
\hat{\Delta}_{\phi}^L = d + O\left(\frac{1}{N}\right)
\label{DeltaL}
\end{align}
for the anomalous dimension of the longitudinal field.

\section{1/\textit{N} Correction to the Correlation Function}
\label{1/N}

There are two contributions to the $O(1/N)$ correction
to the self-energy as depicted in Figs. \ref{SelfEnergy}(a) and (b).
A few more diagrams needed for the extraordinary
transition can be seen in ref.~\cite{PTP2}.
The correlation function at order $1/N$ can be expressed as
\begin{align}
G^{(1)}_{\rho}(y,y') = (yy')^{1-d/2} f(\upsilon)
 - \frac{\eta}{2}\,G^{(0)}_{\rho}(y,y') \log(2yy'\Lambda^2),
\label{G^(1)}
\end{align}
where $f(\upsilon)$ is a function yet to be determined,
$\Lambda$ is a momentum cutoff, and $\eta$ is the bulk critical exponent.
The corresponding contributions to the correlation function
are diagrammatically represented in Figs. \ref{Correlation}(a) and (b).
Since $G_{\rho}^{(0)}(y,y')$ satisfies eq.~(\ref{LG=d}) and the similar
equation in which $y$ is replaced with $y'$,
we retrieve the self-energy
\begin{align}
\Sigma_{\rho}(y,y') = (yy')^{-1-d/2} h(\upsilon).
\label{Sigma=h}
\end{align}
by the operation
\begin{align}
{\cal L\,L}'\,G^{(1)}_{\rho}(y,y') = \Sigma_{\rho}(y,y').
\end{align}
\begin{figure}[!h]
\vskip-1mm
\begin{center}
\includegraphics[width=70mm]{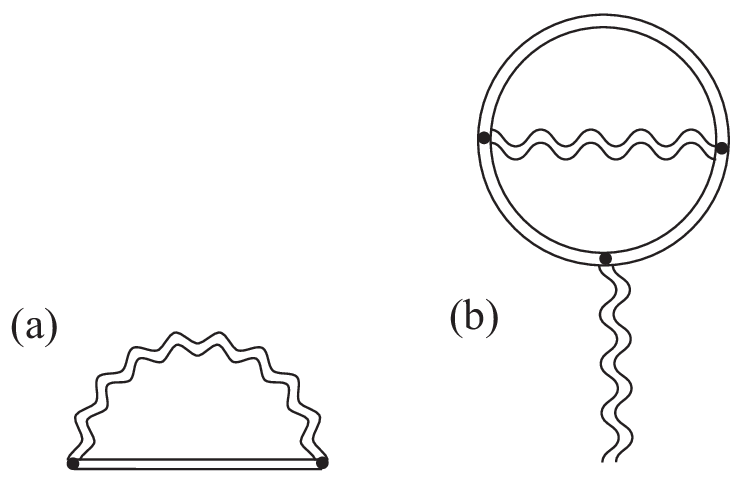}
\caption{Two self-energy diagrams contributing at order $1/N$.}
\label{SelfEnergy}
\end{center}
\end{figure}
\begin{figure}[!h]
\vskip-2mm
\begin{center}
\includegraphics[width=85mm]{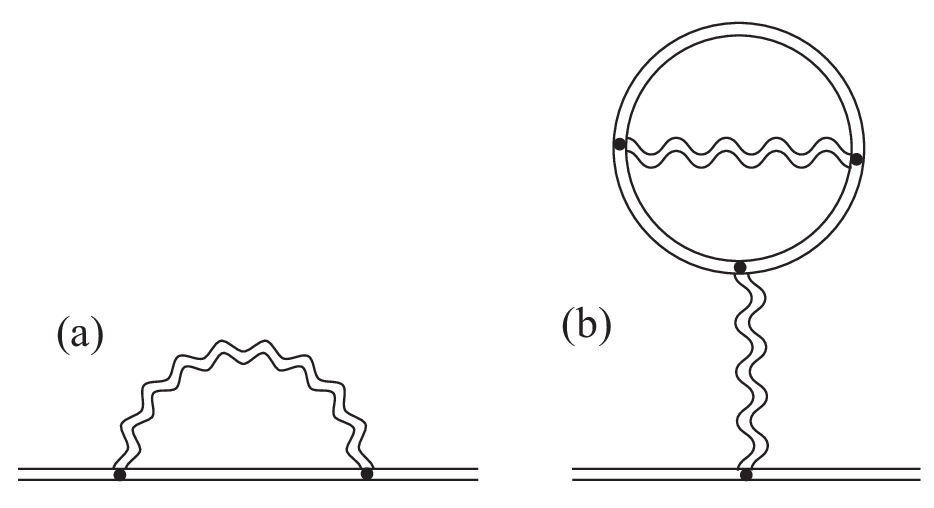}
\caption{Diagrammatic representation of the correlation function
contributing at order $1/N$.}
\label{Correlation}
\end{center}
\end{figure}

\noindent
Note that the second term in eq.~(\ref{G^(1)}) vanishes under this operation.
Therefore, we have
\begin{subequations}
\begin{align}
{\cal D}\, g(\upsilon) &= h(\upsilon),
\label{Dg=h}
\\
{\cal D}\, f(\upsilon) &= g(\upsilon),
\label{Df=g}
\end{align}
\end{subequations}
where $\cal D$ is the differential operator given in eq.~(\ref{D'}).
The homogeneous differential equation of this operator is eq.~(\ref{Dw=0}).
One of the solution is $g^{(0)}(\upsilon)$, while the other independent solution is
$w^{(0)}(\upsilon)$ given by eqs.~(\ref{w_O})-(\ref{w_E}).

The contribution to the self-energy diagram (a) is given by
\begin{align}
\Sigma_{\rho}^{(a)}(y,y') = - \frac{2}{N}\,
\Pi_{\rho}^{-1}(y,y')\, G^{(0)}_{\rho}(y,y')
= (yy')^{-1-d/2} h^{(a)}(\upsilon)
\label{Sigma^a}
\end{align}
with
\begin{align}
h^{(a)}(\upsilon) = - \frac{2}{N}C_f \nu(\upsilon)g^{(0)}(\upsilon)
\label{h^a}
\end{align}
where $g^{(0)}(\upsilon)$ is given by eqs.~(\ref{g_O}) and (\ref{g_S}),
respectively, for the ordinary and special transitions.
Extraordinary transition is discussed later in this Section.
To solve the differential equations (\ref{Dg=h}) and (\ref{Df=g}),
we define functions $\tilde{g}^{(a)}(\upsilon)$ and $\tilde{h}^{(a)}(\upsilon)$ as
\begin{subequations}
\begin{align}
g^{(a)}(\upsilon) &=
% {\color{Lavender}\frac{\tilde{g}^{(a)}(\upsilon)}{(\upsilon^2-1)^{1/2+\mu}},}
% \;\;\;\; 
g^{(0)}(\upsilon)\,\tilde{g}^{(a)}(\upsilon),
\label{tilde^g^a}
\\
h^{(a)}(\upsilon) &=
% {\color{Lavender}\frac{\tilde{h}^{(a)}(\upsilon)}{(\upsilon^2-1)^{1/2+\mu}}.}
% \;\;\;\; 
g^{(0)}(\upsilon)\,\tilde{h}^{(a)}(\upsilon).
\label{tilde^h^a}
\end{align}
\end{subequations}
In the case of the ordinary transition,
the differential operator $\cal D$ for $g^{(0)}(\upsilon)$ is given by eq.~(\ref{D_O}),
%Two independent homogeneous solutions satisfying ${\cal D}w_n(\upsilon)=0$
%($n=1,2$) are
%\begin{subequations}
%\begin{align}
%w_1(\upsilon) &= \frac{P^{1/2+\mu}_{-1/2-\mu}(\upsilon)}{(\upsilon^2-1)^{(1/2+\mu)/2}}
%=\frac{2^{1/2+\mu}}{\Gamma\left(\frac{\textstyle1}{\textstyle2}-\mu\right)}
%\frac{1}{(\upsilon^2-1)^{1/2+\mu}},
%\label{w_1}
%\\
%w_2(\upsilon) &= \frac{Q^{1/2+\mu}_{-1/2-\mu}(\upsilon)}{(\upsilon^2-1)^{(1/2+\mu)/2}}
%=\frac{e^{i(1/2+\mu)\pi}\sqrt{\pi}}{2^{1/2-\mu}\Gamma(1-\mu)}
%\frac{1}{\upsilon}F\left(1,\,\frac{1}{2}\,1-\mu;\,\frac{1}{\upsilon^2}\right)
%\nonumber\\
%&= \frac{e^{i(1/2+\mu)\pi}}{2}
%\left[\,\frac{2^{1/2+\mu}\Gamma\left(\frac{\textstyle1}{\textstyle2}+\mu\right)}
%{(\upsilon^2-1)^{1/2+\mu}}
%+\frac{\Gamma\left(-\frac{\textstyle1}{\textstyle2}-\mu\right)}
%{\Gamma(-2\mu)\,(\upsilon+1)^{1/2+\mu}}
%F\left(\frac{1}{2}+\mu,\,\frac{1}{2}-\mu,\,\frac{3}{2}+\mu;\,\frac{1-\upsilon}{2}\right)\,
%\right],
%\label{w_2}
%\end{align}
%\end{subequations}
%where $P^{1/2+\mu}_{-1/2-\mu}(\upsilon)$ is the Legendre function of the first kind.
and the function $\tilde{g}^{(a)}(\upsilon)$ defined in eq.~(\ref{tilde^g^a}) satisfies
\begin{align}
\left[ (\upsilon^2-1) \frac{d^2}{d\upsilon^2}
 + (1-2\mu) \upsilon \frac{d}{d\upsilon} \right] \tilde{g}^{(a)} = \tilde{h}^{(a)}.
\label{til_g=til_h}
\end{align}
In the case of the special transition,
the differential operator $\cal D$ for $g^{(0)}(\upsilon)$ is given by eq.~(\ref{D_S}),
and the function $\tilde{g}^{(a)}(\upsilon)$ satisfies
\begin{align}
\left[ (\upsilon^2-1)\frac{d}{d\upsilon} - \frac{2}{\upsilon} + (1-2\mu)\upsilon \right]\,
\frac{d}{d\upsilon}\, \tilde{g}^{(a)}(\upsilon) = \tilde{h}^{(a)}(\upsilon).
\label{D^tilde^g=tilde^h_S}
\end{align}
To solve the next differential equation eq.~(\ref{Df=g}), we use
the Wronskian $W\left(g^{(0)},w^{(0)}\right)$ given by eq.~(\ref{Wronskian}).
The general solution of eq.~(\ref{Df=g}) is obtained as
\begin{align}
f(\upsilon) &=
 w^{(0)}(\upsilon) \int_{c}^{\upsilon}
 \frac{g(\upsilon') g^{(0)}(\upsilon')}{(\upsilon'^2-1)W(g^{(0)},w^{(0)})} d\upsilon'
 - g^{(0)}(\upsilon) \int_{c'}^{\upsilon}
 \frac{g(\upsilon') w^{(0)}(\upsilon')}{(\upsilon'^2-1)W(g^{(0)},w^{(0)})} d\upsilon'
\nonumber\\
&= \frac{1}{2\mu}
\left[ w^{(0)}(\upsilon) \int_{c}^{\upsilon} (\upsilon'^2-1)^{d/2-1}
g(\upsilon') g^{(0)}(\upsilon') d\upsilon'
 - g^{(0)}(\upsilon) \int_{c'}^{\upsilon} (\upsilon'^2-1)^{d/2-1}
g(\upsilon') w^{(0)}(\upsilon') d\upsilon' \right]
\label{f=wgg}
\end{align}
for the ordinary and special transitions.
Since $f^{(a)}(\upsilon)$ decreases as $1/v^{d-2}$ for $\upsilon\rightarrow\infty$,
we should set $c=\infty$.
The other parameter $c'$ determines the coefficient for the function $g^{(0)}(\upsilon)$,
which has the same form as that in the large $N$ limit.
Therefore, it is not necessary to determine $c'$, because it does not induce
any logarithmic anomaly.
%
%\begin{align}
%f &= c_1 w_1 + c_2 w_2
%\nonumber\\
%f' &= c'_1 w_1 + c'_2 w_2 + c_1 w'_1 + c_2 w'_2 = c_1 w'_1 + c_2 w'_2
%\nonumber\\
%0 &= c'_1 w_1 + c'_2 w_2
%\nonumber\\
%f'' &= c'_1 w'_1 + c'_2 w'_2 + c_1 w''_1 + c_2 w''_2
%\nonumber\\
%{\cal D} f &= (v2 - 1)( c'_1 w'_1 + c'_2 w'_2 ) = g
%\nonumber\\
%c'_1 w'_1 + c'_2 w'_2 &= g / (v2 - 1)
%\nonumber\\
%0 &= c'_1 w_1 w'_2 + c'_2 w_2 w'_2
%\nonumber\\
%c'_2 w'_2 &= - c'_1 w_1 w'_2 / w_2
%\nonumber\\
%c'_1 w'_1 - c'_1 w_1 w'_2 / w_2 &=  g / (v2 - 1)
%\nonumber\\
%c'_1 w'_1 w_2 - c'_1 w_1 w'_2 &= g / (v2 - 1) w_2
%\nonumber\\
%c'_1 &= g / (v2 - 1) w_2 / ( w'_1 w_2 - w_1 w'_2 )
%\nonumber\\
%0 &= c'_1 w_1 w'_1 + c'_2 w'_1 w_2
%\nonumber\\
%c'_1 w'_1 &= - c'_2 w'_1 w_2 / w_1
%\nonumber\\
%- c'_2 w'_1 w_2 / w_1 + c'_2 w'_2 &= g / (v2 - 1)
%\nonumber\\
%- c'_2 w'_1 w_2 + c'_2 w_1 w'_2 &= g / (v2 - 1) w_1
%\nonumber\\
%c'_2 &= g / (v2 - 1) w_1 / ( w_1 w'_2 - w'_1 w_2 )
%\end{align}
%

By directly integrating eqs.~(\ref{Dg=h}) and (\ref{Df=g}) in this way,
we were able to get to the final results for the boundary anomalous dimensions.
For details, see Appendices \ref{Ordinary} and \ref{Special}
for the ordinary and special transitions, respectively;
the anomalous dimension for the transverse fields in the case of
the extraordinary transition is derived below in this Section.

Here, it is interesting to note that Giombi and Khanchandani \cite{Giombi}
showed that the boundary operator expansion \cite{McAvity}
contains a tower of operators
and succeeded in obtaining the series expansion form for $\hat{\gamma}^O$
in the case of the ordinary transition.
This is related to the fact that $\nu(\upsilon)$ in eq.~(\ref{nu_O})
is represented by a single hypergeometric function.
On the other hand, they could not derive the $1/N$ correction
in the anomalous dimension in the case of the special transition,
because two boundary operators with different dimensions appeared
in the boundary operator expansion \cite{McAvity},
resulting in the occurrence of two towers of operators,
which made difficult to solve the problem.
This is related to the fact that $\nu(\upsilon)$ in eq.~(\ref{nu_S})
is represented by a sum of two hypergeometric functions.

In the case of the special transition,
the $1/N$ correction $\hat{\gamma}^S$ in eq.~(\ref{DeltaS})
is explicitly derived in Appendix \ref{Special} as \cite{PLA2,PTP1}
\begin{align}
\hat{\gamma}^S = \frac{2(4-d)}{\Gamma(d-3)}
\left[\frac{(6-d)\Gamma(2d-6)}{d\Gamma(d-3)} + \frac{1}{\Gamma(5-d)}\right]
\;\;\;\; \textrm{for $3<d<4$},
\label{Eq_S}
\end{align}
which is consistent with the $O(\varepsilon^2)$ result
in the $\epsilon=4-d$ expansion by Reeve \cite{Reeve}.
\begin{align}
\hat{\gamma}^S = 3\varepsilon - \frac{5}{4}\varepsilon^2
+ {\cal O}\left(\varepsilon^3\right).
\label{4-d_S}
\end{align}
Equations (\ref{Eq_S}) and (\ref{4-d_S}) are plotted versus $d$
in Fig.~\ref{gamma_S}. Our result (\ref{Eq_O}) is smooth
between $3\leq d\leq 4$
and matches eq.~(\ref{4-d_S}) near $d=4$.
\begin{figure}[t]
\begin{center}
\includegraphics[width=150mm]{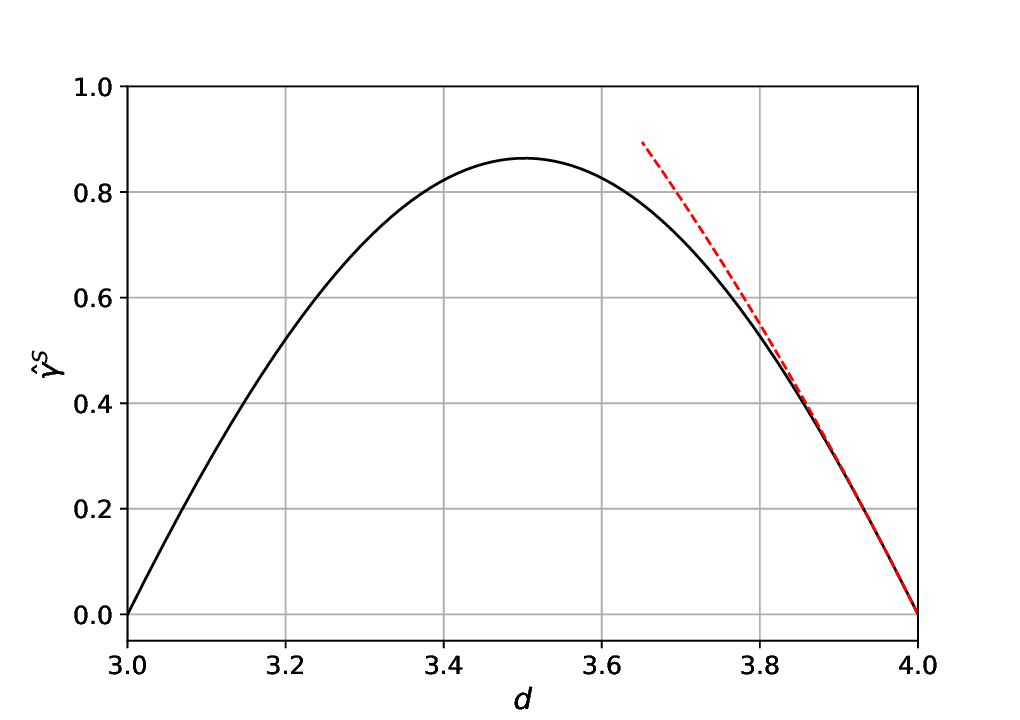}
\caption{Plots of $\hat{\gamma}^S$ versus the space dimension $d$
in the case of the special transition.
Black solid curve is eq.~(\ref{Eq_S}) of the $1/N$ expansion \cite{PLA2,PTP1}
and red dashed curve near $d=4$ is eq.~(\ref{4-d_S})
of the $\epsilon=4-d$ expansion \cite{Reeve}.}
\label{gamma_S}
\end{center}
\end{figure}
Giombi and Khanchandani \cite{Giombi} made a comment on eq.~(\ref{Eq_S})
[see their eq.~(4.60)]
that, in $d = 3$, the anomalous dimension $\hat{\Delta}_{\phi}^{S}$ vanishes,
consistently with the expectation that this should be the lower critical dimension
for the special transition.
Then, they speculated that $\hat{\Delta}_{\phi}^{S} = 0$ to all orders
in the $1/N$ expansion in $d=3$.
Our result (\ref{Eq_S}) is also quoted by Metlitski \cite{Metlitski}
[see their eq.~(47)], who suggested that,
in the special and extraordinary transitions at $d=3$,
there is a critical value $N_c$ for $N$, above which there is no fixed point.
This is related to the fact \cite{Hohenberg}
that the ($d - 1$)-dimensional (surface) system
does not undergo the ferromagnetic phase transition when
$d \leq 3$ and $N \geq 2$.
A possible logarithmic anomaly at the transition point for the $N=2$ case
(the $XY$ model) at $d=3$ was recently discussed by Hu {\it et al.} \cite{Hu}.
A new approach using machine-learning \cite{Shiina}
may be useful in such an analysis.
However, adding the surface magnetic anisotropy would retrieve
the special transition when $d = 3$ and $N \geq 2$.
Such an anisotropy is theoretically studied in some transition
metals \cite{Takayama}.
This new type of special transition is called the anisotropic special transition
and the $1/N$ expansion for the anisotropic special transition
was performed by us \cite{PLA3}.
The single bubble in this case is the same as that of the ordinary transition.
Our final result for the anomalous dimension is again
consistent with the $O(\varepsilon^2)$ result
by Diehl and Eisenriegler \cite{DiehlEisenriegler}.

In the case of the extraordinary transition,
we have considered transverse correlation function to all orders in $1/N$.
Since the singularity of the longitudinal field in
eq.~(\ref{DeltaL}) is weaker than that of the transverse fields in eq.~(\ref{DeltaT}),
we discussed the mixed correlation function
\begin{align}
G_{\rho}(y,y') = \frac{1}{N}\left[ (N-1)\,G^T_{\rho}(y,y')+ G^L_{\rho}(y,y') \right]
\end{align}
instead of the transverse correlation function,
in order to invoke the cancellation theorem below.
At the first order in $1/N$, the self-energy (a) of Fig.~\ref{SelfEnergy}(a)
accompanies an associated diagram
[where $G^{(0)}_{\rho}(y,y')$ is replaced by $M^0(y)M^0(y')$] \cite{PTP2} as
\begin{align}
\Sigma_{\rho}^{(a)}(y,y') = - \frac{2}{N}\, \Pi_{\rho}^{-1}(y,y')
\left[G^{(0)}_{\rho}(y,y') + M^0(y)M^0(y')\right] 
= (yy')^{-1-d/2} h^{(a)}(\upsilon).
\label{Sigma^a_E}
\end{align}
%There was a typo in an equation written below eq.~(4$\cdot$2) of ref.~\cite{PTP2}.
%It should read eq.~(\ref{h^a_E}) without $C_f$.
From eqs.~(\ref{g_E}), (\ref{A}), and (\ref{nu_E}),
the function $h^{(a)}(\upsilon)$ in eq.~(\ref{Sigma^a_E}) is given by
\begin{align}
h^{(a)}(\upsilon) &= - \frac{2}{N}\,\nu(\upsilon)
\left[ C_f g^{(0)}(\upsilon) + A \right]
%\label{h^a_E}
%\end{align}
%where $A$ is given by eq.~(\ref{A}).
%\begin{align}
%h^{(a)}(\upsilon)
%\nonumber\\
%&= - \frac{2}{N}\,\frac{C_F}{C_{\Pi}}\, C_f\, \frac{d-2}{d-1}
%\left[ 
%\frac{1}{\upsilon^{2\mu]}}F\left(\mu,\frac{1}{2}+\mu,1+\mu;\frac{1}{\upsilon^2}\right)
% - \frac{\Gamma(1+\mu)\Gamma\left(\frac{\textstyle1}{\textstyle2}-\mu\right)}
%{\sqrt{\pi}} \right]
%\nonumber\\
%&\times
%\frac{1}{\sqrt{2\pi}2^{\mu-5/2}\Gamma\left(\mu-\frac{\textstyle3}{\textstyle2}\right)}
%\left[ \frac{Q^2_{2\mu}(\upsilon)}{\upsilon^2-1}
% + \frac{1+2\mu}{2\mu-3}\frac{Q^1_{-1+2\mu}(\upsilon)}{(\upsilon^2-1)^{1/2}} \right]
\nonumber\\
&= - \frac{2}{N}\,\frac{C_F}{C_{\Pi}}\, C_f\, \frac{d-2}{2\mu}
\left[ 
\frac{1}{\upsilon^{2\mu}}F\left(\mu,\frac{1}{2}+\mu,1+\mu;\frac{1}{\upsilon^2}\right)
 - \frac{\Gamma(1+\mu)\Gamma\left(\frac{\textstyle1}{\textstyle2}-\mu\right)}
{\sqrt{\pi}} \right]
\nonumber\\
&\times
\frac{\Gamma(3+2\mu)}{2^{3\mu-1}
\Gamma\left(-\frac{\textstyle3}{\textstyle2}+\mu\right)
\Gamma\left(\frac{\textstyle3}{\textstyle2}+2\mu\right)}
\left[ \frac{1}{\upsilon^{3+2\mu}}
F\left(2+\mu,\frac{3}{2}+\mu,\frac{3}{2}+2\mu;\frac{1}{\upsilon^2}\right) \right.
\nonumber\\
&\hskip21mm \left. -\, \frac{\frac{\textstyle1}{\textstyle2}+2\mu}{(2\mu-3)(1+\mu)}
\frac{1}{\upsilon^{1+2\mu}}
F\left(1+\mu,\frac{1}{2}+\mu,\frac{1}{2}+2\mu;\frac{1}{\upsilon^2}\right) \right].
\label{h^a_E}
\end{align}
[There was a trivial typo in the equation below eq.~(4$\cdot$2) of ref.~\cite{PTP2}:
The first factor $g^{(0)}(\upsilon)$ should read $\nu(\upsilon)$.]
The corresponding correlation function is expressed as eq.~(\ref{G^(1)}),
in which $f(\upsilon)$ can be obtained by solving eqs. (\ref{Dg=h}) and (\ref{Df=g})
with $\cal D$ given by eq.~(\ref{D_E}).
Although eq.~(\ref{Dg=h}) has some ambiguity with respect to adding the multiples
of $g^{(0}(\upsilon)$, such an addition cancels exactly with the
corresponding terms occurred in Fig.~\ref{SelfEnergy}(b) and associated
diagrams involving $M^0(y)$ \cite{PTP2}.
This stems from the cancelation theorem proven in the Appendix of ref.~\cite{PTP1}
for a behavior like $U(y)\delta(y-y')$, where $U(y)$ is an arbitrary function of only
$y$ and momentum cutoff $\Lambda$.
Since $h^{(a)}(\upsilon)$ behaves as $\upsilon^{-1-4\mu}, \upsilon^{-1-2\mu}$ for
$\upsilon\rightarrow\infty$, the solution of eq.~(\ref{Df=g}), $f^{(a)}(\upsilon)$,
behaves also as $\upsilon^{-1-4\mu}, \upsilon^{-1-2\mu}$, and does not show
any logarithmic behavior like $\log\upsilon\,/\,\upsilon^{2\mu}$;
it does not show a behavior like $\log(\upsilon-1)$ for $\upsilon\rightarrow 1$, either.
In contrast to the cases of the ordinary and special transitions,
$f^{(a)}(1)\bigr|_{\rm subtracted}/y^{d-2}$ does not contribute to the tadpole diagram (b)
and associated tadpole diagrams involving $M^0(y)$ \cite{PTP2}, because
\begin{align}
\int_0^{\infty}\frac{dy}{y^{d-2}}\Pi^{-1}_q(y,y')
&= C_{\Pi}^{-1}\!\sqrt{y'} \left[ \int_0^{\infty}\frac{dy}{y^{d-5/2}}
\int_0^{\infty} dt\,t^{4-2\mu} J_{1/2+2\mu}(ty)J_{1/2+2\mu}(ty') \right.
\nonumber\\
&\hskip4mm \left. - \frac{1+2\mu}{y'} \int_0^{\infty}\frac{dy}{y^{d-3/2}}
 \int_0^{\infty} dt\,t^{2-2\mu} J_{-1/2+2\mu}(ty)J_{-1/2+2\mu}(ty') \right]
%\nonumber\\
%&= C_{\Pi}^{-1}\!\sqrt{y'}\frac{2^{-d+3/2}\sqrt{\pi}}{\Gamma(d-1)}
% \left[ \int_0^{\infty} dt\,t^{3/2} J_{1/2+2\mu}(ty')
%  - \frac{1+2\mu}{y'} \int_0^{\infty} dt\,t^{1/2} J_{-1/2+2\mu}(ty') \right]
%\nonumber\\
%&= \frac{C_{\Pi}^{-1}}{y'^2}\frac{2^{-d+3/2}\sqrt{\pi}}{\Gamma(d-1)}
%\left[ \frac{d^{3/2}\Gamma\left(\frac{\textstyle d+2}{\textstyle2}\right)}
%{\Gamma\left(\frac{\textstyle d-1}{\textstyle2}\right)}
%- d\frac{2^{1/2}\Gamma\left(\frac{\textstyle d}{\textstyle2}\right)}
%{\Gamma\left(\frac{\textstyle d-1}{\textstyle2}\right)} \right]
\nonumber\\
&=0.
\label{y*Pi^-1=0}
\end{align}
In this way, we found $\hat{\gamma}^T=0$ in eq.~(\ref{DeltaT}).

Next, we discuss the second-order terms in the mixed correlation function.
The diagrams at this order is divided into two groups, (A) and (B).
The first group (A) comprises those composed of two first-order self-energies,
$\Sigma^{(1)}_q(y,y_1)$ and $\Sigma^{(1)}_q(y_2,y')$, as
$\int\Sigma^{(1)}_q(y,y_1)\, G^{(0)}_q(y_1,y_2)\,\Sigma^{(1)}_q(y_2,y')\,dy_1dy_2$.
Again, the homogeneous solution $g^{(0)}(\upsilon)$ of eq.~(\ref{Dg=h})
does not contribute because of the cancellation theorem.
Also, the solution $f^{(a)}(\upsilon)$ of eq.~(\ref{Df=g})
does not produce relevant logarithmic correction.
In particular, eq.~(\ref{y*Pi^-1=0}) guarantees that 
$f^{(a)}(1)\bigr|_{\rm subtracted}/y^{d-2}$ does not contribute
to the corresponding tadpole diagram.
For example,
a second-order contribution to the correlation function $G^{(2)}_q(y,y')$
comes from the product of $\Sigma^{(a)}_q(y,y_1)$
of eq.~(\ref{Sigma^a_E}) and the second term in eq.~(\ref{G^(1)}):
\begin{align}
-\,\frac{\eta}{2N}
\int G^{(0)}_q(y,y_1) \Sigma^{(a)}_q(y_1,y_2)\, G^{(0)}_q(y_2,y')
\log(2y_2y'\Lambda^2)\,dy_1dy_2,
\label{GSigmaGlog}
\end{align}
If we take the logarithmic derivative $\Lambda(\partial/\partial\Lambda)$ at the outset,
eq.~(\ref{GSigmaGlog}) becomes identical to the contribution from
$(-\eta/N)\Sigma^{(a)}_q(y_1,y_2)$.
Therefore, the contribution from this term cancels with the
corresponding tadpole diagram due to the cancellation theorem
or does not contribute to the tadpole diagram due to eq.~(\ref{y*Pi^-1=0}).
The second group (B) comprises the remainder diagrams.
For the diagrams belonging to this group,
essentially the same argument as above holds
and no contribution occurs in $\hat{\gamma}^T$.
Hence, the $O(1/N^2)$ term in $\hat{\Delta}_{\phi}^T$ is zero.
Repeating similar arguments, we found that all higher-order terms
in $\hat{\Delta}_{\phi}^T$ are identically zero and that the relation
\begin{align}
\hat{\Delta}_{\phi}^T = d-1
\label{D^T=d-1}
\end{align}
holds to all orders in the $1/N$ expansion \cite{PTP2}.
In this case, Giombi and Khanchandani \cite{Giombi} discussed that,
at large $N$, the $N - 1$ transverse fields are just free fields in AdS
and that the fields are massless is related to the fact that these are Goldstone modes
for the spontaneously broken $O(N)$ symmetry.
Therefore, they also expected that the relation eq.~(\ref{D^T=d-1})
may hold to all orders in $1/N$.

\section{Summary}

Giombi and Khanchandani extended the boundary operator expansion
by McAvity and Osborn, and succeeded in obtaining the
series expansion result for the anomalous dimension at $O(1/N)$
in the case of the ordinary transition \cite{Giombi}.
Although they could not sum up the series,
we succeeded in proving that their result
coincides with the simple analytic form of ours,
although we found that their final formula expressed by two ${}_3F_2$ functions
is wrong.
Instead, we derived similar correct formulae in Appendices \ref{Appendix A}
and \ref{Appendix B}.

We noticed that the inverse of the single bubble
in semi-finite space,
which is required to perform the $1/N$ expansion,
was correctly evaluated using a method to integrate two point functions
with respect to $\rho$ completely, by McAvity and Osborn \cite{McAvity}
and also by Giombi and Khanchandani \cite{Giombi}
for the ordinary and special transitions,
although they did not analyze for the extraordinary transition.
It would be worthwhile to note that the relationship between
the single bubble and its inverse in the case of the special transition
is reversed in the case of the extraordinary transition.
This is transparent in the Fourier--Bessel integral representation in mixed space.
This was explained explicitly in Section \ref{Large N},
along with the calculation that leads to
the anomalous dimension in the large $N$ limit for the longitudinal field
in the case of the extraordinary transition.
The ($d-1$)-dimensional Fourier transform and the relevant differential operators
(along with their homogeneous solutions) were described in Appendix \ref{Differential}.
In the case of the special transition,
the calculation of solving the differential equation for the inverse
is complicated, and is relegated to Appendix \ref{InverseSpecial}.
It is significant that McAvity and Osborn \cite{McAvity}
proposed a more direct approach of calculating the inverse than ours;
see the Appendix B of their paper. 

However, in the case of the special transition,
Giombi and Khanchandani mentioned in their paper \cite{Giombi}
that they were unable to obtain the final result of the anomalous dimension
at $O(1/N)$, because two towers of operators appeared
in the boundary operator expansion.
This is related to the fact that the inverse of the single bubble
is expressed by a sum of two hypergeometric functions.
In such a situation, it is suitable to directly solve the
differential equations to compute the $1/N$ correction to
the correlation function without resorting the boundary operator expansion.
This method was explained in detail in Section \ref{1/N},
and Appendices \ref{Ordinary} and \ref{Special}
for the ordinary and special transitions, respectively,
which were not written in our earlier papers.
The final part of Section \ref{1/N} was devoted to a discussion of
the anomalous dimension for the transverse fields
to all orders in $1/N$ in the case of the extraordinary transition.
It is a challenge to explore the possibility
to obtain the $O(1/N)$ correction in $\hat{\Delta}_{\phi}^L$
as well as the $O(1/N^2)$ corrections in
$\hat{\Delta}_{\phi}^O$ and $\hat{\Delta}_{\phi}^S$, and left for the future study.

\acknowledgments
This work has been supported by the grant-in-aid (KAKENHI) for Scientific Research B
(Grant No.$\,$JP23K21094)
and Scientific Research C (Grant No.$\,$JP22K03472)
from Japan Society for the Promotion of Science (JSPS).

\appendix

\section{Alternative Form 1}
\label{Appendix A}

$\hat{\gamma}^O$ can be expressed with two ${}_3F_2$ functions
similar to the final formula in eq. (4.57) of ref.~\cite{Giombi}.
The simplest way to derive such a formula is described as follows.
By using
\begin{align}
\frac{\frac{\textstyle3(d-1)}{\textstyle2}+2k}{(d+2k)(2d-3+2k)}
= \frac{1}{4\left(\frac{\textstyle d}{\textstyle2}+k\right)}
+ \frac{1}{4\left(d-\frac{\textstyle3}{\textstyle2}+k\right)},
\label{Fraction}
\end{align}
eq.~(\ref{R2}) can be rewritten as
\begin{align}
R &= \frac{2^{3d-5}}{\sqrt{\pi}} \sum_{k=0}^{\infty}
\left(
\frac{\Gamma\left(\frac{\textstyle d}{\textstyle2}+k\right)}
{\Gamma\left(\frac{\textstyle d}{\textstyle2}+1+k\right)}
+  \frac{\Gamma\left(d-\frac{\textstyle3}{\textstyle2}+k\right)}
{\Gamma\left(d-\frac{\textstyle1}{\textstyle2}+k\right)}
\right)
\frac{\Gamma\left(\frac{\textstyle3(d-1)}{\textstyle2}+k\right)
\Gamma\left(\frac{\textstyle d}{\textstyle2}+k\right)}
{\Gamma\left(d-\frac{\textstyle1}{\textstyle2}+k\right)}\frac{1}{k!}
\nonumber\\
&= \frac{2^{3d-5}}{\sqrt{\pi}}
\frac{\Gamma\left(\frac{\textstyle3(d-1)}{\textstyle2}\right)
\Gamma\left(\frac{\textstyle d}{\textstyle2}\right)}
{\Gamma\left(d-\frac{\textstyle1}{\textstyle2}\right)}\; Z
\label{R10}
\end{align}
with
\begin{align}
Z &=\, \frac{2}{d}\, {}_3F_2\left(\frac{d}{2},\,\frac{3(d-1)}{2},\,\frac{d}{2};\,
\frac{d}{2}+1,\,d-\frac{1}{2};\,1\right)
\nonumber\\
&+\,\frac{2}{2d-3}\, {}_3F_2\left(d-\frac{3}{2},\,\frac{3(d-1)}{2},\,\frac{d}{2};\,
d-\frac{1}{2},\,d-\frac{1}{2};\,1\right).
\label{Z1}
\end{align}
However, the ${}_3F_2(a,b,c;e,f;1)$ function is well defined only for $s=e+f-a-b-c>0$.
When this condition is not satisfied, we can use the analytic continuation
of the ${}_3F_2$ function as written in eq. (2.11) of Gopakumara and Sinha's
paper \cite{Gopakumara} as well as in the footnote \#15
(P. 40) of Giombi and Khanchandan's paper \cite{Giombi}.
This is the mathematical formula given in
Table IIA for $F_p (0)$ (p. 18) of ref.~\cite{HGS}, which reads
\begin{align}
{}_3F_2(a,b,c;\,e,f;\,1) = \frac{\Gamma(e)\Gamma(s)}{\Gamma(e-a)\Gamma(s+a)}
{}_3F_2(a,\,f-b,f-c;\,f,\,s+a;\,1),
\label{AnalContinu}
\end{align}
%\begin{align}
%3F2[a1, a2, a3; b1, b2; 1]
%= \frac{\Gamma(b1)\Gamma(b1+b2-a1-a2-a3)}
%{\Gamma(b1-a1)\Gamma(b1+b2-a2-a3)}
% 3F2F[a1,b2-a2,b2-a3;b2,b1+b2-a2-a3;1].
%\end{align}
where $s = e + f - a - b - c$.
Then, eq.~(\ref{Z1}) is converted to
\begin{align}
Z = \Gamma(2-d) &\left[
\frac{2}{d}\frac{\Gamma\left(\frac{\textstyle d}{\textstyle2}+1\right)}
{\Gamma\left(2-\frac{\textstyle d}{\textstyle2}\right)}\,
{}_3F_2\left(\frac{d}{2},\,1-\frac{d}{2},\,\frac{d}{2}-\frac{1}{2};\,
d-\frac{1}{2},\,2-\frac{d}{2};\,1\right)
\right.
\nonumber\\
&\left. + \frac{2}{2d-3}
\frac{\Gamma\left(d-\frac{\textstyle1}{\textstyle2}\right)}{\sqrt{\pi}}\,
{}_3F_2\left(d-\frac{3}{2},\,1-\frac{d}{2},\,\frac{d}{2}-\frac{1}{2};\,
d-\frac{1}{2},\,\frac{1}{2};\,1\right)
\right].
\label{Z2}
\end{align}
By inserting this $Z$ into $R$ of eq.~(\ref{R10})
and using the resulting $R$ in eq.~(\ref{gammaO1}),
one obtains an expression for $\hat{\gamma}^O$ in terms of two ${}_3F_2$ functions.
We have numerically confirmed that this expression agrees precisely
with our result (\ref{Eq_O}).

\section{Alternative Form 2}
\label{Appendix B}

Another expression for $\hat{\gamma}^O$ involving two ${}_3F_2$ functions
and a constant term, which more closely resembles the final formula in eq. (4.57)
of ref.~\cite{Giombi}, can be derived as follows.
First, as suggested in the footnote \#15 (P. 40) of ref. \cite{Giombi},
we exclude the $k=0$ term from the summation (\ref{R2}) as
\begin{align}
R &= \frac{2^{3d-3}}{\sqrt{\pi}} \left[ \sum_{k=1}^{\infty}
\frac{\left(\frac{\textstyle3(d-1)}{\textstyle2}+2k\right)
\Gamma\left(\frac{\textstyle3(d-1)}{\textstyle2}+k\right)
\Gamma\left(\frac{\textstyle d}{\textstyle2}+k\right)}
{(d+2k)(2d-3+2k)\Gamma\left(d-\frac{\textstyle1}{\textstyle2}+k\right)}\frac{1}{k!}
+ \frac{\left(\frac{\textstyle3(d-1)}{\textstyle2}\right)
\Gamma\left(\frac{\textstyle3(d-1)}{\textstyle2}\right)
\Gamma\left(\frac{\textstyle d}{\textstyle2}\right)}
{d(2d-3)\Gamma\left(d-\frac{\textstyle1}{\textstyle2}\right)} \right]
\nonumber\\
&= \frac{2^{3d-3}}{\sqrt{\pi}} \left[ \sum_{k=0}^{\infty}
\frac{\left(\frac{\textstyle3d+1}{\textstyle2}+2k\right)
\Gamma\left(\frac{\textstyle3d-1}{\textstyle2}+k\right)
\Gamma\left(\frac{\textstyle d}{\textstyle2}+1+k\right)}
{(d+2+2k)(2d-1+2k)(1+k)\Gamma\left(d+\frac{\textstyle1}{\textstyle2}+k\right)}\frac{1}{k!}
+ \frac{\Gamma\left(\frac{\textstyle3d-1}{\textstyle2}\right)
\Gamma\left(\frac{\textstyle d}{\textstyle2}\right)}
{d(2d-3)\Gamma\left(d-\frac{\textstyle1}{\textstyle2}\right)} \right].
\label{R4}
\end{align}
The rational expression except for the Gamma functions inside the summation
can be decomposed into three partial fractions as
\begin{align}
\frac{\frac{\textstyle3d+1}{\textstyle2}+2k}{(d+2+2k)(2d-1+2k)(1+k)}
= \frac{3(d-1)}{2d(2d-3)(1+k)}
-\frac{1}{2d\left(\frac{\textstyle d}{\textstyle2}+1+k\right)}
-\frac{1}{4\left(d-\frac{\textstyle3}{\textstyle2}\right)\left(d-\frac{\textstyle1}{\textstyle2}+k\right)}.
\label{ThreeFractions}
\end{align}
The contribution from the first term
can be evaluated together with the constant term in eq.~(\ref{R4}) as
\begin{align}
T_1 &= \frac{2^{3d-3}}{\sqrt{\pi}} \frac{3(d-1)}{2d(2d-3)}
\left[ \sum_{k=0}^{\infty}
\frac{\Gamma\left(\frac{\textstyle3d-1}{\textstyle2}+k\right)
\Gamma\left(\frac{\textstyle d}{\textstyle2}+1+k\right)}
{\Gamma\left(d+\frac{\textstyle1}{\textstyle2}+k\right)}\frac{1}{(1+k)!}
+ \frac{\Gamma\left(\frac{\textstyle3(d-1)}{\textstyle2}\right)
\Gamma\left(\frac{\textstyle d}{\textstyle2}\right)}
{\Gamma\left(d-\frac{\textstyle1}{\textstyle2}\right)} \right]
\nonumber\\
&= \frac{2^{3d-3}}{\sqrt{\pi}} \frac{3(d-1)}{2d(2d-3)}
\sum_{k=0}^{\infty}
\frac{\Gamma\left(\frac{\textstyle3(d-1)}{\textstyle2}+k\right)
\Gamma\left(\frac{\textstyle d}{\textstyle2}+k\right)}
{\Gamma\left(d-\frac{\textstyle1}{\textstyle2}+k\right)}\frac{1}{k!}
\nonumber\\
&= \frac{2^{3d-3}}{\sqrt{\pi}} \frac{3(d-1)}{2d(2d-3)}
\frac{\Gamma\left(\frac{\textstyle3(d-1)}{\textstyle2}\right)
\Gamma\left(\frac{\textstyle d}{\textstyle2}\right)}
{\Gamma\left(d-\frac{\textstyle1}{\textstyle2}\right)}
F\left(\frac{3(d-1)}{2},\frac{d}{2};d-\frac{1}{2};1\right)
\nonumber\\
&= \frac{2^{3d-3}}{\sqrt{\pi}} \frac{3(d-1)}{2d(2d-3)}
\frac{\Gamma\left(\frac{\textstyle3(d-1)}{\textstyle2}\right)
\Gamma\left(\frac{\textstyle d}{\textstyle2}\right)}
{\Gamma\left(d-\frac{\textstyle1}{\textstyle2}\right)}
\frac{\Gamma\left(d-\frac{\textstyle1}{\textstyle2}\right)\Gamma(1-d)}
{\Gamma\left(\frac{\textstyle d-1}{\textstyle2}\right)
\Gamma\left(1-\frac{\textstyle d}{\textstyle2}\right)}
\nonumber\\
&= \frac{2^{3d-3}}{\sqrt{\pi}} \frac{1}{d(2d-3)}
\frac{\Gamma\left(\frac{\textstyle3d-1}{\textstyle2}\right)
\Gamma\left(\frac{\textstyle d}{\textstyle2}\right)\Gamma(1-d)}
{\Gamma\left(\frac{\textstyle d-1}{\textstyle2}\right)
\Gamma\left(1-\frac{\textstyle d}{\textstyle2}\right)}.
\label{T1}
\end{align}
The contributions from the remaining second and third terms
in eq.~(\ref{ThreeFractions}) are evaluated, respectively, as
\begin{subequations}
\begin{align}
T_2 &= - \frac{2^{3d-3}}{2d\!\sqrt{\pi}}
\sum_{k=0}^{\infty}
\frac{\Gamma\left(\frac{\textstyle3d-1}{\textstyle2}+k\right)
\Gamma\left(\frac{\textstyle d}{\textstyle2}+1+k\right)}
{\left(\frac{\textstyle d}{\textstyle2}+1+k\right)
\Gamma\left(d+\frac{\textstyle1}{\textstyle2}+k\right)}\frac{1}{k!}
\nonumber\\
&= - \frac{2^{3d-3}}{2d\!\sqrt{\pi}}
\frac{\Gamma\left(\frac{\textstyle3d-1}{\textstyle2}\right)
\Gamma\left(\frac{\textstyle d}{\textstyle2}+1\right)}
{\left(\frac{\textstyle d}{\textstyle2}+1\right)
\Gamma\left(d+\frac{\textstyle1}{\textstyle2}\right)}
{}_3F_2\left(\frac{d}{2}+1,\frac{3d-1}{2},\frac{d}{2}+1;\,d+\frac{1}{2},\frac{d}{2}+2;\,1\right)
\label{T2}
\end{align}
and
\begin{align}
T_3 &= - \frac{2^{3d-3}}{4\left(d-\frac{\textstyle 3}{\textstyle2}\right)\!\!\sqrt{\pi}}\,
\sum_{k=0}^{\infty}
\frac{\Gamma\left(\frac{\textstyle3d-1}{\textstyle2}+k\right)
\Gamma\left(\frac{\textstyle d}{\textstyle2}+1+k\right)}
{\left(d-\frac{\textstyle1}{\textstyle2}+k\right)
\Gamma\left(d+\frac{\textstyle1}{\textstyle2}+k\right)}\frac{1}{k!}
\nonumber\\
&= - \frac{2^{3d-3}}{4\left(d-\frac{\textstyle 3}{\textstyle2}\right)\!\!\sqrt{\pi}}
\frac{\Gamma\left(\frac{\textstyle3d-1}{\textstyle2}\right)
\Gamma\left(\frac{\textstyle d}{\textstyle2}+1\right)}
{\left(d-\frac{\textstyle1}{\textstyle2}\right)
\Gamma\left(d+\frac{\textstyle1}{\textstyle2}\right)}
{}_3F_2\left(d-\frac{1}{2},\frac{3d-1}{2},\frac{d}{2}+1;\,d+\frac{1}{2},d+\frac{1}{2};\,1\right).
\label{T3}
\end{align}
\end{subequations}
By using the analytic continuation formula of eq.~(\ref{AnalContinu}),
$T_2$ and $T_3$ are converted, respectively, to
\begin{subequations}
\begin{align}
T_2 &= - \frac{2^{3d-3}}{2d\!\sqrt{\pi}}
\frac{\Gamma\left(\frac{\textstyle3d-1}{\textstyle2}\right)
\Gamma\left(\frac{\textstyle d}{\textstyle2}\right)
\Gamma\left(\frac{\textstyle d}{\textstyle2}\right)\Gamma(1-d)}
{\left(\frac{\textstyle d}{\textstyle2}+1\right)
\Gamma\left(\frac{\textstyle d-1}{\textstyle2}\right)
\Gamma\left(2-\frac{\textstyle d}{\textstyle2}\right)}
{}_3F_2\left(\frac{d}{2}+1,\frac{5}{2}-d,\,1;\,\frac{d}{2}+2,2-\frac{d}{2};\,1\right),
\label{T2'}
\\
T_3 &= - \frac{2^{3d-3}}{4\left(d-\frac{\textstyle 3}{\textstyle2}\right)\!\!\sqrt{\pi}}
\frac{\Gamma\left(\frac{\textstyle3d-1}{\textstyle2}\right)
\Gamma\left(\frac{\textstyle d}{\textstyle2}\right)
\Gamma\left(\frac{\textstyle d}{\textstyle2}\right)\Gamma(1-d)}
{\left(d-\frac{\textstyle1}{\textstyle2}\right)\sqrt{\pi}}
{}_3F_2\left(d-\frac{1}{2},1-\frac{d}{2},\frac{d-1}{2};\,d+\frac{1}{2},\,\frac{1}{2};\,1\right).
\label{T3'}
\end{align}
\end{subequations}
Adding all these three terms, $T_1, T_2$, and $T_3$, and multiplying
the remaining coefficient, we obtain
\begin{align}
\hat{\gamma}^O &= \frac{4-d}{d\Gamma(d-2)}
\frac{2^{2d-2}\sin\left(\frac{\textstyle\pi d}{\textstyle2}\right)
\Gamma\left(\frac{\textstyle d-1}{\textstyle2}\right)
\Gamma\left(\frac{\textstyle3d-1}{\textstyle2}\right)\Gamma(1-d)}{\pi^{3/2}}
\left[\frac{1}{d(2d-3)\Gamma\left(\frac{\textstyle d-1}{\textstyle2}\right)
\Gamma\left(1-\frac{\textstyle d}{\textstyle2}\right)} \right.
\nonumber\\
&- \frac{1}{4\left(\frac{\textstyle d}{\textstyle2}+1\right)
\Gamma\left(\frac{\textstyle d-1}{\textstyle2}\right)
\Gamma\left(2-\frac{\textstyle d}{\textstyle2}\right)}
{}_3F_2\left(\frac{d}{2}+1,\frac{5}{2}-d,\,1;\,\frac{d}{2}+2,2-\frac{d}{2};\,1\right)
\nonumber\\
&\left. - \frac{d}{8\sqrt{\pi}\left(d-\frac{\textstyle3}{\textstyle2}\right)
\left(d-\frac{\textstyle1}{\textstyle2}\right)}
{}_3F_2\left(d-\frac{1}{2},1-\frac{d}{2},\frac{d-1}{2};\,d+\frac{1}{2},\,\frac{1}{2};\,1\right)
\right].
\label{gammaO3}
\end{align}
Again, we have numerically confirmed that this expression agrees precisely
with our result (\ref{Eq_O}).

\section{Differential Operator}
\label{Differential}

Let us consider a differential operator
\begin{align}
{\cal E} = y^2 \left[\, \nabla_{\rho}^2 + \frac{\partial^2}{\partial y^2}
+ \frac{\alpha}{y}\frac{\partial}{\partial y} - \frac{\beta}{y^2} \,\right]
\label{E}
\end{align}
and operate it onto a function $(yy')^{\zeta}f(\upsilon)$.
Here, the parameters, $\alpha$, $\beta$, and $\gamma$,
and the function, $f(\upsilon)$, are arbitrary.
We want to seek the corresponding differential operator $\cal D$,
which is expressed by the $\upsilon$ variable only and satisfies
\begin{align}
{\cal E}\,(yy')^{\zeta} f(\upsilon) = (yy')^{\zeta}\, {\cal D}\, f(\upsilon).
\label{E=D}
\end{align}
From $(\partial/\partial y) \upsilon
%= - (y^2+y'^2+\rho^2)/2y^2y' + 1/y'
= (y^2-y'^2-\rho^2)/2yy'\times(1/y)$,
$\nabla_{\rho}^2 = (d-2)/\rho\times (\partial/\partial\rho)
+(\partial^2/\partial\rho^2)$,
$(\partial/\partial\rho)f = \rho f'/yy'$,
and
$(\partial/\partial\rho)(\rho f'/yy') =  f'/yy' +  \rho^2 f''/(yy')^2$,
the differentiations of $(yy')^{\zeta}f(\upsilon)$ are
\begin{subequations}
\begin{align}
\frac{\partial}{\partial y} (yy')^{\zeta}f(\upsilon)
&= \zeta (yy')^{\zeta} \frac{f}{y}
 + (yy')^{\zeta} \left( \frac{y^2-y'^2-\rho^2}{2yy'} \right) \frac{f'}{y},
\\
\frac{\partial^2}{\partial y^2} (yy')^{\zeta}f(\upsilon)
&= \zeta(\zeta-1) (yy')^{\zeta} \frac{f}{y^2}
 +2(\zeta-1) (yy')^{\zeta} \left( \frac{y^2-y'^2-\rho^2}{2yy'} \right)
  \frac{f'}{y^2}
\nonumber\\
&+ (yy')^{\zeta} \frac{f'}{yy'}
+ (yy')^{\zeta} \left( \frac{y^2-y'^2-\rho^2}{2yy'} \right)^2 \frac{f''}{y^2},
\\
\nabla_{\rho}^2 (yy')^{\zeta}f(\upsilon) &= (d-1) (yy')^{\zeta} \frac{f'}{yy'}
+ (yy')^{\zeta} \rho^2 \frac{f''}{(yy')^2}.
\end{align}
\end{subequations}
Then, using the relations, $(y^2-y'^2-\rho^2)/2yy' = y/y' - \upsilon$ and
%%%%%%%%%%%%%%%%%%%%%%%%%%%%%%%%
%
%$(y^2-y'^2-\rho^2)^2 + 4\rho^2y^2 = (y^2+y'^2+\rho^2)^2 - 4y^2y'^2$
%
%$(y^2-y'^2-\rho^2)^2/(2yy')^2 + \rho^2/y'^2 = \upsilon^2 - 1$
%%%%%%%%%%%%%%%%%%%%%%%%%%%%%%%%
$(y^2-y'^2-\rho^2)^2/(2yy')^2y^2 + \rho^2/(yy')^2 = (\upsilon^2 - 1)/y^2$,
%%%%%%%%%%%%%%%%%%%%%%%%%%%%%%%%
%
%$(y^2+y'^2+\rho^2)^2-4y^2y'^2$
%
%$= (-y^2+y'^2)^2 + 2(y^2+y'^2)\rho^2 + \rho^4$
%
%$= (-y^2+y'^2)^2 + 2(-y^2+y'^2)\rho^2 + \rho^4 + 4y^2\rho^2$
%
%$= (-y^2+y'^2+\rho^2)^2 + 4y^2\rho^2$
%%%%%%%%%%%%%%%%%%%%%%%%%%%%%%%%
we find %$1-2\mu=\zeta$, $-2\mu=\zeta-1$
\begin{align}
{\cal E}\,(yy')^{\zeta} f(\upsilon) &=
(yy')^{\zeta}\,\left[\, (\upsilon^2-1) \frac{d^2}{d\upsilon^2}
- %( \alpha - 4\mu )\,
( \alpha + 2\zeta - 2 )\, \upsilon \frac{d}{d\upsilon} \right.
\nonumber\\
&\left. +\; %\Bigl\{ d - 1 + \alpha + 2(\zeta-1) + 1 \Bigr\}
(d + \alpha + 2\zeta - 2)\,
\frac{y}{y'} \frac{d}{d\upsilon}
 + %(1-2\mu)(\alpha-2\mu)\,
\zeta(\alpha+\zeta-1) - \beta\,
 \right]\, f(\upsilon).
 \label{E=}
\end{align}
%$\alpha=-1+2\mu$, $2\mu=d-1$, $\zeta=1-2\mu$
%$\rightarrow$ $d + \alpha + 2\zeta - 2 = d - 1 + 2\mu + 2 - 4\mu - 2 = d - 1 - 2\mu = 0$
To arrive at eq.~(\ref{E=D}), 
we have to assume that the term proportional to $y/y'$ should vanish
in the right-hand side, which leads to the necessary condition,
\begin{align}
2 - \alpha - 2\zeta = d.
\label{condition}
\end{align}
Under this condition, the differential operator $\cal D$ in eq.~(\ref{E=D})
is identified as
\begin{align}
{\cal D} = (\upsilon^2-1) \frac{d^2}{d\upsilon^2}
+ d\, \upsilon \frac{d}{d\upsilon}
+ \zeta(1-d-\zeta) - \beta.
 \label{D}
\end{align}
%
%$\zeta(1-d-\zeta) - \beta=(d/2-1)(d/2)+\mu^2-1/4=(d/2)^2-(d/2)-\mu^2+1/4
%= (d - 1 - 2\mu)(d - 1 + 2\mu)/4$
%

For the correlation function in the large $N$ limit, eq.~(\ref{G^0}),
which satisfies the differential equation (\ref{LG=d}) with the
differential operator $\cal L$ in eq.~(\ref{L}),
the parameters in eqs.~(\ref{E}) and (\ref{E=D}) are identified to be
$\alpha=0$, $\beta=\mu^2-1/4$, and $\zeta=1-d/2$,
which certainly satisfy the condition (\ref{condition}).
The corresponding operator $\cal D$ in eq.~(\ref{D}) is given by
[see eq.~(3$\cdot$10) in ref.~\cite{PTP2}]
\begin{align}
{\cal D}
% &= (\upsilon^2-1) \frac{d^2}{d\upsilon^2}
%+ d\, \upsilon \frac{d}{d\upsilon}
%+ \frac{d(d-2)}{4} - \frac{(d-n)^2}{4} + \frac{1}{4}
%\nonumber\\
%&= (\upsilon^2-1) \frac{d^2}{d\upsilon^2}
%+ d\, \upsilon \frac{d}{d\upsilon}
%+ \frac{d^2-2d}{4} - \frac{d^2-2nd+n^2}{4} + \frac{1}{4}
%\nonumber\\
%&
%= (\upsilon^2-1) \frac{d^2}{d\upsilon^2}
%+ d\, \upsilon \frac{d}{d\upsilon}
%+ \frac{2(n-1)d+1-n^2}{4}
%\nonumber\\
%&
= (\upsilon^2-1) \frac{d^2}{d\upsilon^2}
+ d\, \upsilon \frac{d}{d\upsilon}
+ \frac{1}{4}(d-2\mu-1)(d+2\mu-1).
\label{D'}
\end{align}
%$(d-2\mu-1)(d+2\mu-1)=(d-d+n-1)(2d-n-1)=(n-1)(2d-n-1)=2(n-1)d+1-n^2$
The last constant term in eq.~(\ref{D'}) is
$d-2=2\mu+1$, $2d-6=4(1+\mu)$, and 0, respectively,
for the ordinary (or anisotropic special) transition with $\mu=(d-3)/2$,
the special transition with $\mu=(d-5)/2$,
and the extraordinary transition with $\mu=(d-1)/2$.
Assuming the solution of
\begin{align}
{\cal D}\, w(\upsilon) = 0
\label{Dw=0}
\end{align}
in a form
\begin{align}
w(\upsilon) = \frac{R(\upsilon)}{(\upsilon^2-1)^{(d-2)/4}},
\end{align}
we find that the function $R(\upsilon)$ satisfies
the associate Legendre differential equation
\begin{align}
\left[\, (\upsilon^2-1)\frac{d^2}{d\upsilon^2}+2\upsilon \frac{d}{d\upsilon}
- \left(- \mu - \frac{1}{2}\right)\left(- \mu + \frac{1}{2}\right) 
- \left(\frac{d}{2}-1\right)^2 \frac{1}{\upsilon^2-1}\, \right] R(\upsilon) = 0.
\end{align}
The two independent solutions of this equation are the first kind 
of the associated Legendre function,
$P_{-\mu-1/2}^{\,d/2-1}(\upsilon)=P_{\mu-1/2}^{\,d/2-1}(\upsilon)$,
and the second kind of the associated Legendre function,
$Q_{-\mu-1/2}^{\,d/2-1}(\upsilon)$ or $Q_{\mu-1/2}^{\,d/2-1}(\upsilon)$.
%
%For the ordinary ($\mu=(d-3)/2$) and special ($\mu=(d-3)/2$) transitions,
%the correlation function is given by
%\begin{align}
%g^{(0)}(\upsilon) &= 
%\frac{\Gamma\left(2-\frac{\textstyle d}{\textstyle2}\right)}{2^{d/2+1}}
%\frac{P_{-\mu-1/2}^{\,d/2-1}(\upsilon)}{(\upsilon^2-1)^{(d-2)/4}},
%\label{P_O}
%\end{align}
%
%They are expressed as
%\begin{subequations}
%\begin{align}
%P_{-\mu-1/2}^{\,d/2-1}(\upsilon) &= P_{\mu-1/2}^{\,d/2-1}(\upsilon) =
%\frac{1}{\Gamma\left(2-\frac{\textstyle d}{\textstyle2}\right)}
%\left(\frac{\upsilon+1}{\upsilon-1}\right)^{(d-2)/4}
%F\left(\frac{1}{2}+\mu,\,\frac{1}{2}-\mu,\,2-\frac{d}{2};\,\frac{1-\upsilon}{2}\right),
%\\
%Q_{-\mu-1/2}^{\,d/2-1}(\upsilon) &= \frac{e^{i\pi(d-2)/2}
%\Gamma\left(\frac{\textstyle d-1}{\textstyle2}-\mu\right)\sqrt{\pi}}
%{2^{1/2-\mu}\Gamma(1-\mu)}
%\frac{(\upsilon^2-1)^{(d-2)/4}}{\upsilon^{(d-1)/2-\mu}}
%F\left(\frac{(d+1)-2\mu}{4},\,\frac{(d-1)-2\mu}{4},\,1-\mu;\,\frac{1}{\upsilon^2}\right),
%\\
%Q_{\mu-1/2}^{\,d/2-1}(\upsilon) &= \frac{e^{i\pi(d-2)/2}
%\Gamma\left(\frac{\textstyle d-1}{\textstyle2}+\mu\right)\sqrt{\pi}}
%{2^{\mu+1/2}\Gamma(1+\mu)}
%\frac{(\upsilon^2-1)^{(d-2)/4}}{\upsilon^{(d-1)/2+\mu}}
%F\left(\frac{(d+1)+2\mu}{4},\,\frac{(d-1)+2\mu}{4},\,1+\mu;\,\frac{1}{\upsilon^2}\right).
%\end{align}
%\end{subequations}
The explicit forms of $g^{(0)}(\upsilon)$
for the ordinary, special, and extraordinary transitions are
given in eqs.~(\ref{g_O}), (\ref{g_S}), and (\ref{g_E}), respectively.
Note that $Q_{\mu-1/2}^{\,d/2-1}(\upsilon)$ has the same form
as $P_{-\mu-1/2}^{\,d/2-1}(\upsilon)$
for the ordinary and special transitions except for the coefficient,
while $Q_{-\mu-1/2}^{\,d/2-1}(\upsilon)$ is an infinite constant proportional to $\Gamma(0)$
for $\mu=(d-1)/2$ in the case of the extraordinary transition.
These functions can be obtained from the correlation function
in mixed space, eq.~(\ref{JJ}), by inversely Fourier transforming as
\begin{align}
G_{\rho}^{(0)}(y,y') = 2^{(d-3)/2}\Gamma\left(\frac{d-1}{2}\right)K_{d-1}
\frac{\sqrt{yy'}}{\rho^{(d-3)/2}}
\int_0^{\infty} dq\,q^{(d-1)/2}\,J_{(d-3)/2}(q\rho)I_{\mu}(qy)K_{\mu}(qy'),
\label{G^0_int}
\end{align}
which is based on the Poisson integral in the ($d-1$)-dimensional Fourier transform,
%\begin{align}
%J_n(y) = \frac{(y/2)^n}{\sqrt{\pi}\Gamma\left(n+\frac{\textstyle1}{\textstyle2}\right)}
%\int_0^{\pi} \cos(y\cos\theta)\sin^{2n}\theta\, d\theta
%\end{align}
%\begin{align}
%\int_0^{\pi} \cos(q\rho\cos\theta)\sin^{d-3}\theta\, d\theta
%= \frac{\sqrt{\pi}\Gamma\left(\frac{\textstyle d}{\textstyle2}-1\right)}
%{\left(\frac{\textstyle q\rho}{\textstyle2}\right)^{(d-3)/2}}
%J_{(d-3)/2}(q\rho)
%\end{align}
%\begin{align}
%\int_0^{\pi} \sin^{\alpha}\theta\, d\theta
% = \frac{\sqrt{\pi}\Gamma\left(\frac{\textstyle\alpha+1}{\textstyle2}\right)}
% {\Gamma\left(\frac{\textstyle\alpha+2}{\textstyle2}\right)}
%\end{align}
%\begin{align}
%\int_0^{\pi} \sin^{d-3}\theta\, d\theta
% = \frac{\sqrt{\pi}\Gamma\left(\frac{\textstyle d}{\textstyle2}-1\right)}
% {\Gamma\left(\frac{\textstyle d-1}{\textstyle2}\right)}
%\end{align}
\begin{align}
K_{d-1}\frac{\displaystyle\int_0^{\pi} \cos(q\rho\cos\theta)\sin^{d-3}\theta\, d\theta}
{\displaystyle\int_0^{\pi} \sin^{d-3}\theta\, d\theta}
= 2^{(d-3)/2}\Gamma\left(\frac{\textstyle d-1}{\textstyle2}\right)K_{d-1}
\frac{J_{(d-3)/2}(q\rho)}{(q\rho)^{(d-3)/2}},
\label{Fourier}
\end{align}
where $K_{d-1}$ is defined in eq.~(\ref{K_d-1}).
The integral (\ref{G^0_int}) actually yields eqs.~(\ref{g_O})-(\ref{g_E})
using the integral formulae (6.578.11) and (6.578.t6) of ref. \cite{Gradshteyn}.

%森口他、数学公式集III, p.119 の2段目の公式とref.~\cite{PTP2}の(3.7)式との比較
%\begin{align}
%&\frac{\Gamma(d-1)}
%{\sqrt{2}2^{d/2}\Gamma\left(\frac{\textstyle d+1}{\textstyle2}\right)}
%\times 2^{(d-3)/2}\Gamma\left(\frac{\textstyle d-1}{\textstyle2}\right)K_{d-1}
%= \frac{2^{(d-3)/2}(d-2)\Gamma(d-2)}
%{\sqrt{2}2^{d/2}\left(\frac{\textstyle d-1}{\textstyle2}\right)}K_{d-1}
%\nonumber\\
%&= \frac{(d-2)\Gamma(d-2)}{2(d-1)}K_{d-1}
%= \left(\frac{d-2}{d-1}\right)C_f
%\end{align}

The other solution of ${\cal D}\,w^{(0)}(\upsilon)=0$,
which is independent of $g^{(0)}(\upsilon)$, is given as follows.
For the ordinary (or anisotropic special) transition, $\mu=(d-3)/2$,
\begin{subequations}
\begin{align}
%P_{-\mu-1/2}^{\,d/2-1}(\upsilon)
%&= \frac{1}{\Gamma\left(2-\frac{\textstyle d}{\textstyle2}\right)}
%\left(\frac{\upsilon+1}{\upsilon-1}\right)^{(d-2)/4}
%F\left(2-\frac{d}{2},\,\frac{d}{2}+1,\,2-\frac{d}{2};\,\frac{1-\upsilon}{2}\right)
%\nonumber\\
%&= \frac{1}{\Gamma\left(2-\frac{\textstyle d}{\textstyle2}\right)}
%\left(\frac{\upsilon+1}{\upsilon-1}\right)^{(d-2)/4}
%\left(1-\frac{1-\upsilon}{2}\right)^{-1-d/2}
%\nonumber\\
%&= \frac{2^{d/2+1}}{\Gamma\left(2-\frac{\textstyle d}{\textstyle2}\right)}
%\frac{1}{(\upsilon^2-1)^{(d-2)/4}},
%\label{P_O}
%\\
%Q_{-\mu-1/2}^{\,d/2-1}(\upsilon) &= \frac{e^{i\pi(d-2)/2}
%\Gamma(1)
%\sqrt{\pi}}
%{2^{2-d/2}\Gamma(1-\mu)}
%\frac{(\upsilon^2-1)^{(d-2)/4}}{\upsilon}
%F\left(1,\,\frac{1}{2},\,1-\mu;\,\frac{1}{\upsilon^2}\right).
%\\
w^{(0)}(\upsilon) &= \frac{1}{\upsilon}
F\left(1,\,\frac{1}{2},\,1-\mu;\,\frac{1}{\upsilon^2}\right)
= \frac{2^{2-d/2}\Gamma(1-\mu)}{e^{i\pi(d-2)/2}\sqrt{\pi}}
\frac{Q_{-\mu-1/2}^{\,d/2-1}(\upsilon)}{(\upsilon^2-1)^{(d-2)/4}}
%\nonumber\\
%&= \frac{1}{\upsilon} \left(1-\frac{1}{\upsilon^2}\right)^{-\mu-1/2}
%F\left(-\mu,\,\frac{1}{2}-\mu,\,1-\mu;\,\frac{1}{\upsilon^2}\right)
%\nonumber\\
%&= \frac{\upsilon^{2\mu+1}}{\upsilon(\upsilon^2-1)^{\mu+1/2}}
%F\left(-\mu,\,\frac{1}{2}-\mu,\,1-\mu;\,\frac{1}{\upsilon^2}\right)
\nonumber\\
& = g^{(0)}(\upsilon)\, \upsilon^{2\mu}
F\left(\frac{1}{2}-\mu,\,-\mu,\,1-\mu;\,\frac{1}{\upsilon^2}\right);
\label{w_O}
%\\
%Q_{\mu-1/2}^{\,d/2-1}(\upsilon)&= \frac{e^{i\pi(d-2)/2}
%\Gamma(d-2)\sqrt{\pi}}{2^{d/2-1}\Gamma\left(\frac{\textstyle d-1}{\textstyle2}\right)}
%\frac{(\upsilon^2-1)^{(d-2)/4}}{\upsilon^{d-2}}
%F\left(\frac{d-1}{2},\,\frac{d-2}{2},\,\frac{d-1}{2};\,\frac{1}{\upsilon^2}\right)
%\nonumber\\
%&= \frac{e^{i\pi(d-2)/2}
%\Gamma(d-2)\sqrt{\pi}}{2^{d/2-1}\Gamma\left(\frac{\textstyle d-1}{\textstyle2}\right)}
%\frac{(\upsilon^2-1)^{(d-2)/4}}{\upsilon^{d-2}}
%\left(1-\frac{1}{\upsilon^2}\right)^{(2-d)/2}
%\nonumber\\
%&= \frac{e^{i\pi(d-2)/2}
%\Gamma(d-2)\sqrt{\pi}}{2^{d/2-1}\Gamma\left(\frac{\textstyle d-1}{\textstyle2}\right)}
%\frac{(\upsilon^2-1)^{(d-2)/4}}{\upsilon^{d-2}}
%\left(\frac{\upsilon^2-1}{\upsilon^2}\right)^{(2-d)/2}
%\nonumber\\
%&= \frac{e^{i\pi(d-2)/2}
%\Gamma(d-2)\sqrt{\pi}}{2^{d/2-1}\Gamma\left(\frac{\textstyle d-1}{\textstyle2}\right)}
%\frac{1}{(\upsilon^2-1)^{(d-2)/4}}.
\end{align}
For the special transition, $\mu=(d-5)/2$,
\begin{align}
%P_{-\mu-1/2}^{\,d/2-1}(\upsilon)
%&= \frac{1}{\Gamma\left(2-\frac{\textstyle d}{\textstyle2}\right)}
%\left(\frac{\upsilon+1}{\upsilon-1}\right)^{(d-2)/4}
%F\left(\frac{d}{2}-2,\,3-\frac{d}{2},\,2-\frac{d}{2};\,\frac{1-\upsilon}{2}\right),
%\nonumber\\
%&= \frac{1}{\Gamma\left(2-\frac{\textstyle d}{\textstyle2}\right)}
%\left(\frac{\upsilon+1}{\upsilon-1}\right)^{(d-2)/4}
%\left(\frac{\upsilon+1}{2}\right)^{2-d/2}
%F\left(\frac{d}{2}-2,\,-1,\,2-\frac{d}{2};\,\frac{\upsilon-1}{\upsilon+1}\right),
%\nonumber\\
%&= \frac{1}{\Gamma\left(2-\frac{\textstyle d}{\textstyle2}\right)}
%\left(\frac{\upsilon+1}{\upsilon-1}\right)^{(d-2)/4}
%\left(\frac{\upsilon+1}{2}\right)^{2-d/2}
%\left(1+\frac{\upsilon-1}{\upsilon+1}\right),
%\nonumber\\
%&= \frac{1}{\Gamma\left(2-\frac{\textstyle d}{\textstyle2}\right)}
%\left(\frac{\upsilon+1}{\upsilon-1}\right)^{(d-2)/4}
%\left(\frac{\upsilon+1}{2}\right)^{2-d/2}
%\left(\frac{2\upsilon}{\upsilon+1}\right),
%\nonumber\\
%&= \frac{2^{d/2-1}}{\Gamma\left(2-\frac{\textstyle d}{\textstyle2}\right)}
%\frac{\upsilon}{(\upsilon^2-1)^{(d-2)/4}},
%\label{P_S}
%\\
%Q_{-\mu-1/2}^{\,d/2-1}(\upsilon) &= \frac{e^{i\pi(d-2)/2}
%\Gamma(2)
%\sqrt{\pi}}
%{2^{3-d/2}\Gamma(1-\mu)}
%\frac{(\upsilon^2-1)^{(d-2)/4}}{\upsilon^2}
%F\left(\frac{3}{2},\,1,\,1-\mu;\,\frac{1-\upsilon}{2}\right).
%\\
w^{(0)}(\upsilon) &= \frac{1}{\upsilon^2}
F\left(\frac{3}{2},\,1,\,1-\mu;\,\frac{1}{\upsilon^2}\right)
= \frac{2^{3-d/2}\Gamma(1-\mu)}{e^{i\pi(d-2)/2}\sqrt{\pi}}
\frac{Q_{-\mu-1/2}^{\,d/2-1}(\upsilon)}{(\upsilon^2-1)^{(d-2)/4}}
\nonumber\\
&= g^{(0)}(\upsilon)\, \upsilon^{2\mu}
F\left(-\frac{1}{2}-\mu,\,-\mu,\,1-\mu;\,\frac{1}{\upsilon^2}\right);
\label{w_S}
\end{align}
For the transverse part of the extraordinary transition, $\mu=(d-1)/2$,
\begin{align}
%P_{\mu-1/2}^{\,d/2-1}(\upsilon) &=
%\frac{1}{\Gamma\left(2-\frac{\textstyle d}{\textstyle2}\right)}
%\left(\frac{\upsilon+1}{\upsilon-1}\right)^{(d-2)/4}
%F\left(\frac{d}{2},\,1-\frac{d}{2},\,2-\frac{d}{2};\,\frac{1-\upsilon}{2}\right)
%
%
%\nonumber\\
%&= \frac{(\upsilon^2-1)^{(d-2)/4}}{\sqrt{\pi}}
%\left[\frac{\sin(d-2)\pi}{2^{d/2}\cos(d/2-1)\pi}
%\frac{\Gamma(d-1)}{\Gamma\left(\frac{\textstyle d+1}{\textstyle2}\right)\upsilon^{d-1}}
%F\left(\frac{d}{2},\,\frac{d-1}{2},\,\frac{d+1}{2};\,\frac{1}{\upsilon^2}\right) \right.
%\nonumber\\
%& \left. +\, 2^{d/2-1}\Gamma\left(\frac{d-1}{2}\right) \frac{\upsilon^{0}}{\Gamma(1)}
%F\left(\frac{1}{2},\,0,\,\frac{3-d}{2};\,\frac{1}{\upsilon^2}\right)
%\right]
%\nonumber\\
w^{(0)}(\upsilon) &= 1
= \frac{1}{2^{d/2-1}\Gamma\left(\frac{\textstyle d-1}{\textstyle2}\right)}
\frac{P_{\mu-1/2}^{\,d/2-1}(\upsilon)}{(\upsilon^2-1)^{(d-2)/4}}
- \textrm{constant}\, \times g^{(0)}(\upsilon).
\label{w_E}
\end{align}
\end{subequations}
In the cases of the ordinary and special transitions,
the Wronskian of the two independent functions is given by
%
%\begin{align}
%W\left(P_{-\mu-1/2}^{\,d/2-1}, Q_{-\mu-1/2}^{\,d/2-1}\right)
%&= P_{-\mu-1/2}^{\,d/2-1}(\upsilon) \frac{d}{d\upsilon} Q_{-\mu-1/2}^{\,d/2-1}(\upsilon)
%- Q_{-\mu-1/2}^{\,d/2-1}(\upsilon) \frac{d}{d\upsilon} P_{-\mu-1/2}^{\,d/2-1}(\upsilon)
%\nonumber\\
%&= - \frac{2^{d-2}e^{i\pi(d-2)}\Gamma\left(\frac{\textstyle -\mu-1/2+d/2-1+1}{\textstyle2}\right)
%\Gamma\left(\frac{\textstyle -\mu-1/2+d/2-1+2}{\textstyle2}\right)}
%{\Gamma\left(\frac{\textstyle -\mu-1/2-d/2+1+1}{\textstyle2}\right)
%\Gamma\left(\frac{\textstyle -\mu-1/2-d/2+1+2}{\textstyle2}\right)\,(\upsilon^2-1)}.
%\nonumber\\
%&= - \frac{2^{d-2}e^{i\pi(d-2)}\Gamma\left(\frac{\textstyle d/2-\mu-1/2}{\textstyle2}\right)
%\Gamma\left(\frac{\textstyle d/2-\mu+1/2}{\textstyle2}\right)}
%{\Gamma\left(\frac{\textstyle 3/2-d/2-\mu}{\textstyle2}\right)
%\Gamma\left(\frac{\textstyle 5/2-d/2-\mu}{\textstyle2}\right)\,(\upsilon^2-1)}.
%\nonumber\\
%&= - \frac{2^{d-2}e^{i\pi(d-2)}\Gamma\left(\frac{\textstyle d-2\mu-1}{\textstyle4}\right)
%\Gamma\left(\frac{\textstyle d-2\mu+1}{\textstyle4}\right)}
%{\Gamma\left(\frac{\textstyle 3-d-2\mu}{\textstyle4}\right)
%\Gamma\left(\frac{\textstyle 5-d-2\mu}{\textstyle4}\right)\,(\upsilon^2-1)},
%\label{Wronskian}
%\end{align}
%where $\mu$ should be changed to $-\mu$ in the case of the transverse
%correlation function for the extraordinary transition,
%because $Q_{-\mu-1/2}^{\,d/2-1}(\upsilon)\propto\Gamma(0)$ is diverging
%at $\mu=(d-1)/2$.
%
\begin{align}
W\left(g^{(0)}, w^{(0)}\right)
&= g^{(0)}(\upsilon) \Bigl[w^{(0)}(\upsilon)\Bigr]'
- \Bigl[g^{(0)}(\upsilon)\Bigr]' w^{(0)}(\upsilon)
\nonumber\\
&= {g^{(0)}}^2 \frac{d}{d\upsilon} \left[
\upsilon^{2\mu}
F\left(\pm\frac{1}{2}-\mu,\,-\mu,\,1-\mu;\,\frac{1}{\upsilon^2}\right) \right]
\nonumber\\
&= 2\mu\, (\upsilon^2-1)^{d/2-2} / (\upsilon^2-1)^{d-2}
\nonumber\\
&= \frac{2\mu}{(\upsilon^2-1)^{d/2}},
\label{Wronskian}
\end{align}
where we used the following relation [eq.~(22) in Sec. 2.8 of ref.~\cite{Erdelyi}]
and put $z=1/\upsilon^2$ and $\gamma=1-\mu$:
%\cite{MoriguchiP61Table3rdLine}
\begin{align}
\frac{d}{dz}z^{\gamma-1}F(\alpha,\beta,\gamma;z)
= (\gamma-1)z^{\gamma-2}F(\alpha,\beta,\gamma-1;z).
\end{align}

\section{Inverse of a Bubble for the Special Transition}
\label{InverseSpecial}

For the special transition, 
to solve eq.~(\ref{Dnu=xi}), we introduce the functions
$\hat{\nu}^{\rm sp}(\upsilon)$ and $\hat{\xi}(\upsilon)$ as
\begin{align}
\nu^{\rm sp}(\upsilon) = \upsilon\,\hat{\nu}^{\rm sp}(\upsilon), \;\;\;\;
\xi(\upsilon) = \upsilon\,\hat{\xi}(\upsilon),
\end{align}
where the superscript `sp' means that this is just
the special solution of eq.~(\ref{Dnu=xi}).
Then, eq.~(\ref{Dnu=xi}) with eq.~(\ref{D_S}) becomes
\begin{align}
\left[(\upsilon^2-1)\frac{d^2}{d\upsilon^2}+(7+2\mu)\upsilon\frac{d}{d\upsilon}
-\frac{2}{\upsilon}\frac{d}{d\upsilon}
+ 3(3+2\mu)\right]\,\hat{\nu}^{\rm sp}(\upsilon)
= \hat{\xi}(\upsilon).
\label{D_S'}
\end{align}
From $\upsilon(d/d\upsilon) = 2\upsilon^2[d/d(\upsilon^2)]$
and $d^2/d\upsilon^2=4\upsilon^2[d^2/d(\upsilon^2)^2]
+2[d/d(\upsilon^2)]$,
eq.~(\ref{D_S'}) can be rewritten as
\begin{align}
\left[4\upsilon^2(\upsilon^2-1)\frac{d^2}{d(\upsilon^2)^2}
+2(\upsilon^2-1)\frac{d}{d(\upsilon^2)}
+2(7+2\mu)\upsilon^2\frac{d}{d(\upsilon^2)}
-4\frac{d}{d(\upsilon^2)}
+ 3(3+2\mu)\right]\,\hat{\nu}^{\rm sp}(\upsilon)
= \hat{\xi}(\upsilon).
\label{D_S''}
\end{align}
Changing the variable $\upsilon$ to $s=\upsilon^2-1$ yields
\begin{align}
\left[4s(s+1)\frac{d^2}{ds^2}
+4(4+\mu)s\frac{d}{ds}
+2(5+2\mu)\frac{d}{ds}
+ 3(3+2\mu)\right]\,\hat{\nu}^{\rm sp}(s)
= \hat{\xi}(s).
\label{D_S'''}
\end{align}
Finally, changing the variable $s$ to $t=1/s=(\upsilon^2-1)^{-1}$ and noting the relations,
$d/ds=-t^2(d/dt)$ and $d^2/ds^2=t^4(d^2/dt^2)+2t^3(d/dt)$, we derived
\begin{align}
\left[4(1+t)t^2\frac{d^2}{dt^2}
-4(2+\mu)t\frac{d}{dt}
-2(1+2\mu)t^2\frac{d}{dt}
+ 3(3+2\mu)\right]\,\hat{\nu}^{\rm sp}(t)
= \hat{\xi}(t).
\label{D_S''''}
\end{align}
On the other hand, the last equality of eq.~(\ref{xi}) was expanded as
\begin{align}
\hat{\xi}(t) &= \frac{C_{\Xi}}{C_f}\,
t^{3+\mu}
F\left(3+\mu,\,\mu,\,\frac{5}{2}+2\mu;\,-t\right)
\nonumber\\
&= \frac{C_{\Xi}}{C_f}\,
t^{3+\mu}
\sum_{n=0}^{\infty}
\frac{(3+\mu)_n(\mu)_n}{\left(\frac{\textstyle5}{\textstyle2}+2\mu\right)_n n!}\,(-t)^n,
\label{xi'}
\end{align}
where $(x)_n$ denotes the conventional symbol $(x)_n=\Gamma(x+n)/\Gamma(x)$.
We assumed a similar series expansion form for $\hat{\nu}(t)$ as
\begin{align}
\hat{\nu}^{\rm sp}(t) &= \frac{C_{\Xi}}{C_f}\,
t^{3+\mu}
\sum_{n=0}^{\infty}
a_n\,(-t)^n.
\label{nu}
\end{align}
Insertion of eqs.~(\ref{xi'}) and (\ref{nu}) into eq.~(\ref{D_S''''}) gives
\begin{align}
&4(3+\mu+n)(2+\mu+n)a_n-4[3+\mu+(n-1)][2+\mu+(n-1)]a_{n-1}
\nonumber\\
-\;& 4(2+\mu)(3+\mu+n)a_n + 2(1+2\mu)[3+\mu+(n-1)]a_{n-1} + 3(2+2\mu)a_n
\nonumber\\
=\;& \frac{(3+\mu)_n(\mu)_n}{\left(\frac{\textstyle5}{\textstyle2}+2\mu\right)_n n!},
\end{align}
which can be simplified as
\begin{align}
\left(n+\frac{3}{2}\right)\left(n+\frac{3}{2}+\mu\right)a_n
-\left(n+\frac{1}{2}\right)(n+2+\mu)a_{n-1}
=\frac{(3+\mu)_n(\mu)_n}{4\left(\frac{\textstyle5}{\textstyle2}+2\mu\right)_n n!}.
\end{align}
It is easy to see that this recurrence formula is satisfied by
\begin{align}
a_n = \frac{1}{4\left(\frac{\textstyle3}{\textstyle2}+\mu\right)}\,
\frac{(3+\mu)_n(1+\mu)_n}{\left(\frac{\textstyle5}{\textstyle2}+2\mu\right)_n
\left(n+\frac{\textstyle3}{\textstyle2}\right)n!}.
\end{align}
From this, we found
\begin{align}
\hat{\nu}^{\rm sp}(t)
%&= \frac{C_{\Xi}}{C_f}\,
%\frac{t^{3+\mu}}{4\left(\frac{\textstyle3}{\textstyle2}+\mu\right)}\,
%\sum_{n=0}^{\infty}
%\frac{(3+\mu)_n(1+\mu)_n}{\left(\frac{\textstyle5}{\textstyle2}+2\mu\right)_n
%\left(n+\frac{\textstyle3}{\textstyle2}\right)
% n!}\,(-t)^n
 %\nonumber\\
&= \frac{C_{\Xi}}{C_f}\,
\frac{t^{3+\mu}}{3(3+2\mu)}\,
\sum_{n=0}^{\infty}
\frac{(3+\mu)_n(1+\mu)_n \left(\frac{\textstyle3}{\textstyle2}\right)_n}
{\left(\frac{\textstyle5}{\textstyle2}+2\mu\right)_n
\left(\frac{\textstyle5}{\textstyle2}\right)_n
 n!}\,(-t)^n
 \nonumber\\
&= \frac{C_{\Xi}}{C_f}\,
\frac{t^{3+\mu}}{3(3+2\mu)}\,
{}_3F_2\left(3+\mu,1+\mu,\frac{3}{2};\,\frac{5}{2}+2\mu,\frac{5}{2};\,-t\right).
\label{nu'}
\end{align}
Thus, we finally obtained
\begin{align}
\nu^{\rm sp}(\upsilon) = 
\frac{C_{\Xi}}{C_f}\frac{1}{3(3+2\mu)}
\frac{\upsilon}{(\upsilon^2-1)^{3+\mu}}\,
{}_3F_2\left(3+\mu,1+\mu,\frac{3}{2};\,\frac{5}{2}+2\mu,\frac{5}{2};\,
\frac{1}{1-\upsilon^2}\right).
\label{nu''}
\end{align}
Of course, there is a possibility to add two homogeneous solutions
of eq.~(\ref{Dnu=xi}) to $\nu(\upsilon)$.
Since $\cal D$ is the same as that for the correlation function in the large $N$ limit,
its two independent homogenous solutions are $g^{(0)}(\upsilon)$
and $w^{(0)}(\upsilon)$ given in eqs.~(\ref{g_S}) and (\ref{w_S}), respectively.
Therefore, $\nu(\upsilon)$ must take the form
\begin{align}
\nu(\upsilon) = \nu^{\rm sp}(\upsilon) + c_1 g^{(0)}(\upsilon) + c_2 w^{(0)}(\upsilon),
\label{nu=nu+c1+c2}
\end{align}
where the coefficients, $c_1$ and $c_2$, are yet to be determined.
Let the corresponding contributions to $\Pi^{-1}_q(y,y')$ be
$\Pi^{-1\,\rm(sp)}_q(y,y')$, $\Pi^{-1\,(1)}_q(y,y')$, and $\Pi^{-1\,(2)}_q(y,y')$.
In the limit $q\rightarrow 0$, they behave like
\begin{subequations}
\begin{align}
\int_0^{\infty}\Pi_q(y,y')\,\Pi^{-1\,\rm(sp)}_q(y',y'')\,dy' &\rightarrow {\rm const.},
\label{(1)_q0}
\\
\int_0^{\infty}\Pi_q(y,y')\,\Pi^{-1\,(1)}_q(y',y'')\,dy' &\rightarrow c_1\times{\rm const.}
\times q^{d-5},
\label{(2)_q0}
\\
\int_0^{\infty}\Pi_q(y,y')\,\Pi^{-1\,(2)}_q(y',y'')\,dy' &\rightarrow c_2\times{\rm const.}
\label{sp_q0}
\end{align}
\end{subequations}
Since the right hand side should be equal to $\delta(y-y'')$, which is
independent of $q$, we found
\begin{align}
c_1=0.
\end{align}
This conclusion can be derived also from the statement that
$\Pi^{-1}_{\rho}(y,y')$ satisfies the homogeneous Neumann boundary condition
in the limit $y'\rightarrow 0$, i.e., $\upsilon\rightarrow\infty$, as
\begin{align}
\Pi^{-1\,(2)}_{\rho}(y,y') = c_2 (yy')^{-2}w^{(0)}(\upsilon)
 \rightarrow \frac{4c_2}{(y^2+\rho^2)^2}, \;\;\;\;
\textrm{for $y'\rightarrow 0$}.
\end{align}
It is easy to see that $\Pi^{-1\,(1)}_{\rho}(y,y')=c_1(yy')^{-2}g^{(0)}(\upsilon)$
is proportional to $c_1 y'^{d-5}$, which diverges similar to eq.~(\ref{(2)_q0}),
and is inappropriate.
The relation between $c_1$ and $c_2$ can be determined from
the behavior in the limit $\upsilon\rightarrow 1$.
To see this behavior, we noticed that $\nu^{\rm sp}(\upsilon)$
in eq.~(\ref{nu''}) is rewritten as
\begin{align}
\nu^{\rm sp}(\upsilon) &= \frac{C_{\Xi}}{C_f}\, \frac{1}{3(3+2\mu)}
\frac{\Gamma\left(\frac{\textstyle5}{\textstyle2}+2\mu\right)
\Gamma\left(\frac{\textstyle5}{\textstyle2}\right)}
{\Gamma(3+\mu)\Gamma(1+\mu)\Gamma\left(\frac{\textstyle3}{\textstyle2}\right)}
\frac{\upsilon}{(\upsilon^2-1)^{3+\mu}}
E\left(3+\mu, 1+\mu,\, \frac{3}{2}\,:\,\frac{5}{2}+2\mu,\, \frac{5}{2}\,:\, \upsilon^2-1\right)
\nonumber\\
&= \frac{C_{\Xi}}{C_f}\, \frac{1}{3(3+2\mu)}
\frac{\Gamma\left(\frac{\textstyle5}{\textstyle2}+2\mu\right)
\Gamma\left(\frac{\textstyle5}{\textstyle2}\right)}
{\Gamma(3+\mu)\Gamma(1+\mu)\Gamma\left(\frac{\textstyle3}{\textstyle2}\right)}
\frac{\upsilon}{(\upsilon^2-1)^{3+\mu}}
\nonumber\\
&\times\,
\left[ \textrm{constant} \times \Gamma(-2)\, (\upsilon^2-1)^{3+\mu}\,
{}_3F_2\left(3+\mu,\,\frac{3}{2}-\mu,\,\frac{3}{2}+\mu;\,3,\,\frac{5}{2}+\mu;\,
1-\upsilon^2\right) \right.
\nonumber\\
&+\, \textrm{constant} \times (\upsilon^2-1)^{1+\mu}\,
{}_3F_2\left(1+\mu,\,-\frac{1}{2}-\mu,\,-\frac{1}{2}+\mu;\,-1,\,\frac{1}{2}+\mu;\,
1-\upsilon^2\right)
\nonumber\\
& \left. +\, \frac{\Gamma\left(\frac{\textstyle3}{\textstyle2}+\mu\right)
\Gamma\left(-\frac{\textstyle1}{\textstyle2}+\mu\right)
\Gamma\left(\frac{\textstyle3}{\textstyle2}\right)}
{\Gamma(1+2\mu)}\,
(\upsilon^2-1)^{3/2} \right],
\label{nu'''}
\end{align}
where $E(\alpha, ...,\, :\,\beta, ...,\,:\, x)$ is the MacRobert's $E$
function related to the GHF
[see eqs. (1) and (2) in Sec. 5.2 of ref.~ \cite{Erdelyi}] as
\begin{align}
E(\alpha_1, ..., \alpha_p\,:\,\beta_1, ...,\beta_q\,:\, x)
&= \frac{\Gamma(\alpha_1)\cdots\Gamma(\alpha_p)}
{\Gamma(\beta_1)\cdots\Gamma(\beta_p)}
{}_pF_q\left(\alpha_1, ..., \alpha_p;\,\beta_1, ...,\beta_q;\,-\frac{1}{x}\right)
\nonumber\\
& =\sum_{r=1}^p
\frac{\displaystyle\prod_{s=1}^{p}\!{}^{\displaystyle'}\,\Gamma(\alpha_s-\alpha_r)}
{\displaystyle\prod_{t=1}^q
\Gamma(\beta_t-\alpha_r)}\Gamma(\alpha_r)x^{\alpha_r}
\nonumber\\
&\hskip-45mm \times\,
{}_{q+1}F_{p-1}\left(\alpha_r, \alpha_r-\beta_1+1, ..., \alpha_r-\beta_q+1;\,
\alpha_r-\alpha_1+1, ..., *, ..., \alpha_r-\alpha_p+1;\,(-1)^{p+q}x\right),
\label{MacRobert}
\end{align}
where the prime in $\prod'$ indicates the omission
of the factor $\Gamma(\alpha_r-\alpha_r)$ and $*$ in ${}_{q+1}F_{p-1}$
the omission of the argument $\alpha_r-\alpha_r+1$.
In eq.~(\ref{nu'''}),
the first and second terms each diverge but their sum converges.
Expansion of these two terms with respect to $\upsilon^2-1$ gives
\begin{align}
\textrm{constant} \times \left[
\frac{\upsilon}{(\upsilon^2-1)^2}
 + \left(\frac{1}{2}-\mu\right)(1+\mu) \frac{\upsilon}{\upsilon^2-1} + \cdots \right],
\end{align}
which are relevant to the bulk anomalies related to the critical exponent $\eta$
and also to the energy density proportional to
$y^{-(1-\alpha)/\nu}\delta(y-y')$ in mixed space.
In contrast, the last term in eq.~(\ref{nu'''}), which is proportional to
$g^{(0)}(\upsilon)$, should cancel with the homogeneous solutions
because this brings an unexpected behavior in the correlation function.
Since the second homogeneous solution $w^{(0)}(\upsilon)$
in eq.~(\ref{w_S}) can be written as
\begin{align}
w^{(0)}(\upsilon) &= \frac{2^{-1/2+\mu}\Gamma(1-2\mu)
\Gamma\left(-\frac{\textstyle3}{\textstyle2}-\mu\right)}
{\Gamma\left(\frac{\textstyle1}{\textstyle2}-\mu\right)\Gamma(-1-2\mu)}
\frac{1}{(1+\upsilon)^{3/2+\mu}}
F\left(\frac{1}{2}-\mu,\,\frac{1}{2}+\mu,\,\frac{5}{2}+\mu;\,\frac{1-\upsilon}{2}\right)
\nonumber\\
&+ \frac{2^{1+2\mu}\Gamma(1-2\mu)
\Gamma\left(\frac{\textstyle3}{\textstyle2}+\mu\right)}
{\Gamma\left(\frac{\textstyle1}{\textstyle2}-\mu\right)}
g^{(0)}(\upsilon),
\label{w_S'}
\end{align}
we should have
\begin{align}
-\,\frac{C_{\Xi}}{C_f}\, \frac{1}{3(3+2\mu)}
\frac{\Gamma\left(\frac{\textstyle5}{\textstyle2}+2\mu\right)
\Gamma\left(\frac{\textstyle5}{\textstyle2}\right)}
{\Gamma(3+\mu)\Gamma(1+\mu)}
\frac{\Gamma\left(\frac{\textstyle3}{\textstyle2}+\mu\right)
\Gamma\left(-\frac{\textstyle1}{\textstyle2}+\mu\right)}
{\Gamma(1+2\mu)}
= c_1 + \frac{2^{1+2\mu}\Gamma(1-2\mu)
\Gamma\left(\frac{\textstyle3}{\textstyle2}+\mu\right)}
{\Gamma\left(\frac{\textstyle1}{\textstyle2}-\mu\right)}\, c_2
\label{C/C=c1+c2}
\end{align}
for the cancellation of the $g^{(0)}(\upsilon)$ terms.
As we already knew that $c_1=0$, eq.~(\ref{C/C=c1+c2}) determines $c_2$ as
\begin{align}
c_2 = \frac{C_{\Xi}}{C_f} C_H,
\label{c_2}
\end{align}
where $C_H$ is given in eq.~(\ref{C_H}),
which is the same as eq.~(5$\cdot$9) in ref.~\cite{PTP2}.
Thus, we succeeded in completely determining $\nu(\upsilon)$,
which is written in eq.~(\ref{nu_S}) and in eq.~(5$\cdot$8) of ref.~\cite{PTP2}.

In the evaluation of diagram (b), we need the Mellin transform
of $\Pi^{-1}_{q=0}(y,y')$,
\begin{align}
\int_0^{\infty} dy' y'^{\sigma-1}\Pi^{-1}_{q=0}(y,y') = D_{\sigma,\,\mu}\, y^{2\mu+\sigma},
\label{Pi^-1=D}
\end{align}
where $D_{\sigma,\,\mu}$ is a constant depending on $\sigma$ and $\mu$.
By using eq.~(\ref{Pi_special}) and the relation (\ref{Bessel}),
$D^{-1}_{\sigma,\,\mu}$ can readily be calculated as
\begin{align}
D^{-1}_{\sigma,\,\mu}\, y^{\sigma-1}
&= \int_0^{\infty} dy' y'^{2\mu+\sigma}\,\Pi_{q=0}(y,y')
\nonumber\\
&= C_{\Pi} \left[ \sqrt{y} \int_0^{\infty} dt\,t^{2+2\mu}\,J_{5/2+2\mu}(ty)
\int_0^{\infty} dy' y'^{1/2+2\mu+\sigma} J_{5/2+2\mu}(ty') \right.
\nonumber\\
&\left. - \; \frac{4(1+\mu)(1+2\mu)}{\sqrt{y}}
\int_0^{\infty} dt\,t^{2\mu}\,J_{3/2+2\mu}(ty)
\int_0^{\infty} dy' y'^{-1/2+2\mu+\sigma} J_{3/2+2\mu}(ty') \right]
\nonumber\\
&= C_{\Pi}\, y^{\sigma-1}
\left[ \frac{2^{1/2+2\mu+\sigma}
\Gamma\left(2+2\mu+\frac{\textstyle\sigma}{\textstyle2}\right)}
{\Gamma\left(\frac{\textstyle3-\sigma}{\textstyle2}\right)}
\frac{2^{1/2-\sigma}
\Gamma\left(2+\mu-\frac{\textstyle\sigma}{\textstyle2}\right)}
{\Gamma\left(\frac{\textstyle3+\sigma}{\textstyle2}+\mu\right)} \right.
\nonumber\\
&\left. - \; 4(1+\mu)(1+2\mu)
\frac{2^{-1/2+2\mu+\sigma}
\Gamma\left(1+2\mu+\frac{\textstyle\sigma}{\textstyle2}\right)}
{\Gamma\left(\frac{\textstyle3-\sigma}{\textstyle2}\right)}
\frac{2^{-1/2-\sigma}
\Gamma\left(1+\mu-\frac{\textstyle\sigma}{\textstyle2}\right)}
{\Gamma\left(\frac{\textstyle3+\sigma}{\textstyle2}+\mu\right)} \right]
\nonumber\\
&= -\, C_{\Pi}\,
\frac{2^{1+2\mu}\left(\frac{\textstyle\sigma}{\textstyle2}\right)
\left(\mu+\frac{\textstyle\sigma}{\textstyle2}\right)
\Gamma\left(1+2\mu+\frac{\textstyle\sigma}{\textstyle2}\right)
\Gamma\left(1+\mu-\frac{\textstyle\sigma}{\textstyle2}\right)}
{\Gamma\left(\frac{\textstyle3-\sigma}{\textstyle2}\right)
\Gamma\left(\frac{\textstyle3+\sigma}{\textstyle2}+\mu\right)}\, y^{\sigma-1}
\label{Mellin^-1}
\end{align}
Thus, we obtained
\begin{align}
D_{\sigma,\,\mu}
 = C_{\Pi}^{-1}\, \frac{\Gamma\left(\frac{\textstyle3-\sigma}{\textstyle2}\right)
\Gamma\left(\frac{\textstyle3+\sigma}{\textstyle2}+\mu\right)}
{2^{1+2\mu}\left(\frac{\textstyle\sigma}{\textstyle2}\right)
\left(\mu+\frac{\textstyle\sigma}{\textstyle2}\right)
\Gamma\left(1+2\mu+\frac{\textstyle\sigma}{\textstyle2}\right)
\Gamma\left(1+\mu-\frac{\textstyle\sigma}{\textstyle2}\right)}
\label{Mellin}
\end{align}
We confirmed that the Mellin transform of $\Pi^{-1}_{q=0}(y,y')$
after the Fourier transform in the limit $q\rightarrow 0$,
i.e., the ($d-1$)-dimensional $\rho$ integration,
of eq.~(\ref{nu=nu+c1+c2}) with $c_1=0$ and $c_2$ given by eq.~(\ref{C/C=c1+c2})
certainly coincides with eq.~(\ref{Mellin}),
although we do not present the detail of this complicated calculation
using MacRobert's $E$ function here. 

\section{\textit{O}(1/\textit{N}) Correction for the Ordinary Transition}
\label{Ordinary}

In the case of the ordinary transition, the function $h^{(a)}(\upsilon)$ in
eq.~(\ref{Sigma^a}) for the diagram (a) is given by eq.~(\ref{h^a})
with $\nu(\upsilon)$ in eq.~(\ref{nu_O}).
%\begin{align}
%h^{(a)}(\upsilon) %&
%= -\frac{2}{N}\,C_f\,\nu(\upsilon)\,g^{(0)}(\upsilon).
%= \frac{2}{N} C_f \frac{2^{2+2\mu}(1-2\mu)\Gamma(\mu)\Gamma(1+2\mu)}
%{\sqrt{\pi}\Gamma(2\mu)
%\left[\Gamma\left(\frac{\textstyle1}{\textstyle2}+\mu\right)\right]^2
%\Gamma\left(\frac{\textstyle1}{\textstyle2}-\mu\right)}
%\frac{Q_{2\mu}^2(\upsilon)}{(\upsilon^2-1)^{3/2+\mu}}
%\nonumber\\
%&= \frac{2^{2+2\mu}(1-2\mu)\Gamma(\mu)\Gamma(1+2\mu)}
%{N\sqrt{\pi}\Gamma(2\mu)\left[\Gamma\left(\frac{\textstyle1}{\textstyle2}+\mu\right)\right]^2\Gamma\left(\frac{\textstyle1}{\textstyle2}-\mu\right)}\Gamma(1+2\mu)K_{d-1}\;
% \frac{Q_{2\mu}^2(\upsilon)}{(\upsilon^2-1)^{3/2+\mu}}
%\nonumber\\
%&= \frac{2^{2+2\mu}(1-2\mu)\Gamma(1+2\mu)}
%{N\sqrt{\pi}\left[\Gamma\left(\frac{\textstyle1}{\textstyle2}+\mu\right)
%\right]^2\Gamma\left(\frac{\textstyle1}{\textstyle2}-\mu\right)}2\Gamma(1+\mu) K_{d-1}\;
%\frac{Q_{2\mu}^2(\upsilon)}{(\upsilon^2-1)^{3/2+\mu}}
%\nonumber\\
%&
%= A\,\frac{Q_{2\mu}^2(\upsilon)}{(\upsilon^2-1)^{3/2+\mu}},
%= -\frac{2}{N}C_X\,\frac{Q_{2\mu}^2(\upsilon)}{(\upsilon^2-1)^{3/2+\mu}},
%\end{align}
Hence, the function $\tilde{h}^{(a)}(\upsilon)$ in eq.~(\ref{tilde^h^a}) reads
\begin{align}
\tilde{h}^{(a)}(\upsilon) 
%&= A\,\frac{Q_{2\mu}^2(\upsilon)}{(\upsilon^2-1)}
%=A \frac{\sqrt{\pi}\Gamma(3+2\mu)}{2^{1+2\mu}
%\Gamma\left(\frac{\textstyle3}{\textstyle2}+2\mu\right)}\frac{1}{\upsilon^{3+2\mu}}
%F\left(2+\mu,\,\frac{3}{2}+\mu,\,\frac{3}{2}+2\mu;\,\frac{1}{\upsilon^2}\right)
&
%= -\frac{2}{N}C_X\,\frac{Q_{2\mu}^2(\upsilon)}{(\upsilon^2-1)}
= -\frac{2}{N}C_X \frac{\sqrt{\pi}\Gamma(3+2\mu)}{2^{1+2\mu}
\Gamma\left(\frac{\textstyle3}{\textstyle2}+2\mu\right)}\frac{1}{\upsilon^{3+2\mu}}
F\left(2+\mu,\,\frac{3}{2}+\mu,\,\frac{3}{2}+2\mu;\,\frac{1}{\upsilon^2}\right)
%\nonumber\\
%&= A \frac{\sqrt{\pi}\Gamma(3+2\mu)}{2^{1+2\mu}
%\Gamma\left(\frac{\textstyle3}{\textstyle2}+2\mu\right)}
%\sum_{n=0}^{\infty} \frac{(2+\mu)_n\left(\frac{\textstyle3}{\textstyle2}+\mu\right)_n}
%{\left(\frac{\textstyle3}{\textstyle2}+2\mu\right)_n n!}\, \upsilon^{-3-2\mu-2n}
\nonumber\\
&= A' \sum_{n=0}^{\infty} \frac{(2+\mu)_n\left(\frac{\textstyle3}{\textstyle2}+\mu\right)_n}
{\left(\frac{\textstyle3}{\textstyle2}+2\mu\right)_n n!}\, \upsilon^{-3-2\mu-2n},
\label{h=sum}
\end{align}
where we put $A'$ as
\begin{align}
A' 
%= A \frac{\sqrt{\pi}\Gamma(3+2\mu)}{2^{1+2\mu}
%\Gamma\left(\frac{\textstyle3}{\textstyle2}+2\mu\right)}
= -\frac{2}{N}C_X \frac{\sqrt{\pi}\Gamma(3+2\mu)}{2^{1+2\mu}
\Gamma\left(\frac{\textstyle3}{\textstyle2}+2\mu\right)}.
\label{A'}
\end{align}
We sought the solution $\tilde{g}(\upsilon)$ in a similar form,
\begin{align}
\tilde{g}^{(a)}(\upsilon) = \sum_{n=0}^{\infty} a_n \upsilon^{-3-2\mu-2n}.
\label{g=sum}
\end{align}
Inserting eqs.~(\ref{h=sum}) and (\ref{g=sum}) into eq.~(\ref{til_g=til_h}) gives
\begin{align}
&\sum_{n=0}^{\infty} a_n\, (-3-2\mu-2n)(-4-2\mu-2n)\, \upsilon^{-3-2\mu-2n}
\nonumber\\
-&\;\sum_{n=1}^{\infty} a_{n-1}\,
\bigl[-3-2\mu-2(n-1)\bigr]\bigl[-4-2\mu-2(n-1)\bigr]\, \upsilon^{-3-2\mu-2n}
\nonumber\\
+&\;\; (1-2\mu) \sum_{n=0}^{\infty} a_n\, (-3-2\mu-2n)\, \upsilon^{-3-2\mu-2n}
\nonumber\\
=&\; A' \sum_{n=0}^{\infty} \frac{(2+\mu)_n\left(\frac{\textstyle3}{\textstyle2}+\mu\right)_n}
{\left(\frac{\textstyle3}{\textstyle2}+2\mu\right)_n n!}\, \upsilon^{-3-2\mu-2n},
\end{align}
which leads to
%\begin{align}
%a_n\, (3+2\mu+2n)(3+4\mu+2n) &= a_{n-1}\,
%\bigl[3+2\mu+2(n-1)\bigr]\bigl[4+2\mu+2(n-1)\bigr]
%\nonumber\\
%&+\; A'\,\frac{(2+\mu)_n\left(\frac{\textstyle3}{\textstyle2}+\mu\right)_n}
%{\left(\frac{\textstyle3}{\textstyle2}+2\mu\right)_n n!}
%\end{align}
%or
\begin{align}
a_n\, \left(\frac{3}{2}+\mu+n\right) = a_{n-1}\,
\frac{(2+\mu+n-1)\left(\frac{\textstyle3}{\textstyle2}+\mu+n-1\right)}
{\frac{\textstyle5}{\textstyle2}+2\mu+n-1}
+ \frac{A'}{4\left(\frac{\textstyle3}{\textstyle2}+2\mu\right)}
\frac{(2+\mu)_n\left(\frac{\textstyle3}{\textstyle2}+\mu\right)_n}
{\left(\frac{\textstyle5}{\textstyle2}+2\mu\right)_n n!}.
\end{align}
Thus, putting
\begin{align}
a_n = A''\,\frac{(2+\mu)_n\left(\frac{\textstyle3}{\textstyle2}+\mu\right)_n}
{\left(\frac{\textstyle5}{\textstyle2}+2\mu\right)_n n!},
\label{a_n}
\end{align}
we obtained
\begin{align}
A'' \left(\frac{3}{2}+\mu+n\right)
 = A'' n + \frac{A' }{4\left(\frac{\textstyle3}{\textstyle2}+2\mu\right)},
\end{align}
and in turn
\begin{align}
A'' = \frac{A' }{2(3+2\mu)\left(\frac{\textstyle3}{\textstyle2}+2\mu\right)}
%= \frac{A}{3+2\mu}
%\frac{\sqrt{\pi}}{2^{2+2\mu}}
%\frac{\Gamma(3+2\mu)}{\Gamma\left(\frac{\textstyle5}{\textstyle2}+2\mu\right)}.
= -\frac{2}{N}\frac{C_X}{3+2\mu}
\frac{\sqrt{\pi}}{2^{2+2\mu}}
\frac{\Gamma(3+2\mu)}{\Gamma\left(\frac{\textstyle5}{\textstyle2}+2\mu\right)}.
\end{align}
Equation (\ref{g=sum}) with $a_n$ given by eq.~(\ref{a_n}) is no other than
the hypergeometric function of $1/\upsilon^2$:
\begin{align}
\tilde{g}^{(a)}(\upsilon) = \frac{A''}{\upsilon^{3+2\mu}}
F\left(2+\mu,\,\frac{3}{2}+\mu,\,\frac{5}{2}+2\mu;\,\frac{1}{\upsilon^2}\right).
\end{align}
Thus, the solution of $g(\upsilon)$ is given by \cite{PTP1}
\begin{align}
g^{(a)}(\upsilon) &= \frac{\tilde{g}^{(a)}(\upsilon)}{(\upsilon^2-1)^{1/2+\mu}}
%
%\nonumber\\
%&= -\frac{2}{N}\frac{C_X}{3+2\mu}
%\frac{\sqrt{\pi}}{2^{2+2\mu}}
%\frac{\Gamma(3+2\mu)}{\Gamma\left(\frac{\textstyle5}{\textstyle2}+2\mu\right)}
%\frac{1}{\upsilon^{3+2\mu}(\upsilon^2-1)^{1/2+\mu}}
%F\left(2+\mu,\,\frac{3}{2}+\mu,\,\frac{5}{2}+2\mu;\,\frac{1}{\upsilon^2}\right)
%\end{align}
%\begin{align}
%= -\; \frac{2^{2+2\mu}}{\sqrt{\pi}}
%\frac{\Gamma\left(\frac{\textstyle5}{\textstyle2}+2\mu\right)}{\Gamma(3+2\mu)}
%A''
%\frac{Q^{1}_{1+2\mu}(\upsilon)}{(\upsilon^2-1)^{\,1+\mu}}
%\nonumber\\
%&= -\; \frac{2^{2+2\mu}}{\sqrt{\pi}}
%\frac{\Gamma\left(\frac{\textstyle5}{\textstyle2}+2\mu\right)}{\Gamma(3+2\mu)}
%A \frac{\sqrt{\pi}\Gamma(3+2\mu)}{2^{1+2\mu}
%\Gamma\left(\frac{\textstyle3}{\textstyle2}+2\mu\right)}
%\frac{1}{4\left(\frac{\textstyle3}{\textstyle2}+\mu\right)
%\left(\frac{\textstyle3}{\textstyle2}+2\mu\right)}
%\frac{Q^{1}_{1+2\mu}(\upsilon)}{(\upsilon^2-1)^{\,1+\mu}}
%\nonumber\\
%&
%= -\; \frac{A}{3+2\mu}
%\frac{Q^{1}_{1+2\mu}(\upsilon)}{(\upsilon^2-1)^{\,1+\mu}},
= \frac{2}{N} \frac{C_X}{3+2\mu}
\frac{Q^{1}_{1+2\mu}(\upsilon)}{(\upsilon^2-1)^{\,1+\mu}},
\label{g^a}
\end{align}
where we used the relation
%\begin{align}
%Q^{\,\mu}_{\nu}(\upsilon) = \frac{\sqrt{\pi}e^{\,\mu\pi i}}{2^{\nu+1}}
%\frac{\Gamma(\mu+\nu+1)}{\Gamma\left(\nu+\frac{\textstyle3}{\textstyle2}\right)}
%\frac{(\upsilon^2-1)^{\,\mu/2}}{\upsilon^{\,\mu+\nu+1}}
%F\left(\frac{\mu+\nu+1}{2},\,\frac{\mu+\nu+2}{2},\,\nu+\frac{3}{2};\,\frac{1}{\upsilon^2}\right).
%\end{align}
%\begin{align}
%Q^{1}_{1+2\mu}(\upsilon) = -\; \frac{\sqrt{\pi}}{2^{2+2\mu}}
%\frac{\Gamma(3+2\mu)}{\Gamma\left(\frac{\textstyle5}{\textstyle2}+2\mu\right)}
%\frac{(\upsilon^2-1)^{\,1/2}}{\upsilon^{\,3+2\mu}}
%F\left(\frac{3}{2}+\mu,\,2+\mu,\,\frac{5}{2}+2\mu;\,\frac{1}{\upsilon^2}\right).
%\end{align}
\begin{align}
\frac{1}{\upsilon^{\,3+2\mu}}
F\left(\frac{3}{2}+\mu,\,2+\mu,\,\frac{5}{2}+2\mu;\,\frac{1}{\upsilon^2}\right)
 = -\; \frac{2^{2+2\mu}}{\sqrt{\pi}}
\frac{\Gamma\left(\frac{\textstyle5}{\textstyle2}+2\mu\right)}{\Gamma(3+2\mu)}
\frac{Q^{1}_{1+2\mu}(\upsilon)}{(\upsilon^2-1)^{\,1/2}}.
\end{align}
The second homogeneous solution, $w^{(0)}(\upsilon)$, cannot be added here,
because it violates the boundary condition that
$f^{(a)}(\upsilon)\sim 1/v^{1+2\mu}$ for $\upsilon\rightarrow\infty$.
In contrast, the first homogeneous solution, $g^{(0)}(\upsilon)$,
can be added to $g^{(a)}(\upsilon)$.
However, this contribution stems from the $U(y)\delta(y-y')$ term in the self-energy,
which exactly cancels with the corresponding tadpole diagram (b)
due to eq.~(\ref{Bessel}) or diagrammatically Fig.~\ref{Bubble}(b);
see the Appendix of ref.~\cite{PTP1},
which is also noted in ref.~\cite{Metlitski}.
Due to this cancelation, a term proportional to $g^{(0)}(\upsilon)$
is not necessary to be added to eq.~(\ref{g^a}).

Then, the solution $f^{(a)}(\upsilon)$
of the next differential equation, eq.~(\ref{Df=g}),
is given by the integral eq.~(\ref{f=wgg}).
However, It is easily seen that the asymptotic behavior of eq.~(\ref{f=wgg})
in $\upsilon\rightarrow\infty$ does not have any logarithmic correction
to the $1/\upsilon^{d-2}$ behavior.
Therefore, there is no contribution to $\eta_{/\!/}$ from the diagram (a).
except for the contribution to the bulk critical exponent $\eta$.

Next, we consider the diagram (b), which is evaluated as
$f^{(a)}(1)\bigr|_{\rm subtracted} / y''^{d-2}$.
Here, the subscript `subtracted' means the subtraction of the bulk divergent value.
Since the second homogeneous solution
$w^{(0)}(\upsilon)$ in eq.~(\ref{w_O}) can be rewritten as
\begin{align}
w^{(0)}(\upsilon) &= \frac{1}{\upsilon}
F\left(1,\,\frac{1}{2},\,1-\mu;\,\frac{1}{\upsilon^2}\right)
= \frac{2^{2-d/2}\Gamma(1-\mu)}{e^{i\pi(d-2)/2}\sqrt{\pi}}
\frac{Q_{-\mu-1/2}^{\,d/2-1}(\upsilon)}{(\upsilon^2-1)^{(d-2)/4}}
%
%\nonumber\\
%&= \frac{2^{2-d/2}\Gamma(1-\mu)}{e^{i\pi(d-2)/2}\sqrt{\pi}}
%\frac{e^{i\pi(d-2)/2}}{2(\upsilon^2-1)^{(d-2)/4}}
%\left[ \Gamma\left(\frac{d}{2}-1\right)
%\left(\frac{\upsilon+1}{\upsilon-1}\right)^{(d-2)/4}
%F\left(\frac{1}{2}+\mu,\,\frac{1}{2}-\mu,\,\frac{1}{2}-\mu;\,\frac{1-\upsilon}{2}\right) \right.
%\nonumber\\
%&\left. + \frac{\Gamma\left(1-\frac{\textstyle d}{\textstyle2}\right)
%\Gamma\left(\frac{\textstyle d-1}{\textstyle2}-\mu\right)}
%{\Gamma\left(\frac{\textstyle 3-d}{\textstyle2}-\mu\right)}
%\left(\frac{\upsilon-1}{\upsilon+1}\right)^{(d-2)/4}
%F\left(\frac{1}{2}+\mu,\,\frac{1}{2}-\mu,\,\frac{3}{2}+\mu;\,\frac{1-\upsilon}{2}\right)
%\right]
%\nonumber\\
%&= \frac{2^{1-d/2}\Gamma(1-\mu)}{\sqrt{\pi}(\upsilon^2-1)^{(d-2)/4}}
%\left[ \Gamma\left(\frac{1}{2}+\mu\right)
%\left(\frac{\upsilon+1}{\upsilon-1}\right)^{(d-2)/4}
%\left(\frac{2}{\upsilon+1}\right)^{1/2+\mu} \right.
%\nonumber\\
%&\left. + \frac{\Gamma\left(-\frac{\textstyle 1}{\textstyle2}-\mu\right)
%\Gamma(1)}{\Gamma(-2\mu)}
%\left(\frac{\upsilon-1}{\upsilon+1}\right)^{(d-2)/4}
%F\left(\frac{1}{2}+\mu,\,\frac{1}{2}-\mu,\,\frac{3}{2}+\mu;\,\frac{1-\upsilon}{2}\right)
%\right]
%\nonumber\\
%&= \frac{2^{1-d/2}\Gamma(1-\mu)}{\sqrt{\pi}(\upsilon^2-1)^{(d-2)/4}}
%\left[ 2^{1/2+\mu}\Gamma\left(\frac{1}{2}+\mu\right)
%\frac{1}{(\upsilon^2-1)^{1/2+\mu}} \right.
%\nonumber\\
%&\left. + \frac{\Gamma\left(-\frac{\textstyle 1}{\textstyle2}-\mu\right)}{\Gamma(-2\mu)}
%\left(\frac{\upsilon-1}{\upsilon+1}\right)^{(d-2)/4}
%F\left(\frac{1}{2}+\mu,\,\frac{1}{2}-\mu,\,\frac{3}{2}+\mu;\,\frac{1-\upsilon}{2}\right)
%\right]
%
\nonumber\\
&= \frac{2^{-1/2-\mu}\Gamma(1-\mu)}{\sqrt{\pi}}
\left[ 2^{1/2+\mu}\Gamma\left(\frac{1}{2}+\mu\right)
g^{(0)}(\upsilon) \right.
\nonumber\\
&\left. + \frac{\Gamma\left(-\frac{\textstyle 1}{\textstyle2}-\mu\right)}{\Gamma(-2\mu)}
\frac{1}{(\upsilon+1)^{1/2+\mu}}
F\left(\frac{1}{2}+\mu,\,\frac{1}{2}-\mu,\,\frac{3}{2}+\mu;\,\frac{1-\upsilon}{2}\right)
\right],
\end{align}
%Looking at the second term of $w_2(\upsilon)$ in eq.~(\ref{w_2}),
we found that the $f^{(a)}(1)\bigr|_{\rm subtracted}$ contribution can
only come from the integral of the first term in eq.~(\ref{f=wgg}).
Taking into account the factor of $-1$
of one loop diagram, we obtained \cite{PTP1}
\begin{align}
f^{(a)}(1)\bigr|_{\rm subtracted}
%
%&= \frac{\Gamma\left(\frac{\textstyle1}{\textstyle2}-\mu\right)\Gamma(-\mu)}
%{e^{i(1/2+\mu)\pi}2^{1+2\mu}\!\sqrt{\pi}}
%\frac{e^{i(1/2+\mu)\pi}}{2}
%\frac{\Gamma\left(-\frac{\textstyle1}{\textstyle2}-\mu\right)}
%{\Gamma(-2\mu)\,2^{1/2+\mu}}
%\frac{2^{1/2+\mu}}{\Gamma\left(\frac{\textstyle1}{\textstyle2}-\mu\right)}
%\frac{A}{3+2\mu}
%\frac{\sqrt{\pi}}{2^{2+2\mu}}
%\frac{\Gamma(3+2\mu)}{\Gamma\left(\frac{\textstyle5}{\textstyle2}+2\mu\right)}
%この最初の式はノートの古い表式に基づくもので、これ以下は新しい表式
%\nonumber\\
%&\times \int_1^{\infty} \frac{d\upsilon}{\upsilon^{3+2\mu}}
%\frac{1}{(\upsilon^2-1)^{1/2+\mu}}
%F\left(2+\mu,\,\frac{3}{2}+\mu,\,\frac{5}{2}+2\mu;\,\frac{1}{\upsilon^2}\right)
%\nonumber\\
%&= -\,\frac{1}{2\mu} \frac{2^{-1/2-\mu}\Gamma(1-\mu)}{\sqrt{\pi}}
%\frac{\Gamma\left(-\frac{\textstyle1}{\textstyle2}-\mu\right)}
%{\Gamma(-2\mu)\,2^{1/2+\mu}}
%\left(-\frac{2}{N}C_X\right)
%\frac{1}{3+2\mu}
%\frac{\sqrt{\pi}}{2^{2+2\mu}}
%\frac{\Gamma(3+2\mu)}{\Gamma\left(\frac{\textstyle5}{\textstyle2}+2\mu\right)}
%\nonumber\\
%&= \frac{2^{-1/2-\mu}\Gamma(-\mu)}{2}
%\frac{\Gamma\left(-\frac{\textstyle1}{\textstyle2}-\mu\right)}
%{\Gamma(-2\mu)\,2^{1/2+\mu}}
%\left(-\frac{2}{N}C_X\right)
%\frac{1}{3+2\mu}
%\frac{1}{2^{2+2\mu}}
%\frac{\Gamma(3+2\mu)}{\Gamma\left(\frac{\textstyle5}{\textstyle2}+2\mu\right)}
%\nonumber\\
%
&= %A
-\,\frac{2}{N}C_X
\,\frac{\Gamma(-\mu)\Gamma\left(-\frac{\textstyle1}{\textstyle2}-\mu\right)
\Gamma(3+2\mu)}
{2^{4+4\mu}(3+2\mu)\Gamma(-2\mu)
\Gamma\left(\frac{\textstyle5}{\textstyle2}+2\mu\right)}
\nonumber\\
&\times \int_1^{\infty} \frac{d\upsilon}{\upsilon^{3+2\mu}}
\frac{1}{(\upsilon^2-1)^{1/2+\mu}}
F\left(2+\mu,\,\frac{3}{2}+\mu,\,\frac{5}{2}+2\mu;\,\frac{1}{\upsilon^2}\right)
\nonumber\\
&= %A
-\,\frac{2}{N}C_X\,\frac{\Gamma(-\mu)\Gamma\left(-\frac{\textstyle1}{\textstyle2}-\mu\right)
\Gamma(3+2\mu)}
{2^{5+4\mu}(3+2\mu)\Gamma(-2\mu)
\Gamma\left(\frac{\textstyle5}{\textstyle2}+2\mu\right)}
\nonumber\\
&\times \int_0^1 dt\, t^{1/2+2\mu} (1-t)^{-1/2-\mu}
F\left(2+\mu,\,\frac{3}{2}+\mu,\,\frac{5}{2}+2\mu;\,t\right)
\nonumber\\
&= %A
-\,\frac{2}{N}C_X\,\frac{\Gamma(-\mu)\Gamma\left(-\frac{\textstyle1}{\textstyle2}-\mu\right)
\Gamma(3+2\mu)}
{2^{5+4\mu}(3+2\mu)\Gamma(-2\mu)
\Gamma\left(\frac{\textstyle5}{\textstyle2}+2\mu\right)}
\frac{\Gamma\left(\frac{\textstyle3}{\textstyle2}+2\mu\right)
\Gamma\left(\frac{\textstyle1}{\textstyle2}-\mu\right)}{\Gamma(2+\mu)}
\nonumber\\
&\times
{}_3F_2\left(2+\mu,\,\frac{3}{2}+\mu,\,\frac{3}{2}+2\mu;\,
\frac{5}{2}+2\mu,\,2+\mu;\,1\right)
\nonumber\\
&= %A
-\,\frac{2}{N}C_X\,\frac{\Gamma(-\mu)\Gamma\left(-\frac{\textstyle1}{\textstyle2}-\mu\right)
\Gamma(3+2\mu)}
{2^{5+4\mu}(3+2\mu)\Gamma(-2\mu)}
\frac{\Gamma\left(\frac{\textstyle3}{\textstyle2}+2\mu\right)
\Gamma\left(\frac{\textstyle1}{\textstyle2}-\mu\right)}{\Gamma(2+\mu)}
\frac{\Gamma\left(-\frac{\textstyle1}{\textstyle2}-\mu\right)}
{\Gamma(1+\mu)}
%\nonumber\\
%&= A\,\frac{\left[\Gamma\left(-\frac{\textstyle1}{\textstyle2}-\mu\right)\right]^2
%\Gamma(3+2\mu)\Gamma\left(\frac{\textstyle3}{\textstyle2}+2\mu\right)}
%{2^{5+4\mu}(3+2\mu)\Gamma(1+\mu)\Gamma(2+\mu)}
%\frac{\Gamma(-\mu)\Gamma\left(\frac{\textstyle1}{\textstyle2}-\mu\right)}{\Gamma(-2\mu)}
%\nonumber\\
%&= A\,\frac{\left[\Gamma\left(-\frac{\textstyle1}{\textstyle2}-\mu\right)\right]^2
%\Gamma(3+2\mu)\Gamma\left(\frac{\textstyle3}{\textstyle2}+2\mu\right)}
%{2^{5+4\mu}(3+2\mu)\Gamma(1+\mu)\Gamma(2+\mu)}
%\frac{2\sqrt{\pi}}{2^{-2\mu}}
\nonumber\\
&= %A
-\,\frac{2}{N}C_X\,\frac{\sqrt{\pi}\left[\Gamma\left(-\frac{\textstyle1}{\textstyle2}-\mu\right)\right]^2
\Gamma(3+2\mu)\Gamma\left(\frac{\textstyle3}{\textstyle2}+2\mu\right)}
{2^{4+2\mu}(3+2\mu)\Gamma(1+\mu)\Gamma(2+\mu)},
\label{f(1)^a}
\end{align}
where we used the gamma function relation (\ref{Gam(2x)}) with $x=-\mu$.
%\begin{align}
%\frac{2^{2+2\mu}}{\sqrt{\pi}}
%\frac{\Gamma\left(\frac{\textstyle5}{\textstyle2}+2\mu\right)}{\Gamma(3+2\mu)}
%A''
%= \frac{A}{3+2\mu}
%\end{align}
Then, the self-energy of Figure (b) is evaluated by using eq.~(\ref{Pi^-1})
in mixed space as \cite{PTP1}
\begin{align}
\Sigma^{(b)}_q(y,y') &= -\,f^{(a)}(1)\bigr|_{\rm subtracted}\, \delta(y - y')
\int_0^{\infty}\Pi^{-1}_{q=0}(y,y'')\frac{dy''}{y''^{d-2}}
\nonumber\\
%&
%\int_0^{\infty} \Pi_{q=0}(y,y'')\frac{dy''}{y''^{d-2}}
&= -\,f^{(a)}(1)\bigr|_{\rm subtracted}\, \delta(y - y')\,
C^{-1}_{\Pi} \int_0^{\infty} \frac{dy''}{y''^{d-2}} \sqrt{y'y''}
\int_0^{\infty} dt\,t^{2-2\mu}\,J_{1/2+2\mu}(ty)J_{1/2+2\mu}(ty'')
\nonumber\\
&= -\,f^{(a)}(1)\bigr|_{\rm subtracted}\,
C^{-1}_{\Pi}
\frac{2^{1-2\mu}\sqrt{\pi}\Gamma\left(\frac{\textstyle3}{\textstyle2}+\mu\right)}
{\Gamma(1+2\mu)\Gamma(\mu)}\,\frac{1}{y^2}\,\delta(y - y')
%
%\nonumber\\
%&= \frac{2^{2+2\mu}}{\Gamma\left(\frac{\textstyle1}{\textstyle2}+\mu\right)
%\Gamma\left(\frac{\textstyle1}{\textstyle2}-\mu\right)K_{d-1}}
%\frac{2^{1-2\mu}\sqrt{\pi}\Gamma\left(\frac{\textstyle3}{\textstyle2}+\mu\right)}
%{\Gamma(1+2\mu)\Gamma(\mu)}\frac{1}{y^2}
%\nonumber\\
%&= \frac{2^3\sqrt{\pi}\Gamma\left(\frac{\textstyle3}{\textstyle2}+\mu\right)}
%{\Gamma\left(\frac{\textstyle1}{\textstyle2}+\mu\right)
%\Gamma\left(\frac{\textstyle1}{\textstyle2}-\mu\right)
%\Gamma(1+2\mu)\Gamma(\mu)K_{d-1}}
%\frac{1}{y^2}
%\nonumber\\
%&= \frac{2^2\sqrt{\pi}\Gamma\left(\frac{\textstyle3}{\textstyle2}+\mu\right)}
%{\Gamma\left(\frac{\textstyle1}{\textstyle2}+\mu\right)
%\Gamma\left(\frac{\textstyle1}{\textstyle2}-\mu\right)
%\Gamma(\mu)C_f}
%\frac{1}{y^2}
%
\nonumber\\
&= -\,f^{(a)}(1)\bigr|_{\rm subtracted}\,
\frac{2^{1+2\mu}\Gamma\left(\frac{\textstyle3}{\textstyle2}+\mu\right)}
{\Gamma\left(\frac{\textstyle1}{\textstyle2}-\mu\right)\Gamma(2\mu)C_f}\,
\frac{1}{y^2}\, \delta(y - y')
= \frac{A}{C_f}\,\frac{1}{y^2}\, \delta(y - y'),
\label{Sigma^b}
\end{align}
where we used the integral formula (6.561.14) of ref.~\cite{Gradshteyn} twice
and the gamma function relation (\ref{Gam(2x)}) with $x=\mu$.
%
%\begin{align}
%C_{\Pi} &= \Gamma\left(\frac{\textstyle d}{\textstyle2}-1\right)
%\Gamma\left(2-\frac{\textstyle d}{\textstyle2}\right)
% \frac{K_{d-1}}{2^{d-1}}
% = \Gamma\left(\frac{\textstyle1}{\textstyle2}+\mu\right)
%\Gamma\left(\frac{\textstyle1}{\textstyle2}-\mu\right)
% \frac{K_{d-1}}{2^{2+2\mu}}\nonumber
%\end{align}
%
%\begin{align}
%C_f = \frac{1}{2}K_{d-1}\Gamma(d-2) =  \frac{1}{2}K_{d-1}\Gamma(1+2\mu)
%\end{align}
%
Finally, the contribution to the correlation function from this diagram
is obtained by integrating eqs.~(\ref{Dg=h}) and (\ref{Df=g}).
Integrating eq.~(\ref{Dg=h}) is trivial due to the delta function in eq.~(\ref{Sigma^b}),
which leads to
\begin{align}
g^{(b)}(\upsilon) = \frac{A}{C_f}C_f g^{(0)}(\upsilon)  = \frac{A}{(\upsilon^2-1)^{d/2-1}}
 \;\;\;\; \leftrightarrow \;\;\;\;  \tilde{g}^{(b)}(\upsilon) = A.
\end{align}
The contribution to the correlation function, $f^{(b)}(\upsilon)$,
can be calculated using eq.~(\ref{f=wgg}).
The anomaly arises only from the second term.
Using eq.~(\ref{g_O}) and the first equality in eq.~(\ref{w_O}) yields 
\begin{align}
f^{(b)}(\upsilon) &=  
%\frac{\Gamma\left(\frac{\textstyle1}{\textstyle2}-\mu\right)\Gamma(-\mu)}
%{e^{i(1/2+\mu)\pi}2^{1+2\mu}\!\sqrt{\pi}}\,
-\,\frac{1}{2\mu}
g^{(0)}(\upsilon) \int_{c'}^{\upsilon} 
\tilde{g}^{(b)}(\upsilon') w^{(0)}(\upsilon') d\upsilon'
%\nonumber\\
%&= \frac{\Gamma\left(\frac{\textstyle1}{\textstyle2}-\mu\right)\Gamma(-\mu)}
%{e^{i(1/2+\mu)\pi}2^{1+2\mu}\!\sqrt{\pi}}
%\frac{2^{1/2+\mu}}{\Gamma\left(\frac{\textstyle1}{\textstyle2}-\mu\right)}
%\frac{e^{i(1/2+\mu)\pi}\sqrt{\pi}}{2^{1/2-\mu}\Gamma(1-\mu)}
%\frac{A}{(\upsilon^2-1)^{1/2+\mu}}
%\int_{c'}^{\upsilon}
%\frac{1}{\upsilon'}F\left(1,\,\frac{1}{2}\,1-\mu;\,\frac{1}{\upsilon'^2}\right)d\upsilon'
\nonumber\\
&= -\,\frac{A}{2\mu\,(\upsilon^2-1)^{d/2-1}} \int_{c'}^{\upsilon}
\frac{1}{\upsilon'}F\left(1,\,\frac{1}{2}\,1-\mu;\,\frac{1}{\upsilon'^2}\right)d\upsilon'
\nonumber\\
&\rightarrow  -\,\frac{A}{2\mu\,(\upsilon^2-1)^{d/2-1}} \log\upsilon,
\;\;\;\; \textrm{for $\upsilon\rightarrow\infty$}.
\end{align}
Thus, we found \cite{PTP1}
\begin{align}
\hat{\gamma}^O &= - \frac{NA}{C_f}\,\frac{1}{2\mu}
\nonumber\\
&= \frac{N}{C_f}\,\frac{1}{2\mu}\, f^{(a)}(1)\bigr|_{\rm subtracted}\,
\frac{2^{1+2\mu}\Gamma\left(\frac{\textstyle3}{\textstyle2}+\mu\right)}
{\Gamma\left(\frac{\textstyle1}{\textstyle2}-\mu\right)\Gamma(2\mu)}
\nonumber\\
&= \frac{(1-2\mu)}
{\Gamma(1+2\mu)}
\frac{\Gamma(3+4\mu)}
{(3+2\mu)}
\frac{1}
{\Gamma(2+2\mu)}
%
%\nonumber\\
%&= \frac{2(4-d)}
%{\Gamma(d-2)}
%\frac{\Gamma(2d-4)}
%{d}
%\frac{1}
%{\Gamma(d-2)}
%\nonumber\\
%&= \frac{(4-d)}
%{\Gamma(d-2)}
%\frac{\Gamma(2d-3)}
%{d(d-2)}
%\frac{1}
%{\Gamma(d-2)}
%\nonumber\\
%&= \frac{(4-d)}
%{\Gamma(d-2)}
%\frac{\Gamma(2d-3)}
%{d}
%\frac{1}
%{\Gamma(d-1)},
\end{align}
which is identical to eq.~(\ref{Eq_O}).

\section{\textit{O}(1/\textit{N}) Correction for the Special Transition}
\label{Special}

To derive the $O(1/N)$ correction to the correlation function
in the case of the special transition,
we have to solve the differential equations (\ref{Dg=h}) and (\ref{Df=g}),
where the differential operator $\cal D$ is given by eq.~(\ref{D_S}).
The self-energy diagram (a) is represented by the function $h^{(a)}(\upsilon)$
in eq.~(\ref{h^a}).
Here, $\nu(\upsilon)$ of eq.~(\ref{nu_S}) is composed of two terms:
one is the special solution $\nu^{\rm sp}(\upsilon)$ and the other
is the homogeneous solution $c_2 w^{(0)}(\upsilon)$,
where $c_2=(C_{\Xi}/C_f)C_H$; see eqs.~(\ref{C_H}) and (\ref{c_2}).
The function $h^{(a)}(\upsilon)$ is scaled as $\tilde{h}^{(a)}(\upsilon)$
in eq.~(\ref{tilde^h^a}).
Similarly, the corresponding solution of eq.~(\ref{Dg=h}), $g^{(a)}(\upsilon)$,
is scaled as $\tilde{g}^{(a)}(\upsilon)$ in eq.~(\ref{tilde^g^a}). 
Then, the differential equation (\ref{Dg=h}) becomes eq.~(\ref{D^tilde^g=tilde^h_S}).
Using the method of variation of constant, we put
\begin{align}
\frac{d}{d\upsilon} \tilde{g}(\upsilon)
= \frac{(\upsilon^2-1)^{1/2+\mu}}{\upsilon^2}C(\upsilon)
\label{d_tilde^g=C}
\end{align}
into eq.~(\ref{D^tilde^g=tilde^h_S}) to get
\begin{align}
\frac{(\upsilon^2-1)^{3/2+\mu}}{\upsilon^2}\frac{d}{d\upsilon}C(\upsilon)
= \tilde{h}(\upsilon).
\label{d_C=tilde^h}
\end{align}
Let $\tilde{h}^{\rm sp}(\upsilon)$ be the function $h^{(a)}(\upsilon)$
that corresponds to the special solution $\nu^{\rm sp}(\upsilon)$ in eq.~(\ref{nu''}).
Integration of eq.~(\ref{d_C=tilde^h}) for this $\tilde{h}^{\rm sp}(\upsilon)$ gives
\begin{align}
C(\upsilon) &= \int_{\infty}^{\upsilon} d\upsilon'
\frac{\upsilon'^2}{(\upsilon'^2-1)^{3/2+\mu}}\tilde{h}^{\rm sp}(\upsilon')
\nonumber\\
&= \frac{C_{\Xi}}{3(3+2\mu)} \int_{\infty}^{\upsilon}
\frac{\upsilon'^3 d\upsilon'}{(\upsilon'^2-1)^{9/2+3\mu}}
{}_3F_2\left(3+\mu,1+\mu,\,\frac{3}{2};\,\frac{5}{2}+2\mu,\,\frac{5}{2};\,
\frac{1}{1-\upsilon'^2}\right)
\nonumber\\
&= -\,\frac{C_{\Xi}}{6(3+2\mu)} \int_{\infty}^{1/(\upsilon^2-1)}dt\,t^{5/2+2\mu}(t^{-1}+1)\,
{}_3F_2\left(3+\mu,1+\mu,\,\frac{3}{2};\,\frac{5}{2}+2\mu,\,\frac{5}{2};\,-t\right)
\nonumber\\
&= -\,\frac{C_{\Xi}}{3(3+2\mu)(5+4\mu)} t^{5/2+2\mu}
{}_4F_3\left(3+\mu,1+\mu,\,\frac{3}{2},\,\frac{5}{2}+2\mu;\,
\frac{5}{2}+2\mu,\,\frac{5}{2},\,\frac{7}{2}+2\mu;\,-t\right)
\nonumber\\
&-\,\frac{C_{\Xi}}{3(3+2\mu)(7+4\mu)} t^{7/2+2\mu}
{}_4F_3\left(3+\mu,1+\mu,\,\frac{3}{2},\,\frac{7}{2}+2\mu;\,
\frac{5}{2}+2\mu,\,\frac{5}{2},\,\frac{9}{2}+2\mu;\,-t\right),
\label{C=}
\end{align}
where we put $t=1/(\upsilon^2-1)$.
Thus, we obtained
\begin{align}
\tilde{g}^{\rm sp}(\upsilon)
&= -\,\frac{C_{\Xi}}{3(3+2\mu)(5+4\mu)}
\int_{\infty}^{\upsilon} \frac{d\upsilon'}{\upsilon'^2} t^{2+\mu}
{}_3F_2\left(3+\mu,1+\mu,\,\frac{3}{2};\,\frac{5}{2},\,\frac{7}{2}+2\mu;\,-t\right)
\nonumber\\
&-\,\frac{C_{\Xi}}{3(3+2\mu)(7+4\mu)}
\int_{\infty}^{\upsilon} \frac{d\upsilon'}{\upsilon'^2} t^{3+\mu}
{}_4F_3\left(3+\mu,1+\mu,\,\frac{3}{2},\,\frac{7}{2}+2\mu;\,
\frac{5}{2}+2\mu,\,\frac{5}{2},\,\frac{9}{2}+2\mu;\,-t\right).
\label{tilde^g^sp}
\end{align}

The solution of eq.~(\ref{Df=g}), $f^{(a)}(\upsilon)$, an be
obtained by using eq.~(\ref{f=wgg}) with $c=\infty$.
As in the case of the ordinary transition,
$f^{(a)}(\upsilon)$ does not directly
bring any logarithmic correction to $g^{(0)}(\upsilon)$ for $\upsilon\rightarrow\infty$.
Only possibility is the contribution to diagram (b)
through $f^{(a)}(1)\bigr|_{\rm subtracted}$.
Similar to the ordinary transition, $f^{(a)}(1)\bigr|_{\rm subtracted}$
contribution comes only from the integral of the first term in eq.~(\ref{f=wgg}).
The finite $w^{(0)}(\upsilon)$ in the limit
$\upsilon\rightarrow 1$ in eq.~(\ref{w_S'}) is
\begin{align}
w^{(0)}(1) &= \frac{2^{-1/2+\mu}\Gamma(1-2\mu)
\Gamma\left(-\frac{\textstyle3}{\textstyle2}-\mu\right)}
{\Gamma\left(\frac{\textstyle1}{\textstyle2}-\mu\right)\Gamma(-1-2\mu)}
\frac{1}{2^{3/2+\mu}}
= \frac{\mu}{\frac{\textstyle3}{\textstyle2}+\mu}.
\end{align}
Taking into account the factor of $-1$ of one loop diagram,
we obtained
\begin{align}
&f^{(a)}(1)\bigr|_{\rm subtracted}
= -\,\frac{w^{(0)}(1)}{2\mu}
\int_{\infty}^1 \frac{\upsilon^2}{(\upsilon^2-1)^{3/2+\mu}}\tilde{g}^{(a)}(\upsilon)d\upsilon
\nonumber\\
&= -\,\frac{w^{(0)}(1)}{2\mu}
\left[ \tilde{g}^{(a)}(1) 
\int_{\infty}^1 \frac{\upsilon^2 d\upsilon}{(\upsilon^2-1)^{3/2+\mu}}
%\tilde{g}^{(a)}(\upsilon)
- \int_{\infty}^1 \left\{\tilde{g}^{(a)}(\upsilon)\right\}' d\upsilon
\int_{\infty}^{\upsilon} \frac{\upsilon'^2 d\upsilon'}{(\upsilon'^2-1)^{3/2+\mu}} \right]
\nonumber\\
&= \frac{w^{(0)}(1)}{(2\mu)^2}
\left[ \tilde{g}^{(a)}(1) F\left(\frac{3}{2}+\mu, \mu, 1+\mu; 1\right)
- \int_{\infty}^1 \left\{\tilde{g}^{(a)}(\upsilon)\right\}'
\upsilon^{-2\mu}
F\left(\frac{3}{2}+\mu, \mu, 1+\mu; \frac{1}{\upsilon^2}\right) d\upsilon \right] 
\nonumber\\
&= \frac{w^{(0)}(1)}{(2\mu)^2}
\left[ \frac{\Gamma(1+\mu)\Gamma\left(-\frac{\textstyle1}{\textstyle2}-\mu\right)}
{\Gamma\left(-\frac{\textstyle1}{\textstyle2}\right)} \tilde{g}^{(a)}(1)
- \int_{\infty}^1 \frac{\left\{\tilde{g}^{(a)}(\upsilon)\right\}' \upsilon^3}
{(\upsilon^2-1)^{3/2+\mu}}
F\left(\frac{3}{2}+\mu, 1, 1+\mu; \frac{1}{1-\upsilon^2}\right)  d\upsilon \right]
\nonumber\\
&= -\, \frac{w^{(0)}(1)}{(2\mu)^2} \int_{\infty}^1
\upsilon^3 \left\{\tilde{g}^{(a)}(\upsilon)\right\}'
\left[ - \frac{\Gamma(1+\mu)\Gamma\left(-\frac{\textstyle1}{\textstyle2}-\mu\right)}
{\Gamma\left(-\frac{\textstyle1}{\textstyle2}\right)}
\frac{1}{\upsilon^3} \right.
\nonumber\\
&\hskip50mm \left.
+\, \frac{1}{(\upsilon^2-1)^{3/2+\mu}}
F\left(\frac{3}{2}+\mu, 1, 1+\mu; \frac{1}{1-\upsilon^2}\right) \right] d\upsilon
\nonumber\\
&= -\, \frac{w^{(0)}(1)}{2\mu(1+2\mu)} \int_{\infty}^1
\frac{\upsilon^3}{(\upsilon^2-1)^{3/2+\mu}}
\left\{\tilde{g}^{(a)}(\upsilon)\right\}'
F\left(1, 1-\mu, \frac{1}{2}-\mu; 1-\upsilon^2\right) d\upsilon
\nonumber\\
&= \frac{w^{(0)}(1)}{4\mu(1+2\mu)} \int_0^{\infty}
t^{-3/2+\mu} dt\,\upsilon^2
\left\{\tilde{g}^{(a)}(\upsilon)\right\}'
F\left(1, 1-\mu, \frac{1}{2}-\mu; -\frac{1}{t}\right).
\label{f(1)_sp}
\end{align}
Here we ignored the necessary factor of $-2/N$ till eq.~(\ref{f^a_S}).
Inserting eq.~(\ref{tilde^g^sp}) into eq.~(\ref{f(1)_sp}) gives
\begin{align}
f^{\rm sp}(1)\bigr|_{\rm subtracted}
&= -\,\frac{C_{\Xi}\,w^{(0)}(1)}{12\mu(1+2\mu)(3+2\mu)}
 \int_0^{\infty} dt\,t^{1/2+2\mu}
 F\left(1, 1-\mu, \frac{1}{2}-\mu; -\frac{1}{t}\right)
\nonumber\\
&\times \left[ \frac{1}{5+4\mu}
{}_3F_2\left(3+\mu,1+\mu,\,\frac{3}{2};\,\frac{5}{2},\,\frac{7}{2}+2\mu;\,-t\right) \right.
\nonumber\\
&\hskip2mm \left. +\, \frac{t}{7+4\mu}
{}_4F_3\left(3+\mu,1+\mu,\,\frac{3}{2},\,\frac{7}{2}+2\mu;\,
\frac{5}{2}+2\mu,\,\frac{5}{2},\,\frac{9}{2}+2\mu;\,-t\right) \right]
\nonumber\\
&= -\,\frac{C_{\Xi}\,w^{(0)}(1)}{12\mu(1+2\mu)(3+2\mu)}
\frac{\Gamma\left(\frac{\textstyle1}{\textstyle2}-\mu\right)
\Gamma\left(\frac{\textstyle5}{\textstyle2}\right)
\Gamma\left(\frac{\textstyle5}{\textstyle2}+2\mu\right)}
{2\Gamma(1-\mu)\Gamma(3+\mu)\Gamma(1+\mu)
\Gamma\left(\frac{\textstyle3}{\textstyle2}\right)}
\nonumber\\
& \times \int_0^{\infty} dt\,t^{(3/2+2\mu)-1} E\left(1,\, 1-\mu\,:\, \frac{1}{2}-\mu\,:\, t\right)
\left[ E\left(3+\mu,1+\mu,\,\frac{3}{2}\,:\,\frac{5}{2},\,\frac{7}{2}+2\mu\,:\,\frac{1}{t}\right)
\right.
\nonumber\\
&\hskip10mm \left. +\,t\, E\left(3+\mu,1+\mu,\,\frac{3}{2},\,\frac{7}{2}+2\mu\,:\,
\frac{5}{2}+2\mu,\,\frac{5}{2},\,\frac{9}{2}+2\mu\,:\,\frac{1}{t}\right) \right],
\label{f^sp(1)}
\end{align}
where we used MacRobert's $E$ function, eq.~(\ref{MacRobert}).
The integrals appearing in eq.~(\ref{f^sp(1)}) are evaluated in Eq (2)
of Ragab's paper \cite{Ragab} as
\begin{align}
&\int_0^{\infty} dt\,t^{k-1} 
E(\alpha_1, ...,\,\alpha_p\,:\,\beta_1, ...,\,\beta_q\,:\, t)
E\left(\sigma_1, ...,\,\sigma_n\,:\,\tau_1, ...,\,\tau_m\,:\, \frac{z}{t}\right)
\nonumber\\
&=\frac{\pi}{\sin k\pi} \left[ z^k
E(\alpha_1, ...,\,\alpha_p,\,\sigma_1-k, ...,\,\sigma_n-k\,:\,
1-k,\,\beta_1, ...,\,\beta_q,\,\tau_1-k, ...,\,\tau_m-k\,:\, -z) \right.
\nonumber\\
&\left. -\, E(\alpha_1+k, ...,\,\alpha_p+k,\,\sigma_1, ...,\,\sigma_n\,:\,
1+k,\,\beta_1+k, ...,\,\beta_q+k,\,\tau_1, ...,\,\tau_m\,:\, -z) \right].
\label{Ragab}
\end{align}
%
%\begin{align}
%\frac{\Gamma(\alpha_1)\cdots\Gamma(\alpha_p)}
%{\Gamma(\beta_1)\cdots\Gamma(\beta_p)}
%{}_pF_q\left(\alpha_1, ..., \alpha_p;\,\beta_1, ...,\beta_q;\,-\frac{1}{x}\right),
%\label{MacRobert}
%\end{align}
%
Thus, eq.~(\ref{f^sp(1)}) becomes
\begin{align}
f^{\rm sp}(1)\bigr|_{\rm subtracted}
&= -\,\frac{C_{\Xi}\,w^{(0)}(1)}{12\mu(1+2\mu)(3+2\mu)}
\frac{\Gamma\left(\frac{\textstyle1}{\textstyle2}-\mu\right)
\Gamma\left(\frac{\textstyle5}{\textstyle2}\right)
\Gamma\left(\frac{\textstyle5}{\textstyle2}+2\mu\right)}
{2\Gamma(1-\mu)\Gamma(3+\mu)\Gamma(1+\mu)
\Gamma\left(\frac{\textstyle3}{\textstyle2}\right)}
\frac{\pi}{\sin\left(\frac{\textstyle3}{\textstyle2}+2\mu\right)\pi}
\nonumber\\
&\times \Bigl( A_1 + A_2 + A_3 + A_4 \Bigr),
\end{align}
where $A_1, A_2, A_3$ and $A_4$ are given by
\begin{subequations}
\begin{align}
A_1 &= E\left(1,\,1-\mu, \frac{3}{2}-\mu, -\frac{1}{2}-\mu,\,-2\mu\,:\,
-\frac{1}{2}-2\mu, \frac{1}{2}-\mu, 1-2\mu,\,2\,:\, -1\right),
\label{A1}
\\
A_2 &= - \,E\left(1,\,1-\mu, \frac{1}{2}-\mu, -\frac{3}{2}-\mu, -1-2\mu,\,1\,:\,
-\frac{3}{2}-2\mu, \frac{1}{2}-\mu,\,0,\,-2\mu,\,2\,:\, -1\right),
\label{A2}
\\
A_3 &= -\, E\left(\frac{5}{2}+2\mu, \frac{5}{2}+\mu, 3+\mu, 1+\mu,\,\frac{3}{2}\,:\,
\frac{5}{2}+2\mu, 2+\mu,\,\frac{5}{2},\,\frac{7}{2}+2\mu\,:\, -1\right),
\label{A3}
\\
A_4 &= \, E\left(\frac{7}{2}+2\mu, \frac{7}{2}+\mu, 3+\mu, 1+\mu,\,\frac{3}{2},\,
\frac{7}{2}+2\mu\,:\,
\frac{7}{2}+2\mu, 3+\mu, \frac{5}{2}+2\mu,\,\frac{5}{2},\,\frac{9}{2}+2\mu\,:\, -1\right),
\label{A4}
\end{align}
\end{subequations}
where we used $\sin(5/2+2\mu)\pi=-\sin(3/2+2\mu)\pi$.
If the two arguments $\alpha_i$ and $\beta_j$ are equal,
they can simply be erased from the arguments in the $E$ function,
similar to the GHF.
There are several recurrence formulae for the $E$ function as follows
[see eqs. 1 and 2 in Sec. 9.42 of ref.~\cite{Gradshteyn}]:
\begin{subequations}
\begin{align}
\alpha_1 x E(\alpha_1, ...,\, \alpha_p\,:\,\beta_1, ...,\,\beta_q\,:\, x)
&= x E(\alpha_1+1,\, \alpha_2, ...,\, \alpha_p\,:\,\beta_1, ...,\,\beta_q\,:\, x)
\nonumber\\
&+\, E(\alpha_1+1, ...,\, \alpha_p+1\,:\,\beta_1+1, ...,\,\beta_q+1\,:\, x),
\label{E_1}
\\
(\beta_1-1) x E(\alpha_1, ...,\, \alpha_p\,:\,\beta_1, ...,\,\beta_q\,:\, x)
&= x E(\alpha_1, ...,\,\alpha_p\,:\,\beta_1-1,\, \beta_2, ...,\,\beta_q\,:\, x)
\nonumber\\
&+\, E(\alpha_1+1, ...,\,\alpha_p+1\,:\,\beta_1+1, ...,\,\beta_q+1\,:\, x),
\label{E_2}
\end{align}
\vskip-13mm
\begin{align}
E(\alpha_1+1,\, \alpha_2, ...,\, \alpha_p\,:\,\beta_1, ...,\,\beta_q\,:\, x)
- E(\alpha_1, ...,\,\alpha_p\,:\,\beta_1-1,\, \beta_2,...,\,\beta_q\,:\, x)
\nonumber\\
= \left[\alpha_1-(\beta_1-1)\right]
E(\alpha_1, ...,\,\alpha_p\,:\,\beta_1, ...,\,\beta_q\,:\, x).
\label{E_3}
\end{align}
Equation~(\ref{E_3}) is very easily obtained by subtracting
eq.~(\ref{E_2}) from eq.~(\ref{E_1}).
As special cases, eqs.~(\ref{E_1}) and (\ref{E_3}) reduces to
\begin{align}
E(1,\, \alpha_2, ...,\, \alpha_p\,:\,2,\, \beta_2, ...,\, \beta_q\,:\, -t)
&= t E(\alpha_2-1, ...,\, \alpha_p-1\,:\,\beta_2-1, ...,\,\beta_q-1\,:\, -t)
\nonumber\\
&-\,t\frac{\Gamma(\alpha_2-1)\cdots\Gamma(\alpha_p-1)}
{\Gamma(\beta_2-1)\cdots\beta(\alpha_p-1)},
\label{E_4}
\\
E(1,\, 1,\, \alpha_3, ...,\, \alpha_p\,:\,0,\, 2,\, \beta_3, ...,\,\beta_q\,:\, -t)
&= E(\alpha_3, ...,\, \alpha_p\,:\,\beta_2, ...,\,\beta_q\,:\, -t)
\nonumber\\
&-\,t E(\alpha_3-1, ...,\, \alpha_p-1\,:\,\beta_3-1, ...,\,\beta_q-1\,:\, -t)
\nonumber\\
&+\, t\frac{\Gamma(\alpha_3-1)\cdots\Gamma(\alpha_p-1)}
{\Gamma(\beta_3-1)\cdots\beta(\alpha_p-1)}.
\label{E_5}
\end{align}
\end{subequations}
Moreover, from eq. (5) in P. 20 of ref.~\cite{HGS} for integer $s=e+f-a-b-c$, \begin{align}
&\frac{F_p(0;4,5)}
{\Gamma(1-c)\Gamma(1-b)\Gamma(1-a)\Gamma(1-f+a)\Gamma(1-f+c)}
-\frac{F_p(5;0,4)}
{\Gamma(1-a)\Gamma(a)\Gamma(1-b)\Gamma(b)\Gamma(1-c)\Gamma(c)}
\nonumber\\
&\hskip30mm = -\frac{2-f}{\pi\Gamma(s)\Gamma(e-s)\Gamma(e-b)\Gamma(e-a)}
\begin{cases}
F_n(0;2,3), \\
F_n(0;4,5)
\end{cases}
\label{Fp-Fp=Fn}
\end{align}
%[were we used the relation $\Gamma(1-x)\Gamma(x) = \pi\,/\sin\pi x$],
with
\begin{subequations}
\begin{align}
F_p(0;4,5) &= \frac{{}_3F_2(a,b,c;e,f;1)}{\Gamma(s)\Gamma(e)\Gamma(f)},
\\
F_p(5;0,4) &= \frac{{}_3F_2(1-f+a,1-f+b,1-f+c;2-f,1-f+e;1)}
{\Gamma(s)\Gamma(2-f)\Gamma(1-f+e)},
\\
F_n(0;2,3) &= \frac{{}_3F_2(1-s,1-e+a,1-f+a;2-s-b,2-s-c;1)}
{\Gamma(1-a)\Gamma(2-s-b)\Gamma(2-s-c)},
\\
F_n(0;4,5) &= \frac{{}_3F_2(1-a,1-b,1-c;2-e,2-f;1)}
{\Gamma(1-s)\Gamma(2-e)\Gamma(2-f)},
\end{align}
\end{subequations}
one can derive
\begin{subequations}
\begin{align}
&E(a, b, c\,:\, e, f\,:\, -1) -\, E(1-f+a, 1-f+b, 1-f+c\,:\, 2-f, 1-f+e\,:\, -1)
\nonumber\\
&= -\, \frac{\sin\pi(2-f) \Gamma(a)\Gamma(b)\Gamma(c)
\Gamma(1-f+a)\Gamma(1-f+b)\Gamma(1-f+c)}
{\Gamma(e-a)\Gamma(e-b)\Gamma(e-c)} \times H
\label{E-E_1}
\end{align}
with $H=1$ from $F_n(0;2,3)$ of eq.~(\ref{Fp-Fp=Fn})
for $s=e+f-a-b-c=1$, and
\begin{align}
H = \frac{\Gamma(1-a)\Gamma(1-b)\Gamma(1-c)}{\Gamma(2-e)\Gamma(2-f)}
{}_3F_2(1-a, 1-b, 1-c; 2-e, 2-f; 1)
\label{E-E_2}
\end{align}
from $F_n(0;4,5)$  of eq.~(\ref{Fp-Fp=Fn}) for $s=e+f-a-b-c=0$,
where both sides of eq.~(\ref{Fp-Fp=Fn}) were multiplied by $\Gamma(s)$.
\end{subequations}
Using all these relations, we obtained
\begin{align}
A_1 + A_2 + A_3 + A_4
&=-\,\frac{\Gamma(-\mu)
\Gamma\left(\frac{\textstyle1}{\textstyle2}-\mu\right)
\Gamma\left(-\frac{\textstyle3}{\textstyle2}-\mu\right)
\Gamma(-1-2\mu)}
{\Gamma\left(-\frac{\textstyle3}{\textstyle2}-2\mu\right)
\Gamma\left(-\frac{\textstyle1}{\textstyle2}-\mu\right)
\Gamma(-2\mu)}
\nonumber\\
&-\,\frac{\Gamma(-\mu)
\Gamma\left(-\frac{\textstyle5}{\textstyle2}-\mu\right)
\Gamma(-2-2\mu)}
{\Gamma\left(-\frac{\textstyle5}{\textstyle2}-2\mu\right)
\Gamma(-1-2\mu)}
\nonumber\\
&= \frac{\Gamma(-\mu)
\Gamma\left(-\frac{\textstyle5}{\textstyle2}-\mu\right)\mu(3+2\mu)}
{2\Gamma\left(-\frac{\textstyle3}{\textstyle2}-2\mu\right)(2+2\mu)}.
\label{A1A2A3A4}
\end{align}

In addition to this special solution part,
the homogeneous $c_2w^{(0)}(\upsilon)$ part in $\nu(\upsilon)$
in eq.~(\ref{nu=nu+c1+c2}) should be considered also.
For this purpose, eqs.~(\ref{d_tilde^g=C})-(\ref{C=}) can be used
to determine the corresponding contribution to $\tilde{g}(\upsilon)$,
which is denoted as $\tilde{g}^0(\upsilon)$.
By the replacement of $\tilde{h}^{\rm sp}(\upsilon)$ with $c_2w^{(0)}(\upsilon)$,
eq.~(\ref{C=}) becomes
\begin{align}
C(\upsilon) &= c_2 \int_{\infty}^{\upsilon} d\upsilon'
\frac{\upsilon'^2}{(\upsilon'^2-1)^{3/2+\mu}}w^{(0)}(\upsilon')
\nonumber\\
&= c_2 \int_{\infty}^{\upsilon} d\upsilon'
\frac{\upsilon'^2}{(\upsilon'^2-1)^{3/2+\mu}} \frac{1}{\upsilon'^2}
F\left(\frac{3}{2},\,1,\,1-\mu;\,\frac{1}{\upsilon'^2}\right)
\nonumber\\
&= -\,\frac{c_2}{2} \int_{\infty}^{\frac{\scriptstyle1}{\scriptstyle\upsilon^2-1}} dt\,
t^{\,\mu} (1+t)^{1/2}
F\left(-\frac{1}{2}-\mu,\,1,\,1-\mu;\,-t\right)
\nonumber\\
&= -\,\frac{c_2}{2} \int_{\infty}^{\frac{\scriptstyle1}{\scriptstyle\upsilon^2-1}} dt\,
t^{\,\mu} (1+t)
F\left(\frac{3}{2},\,-\mu,\,1-\mu;\,-t\right)
\nonumber\\
&= -\,\frac{c_2\, t^{1+\mu}}{2(1+\mu)}
{}_3F_2\left(\frac{3}{2},\,-\mu,\,1+\mu; 1-\mu,2+\mu;\,-t\right)
\nonumber\\
&\hskip4mm -\,\frac{c_2\, t^{2+\mu}}{2(2+\mu)}
{}_3F_2\left(\frac{3}{2},\,-\mu,\,2+\mu; 1-\mu,3+\mu;\,-t\right),
\end{align}
where we put $t=1/(\upsilon^2-1)$ as usual.
Then, $\tilde{g}(\upsilon)$ is written, similar to eq.~(\ref{tilde^g^sp}), as
\begin{align}
\tilde{g}^0(\upsilon)
= -\,\frac{c_2}{2}
\int_{\infty}^{\upsilon} \frac{d\upsilon'}{\upsilon'^2} \frac{1}{t^{1/2+\mu}}
&\left[ \frac{t^{1+\mu}}{1+\mu}
{}_3F_2\left(\frac{3}{2},\,-\mu,\,1+\mu; 1-\mu,2+\mu;\,-t\right) \right.
\nonumber\\
&\hskip-2mm \left. +\,\frac{t^{2+\mu}}{2+\mu}
{}_3F_2\left(\frac{3}{2},\,-\mu,\,2+\mu; 1-\mu,3+\mu;\,-t\right) \right].
\label{tilde^g^0}
\end{align}
To obtain the corresponding $f^0(1)\bigr|_{\rm subtracted}$,
$\tilde{g}^0(\upsilon)$ is inserted into eq.~(\ref{f(1)_sp}),
in place of $\tilde{g}^{(a)}$, as
\begin{align}
f^0(1)\bigr|_{\rm subtracted}
&= -\, \frac{c_2\,w^{(0)}(1)}{8\mu(1+2\mu)}
 \int_0^{\infty} dt\,t^{\,\mu-1}
 F\left(1, 1-\mu, \frac{1}{2}-\mu; -\frac{1}{t}\right)
\nonumber\\
&\times \left[ \frac{1}{1+\mu}
{}_3F_2\left(\frac{3}{2},\,-\mu,\,1+\mu; 1-\mu,\,2+\mu;\,-t\right) \right.
\nonumber\\
&\hskip2mm \left. +\,\frac{t}{2+\mu}
{}_3F_2\left(\frac{3}{2},\,-\mu,\,2+\mu; 1-\mu,\,3+\mu;\,-t\right) \right]
\nonumber\\
&= -\,\frac{c_2\,w^{(0)}(1)}{8\mu(1+2\mu)}
\frac{\Gamma\left(\frac{\textstyle1}{\textstyle2}-\mu\right)}
{\Gamma\left(\frac{\textstyle3}{\textstyle2}\right)\Gamma(-\mu)}
\int_0^{\infty} dt\,t^{\,\mu-1} E\left(1, 1-\mu\,:\, \frac{1}{2}-\mu\,:\, t\right)
\nonumber\\
& \times
\left[ E\left(\frac{3}{2},\,-\mu,\,1+\mu\,:\,1-\mu,\,2+\mu\,:\,\frac{1}{t}\right)
\right.
\nonumber\\
&\left. +\,t\, E\left(\frac{3}{2},\,-\mu,\,2+\mu\,:\,
1-\mu,\,3+\mu\,:\,\frac{1}{t}\right) \right]
\nonumber\\
&= -\,\frac{c_2\,w^{(0)}(1)}{8\mu(1+2\mu)}
\frac{\Gamma\left(\frac{\textstyle1}{\textstyle2}-\mu\right)}
{\Gamma\left(\frac{\textstyle3}{\textstyle2}\right)\Gamma(-\mu)}\,
\frac{\pi}{\sin\mu\pi}\, (B_1 + B_2 + B_3 + B_4),
\label{f^0(1)}
\end{align}
where we used Ragab's formula (\ref{Ragab}) again to obtain
$B_1, B_2, B_3$ and $B_4$ as
\begin{subequations}
\begin{align}
B_1 &= E\left(1,\, 1-\mu, \frac{3}{2}-\mu,\,-2\mu,\, 1\,:\,
1-\mu, \frac{1}{2}-\mu, 1-2\mu,\, 2\,:\, -1\right),
\label{B1}
\\
B_2 &= - \,E\left(1,\, 1-\mu, \frac{1}{2}-\mu, -1-2\mu,\, 1\,:\,
-\mu,\, \frac{1}{2}-\mu,\, -2\mu,\, 2\,:\, -1\right),
\label{B2}
\\
B_3 &= -\, E\left(1+\mu,\, 1,\, \frac{3}{2},\, -\mu,\, 1+\mu\,:\,
1+\mu,\, \frac{1}{2},\, 1-\mu, 2+\mu\,:\, -1\right),
\label{B3}
\\
B_4 &= \, E\left(2+\mu,\, 2,\, \frac{3}{2},\, -\mu,\, 2+\mu\,:\,
2+\mu,\, \frac{3}{2},\, 1-\mu, 3+\mu\,:\, -1\right).
\label{B4}
\end{align}
\end{subequations}
Again, using eqs.~(\ref{E_1})-(\ref{E_5}),
(\ref{E-E_1}), and (\ref{E-E_2}), we obtained
\begin{align}
B_1 + B_2 + B_3 + B_4 &= -\frac{1}{4\mu}
\left[F(-2\mu,1,1-2\mu;1)+2\mu F(1, 1, 2; 1)\right]
\nonumber\\
&+\, \frac{1}{2(1+2\mu)}
\left[F(-1-2\mu,1,-2\mu;1)+(1+2\mu) F(1, 1, 2; 1)\right]
\nonumber\\
&+\, \frac{1}{2\mu(1+\mu)}
\left[(1+\mu)F(-\mu,1,1-\mu;1)+\mu F(1+\mu, 1, 2+\mu; 1)\right]
\nonumber\\
&-\, \frac{1}{2(1+\mu)^2}
\left[(1+\mu)F(-1-\mu,1,-\mu;1)+(1+\mu) F(1+\mu, 1, 2+\mu; 1)\right]
\nonumber\\
&=-\, \frac{\mu}{2(1+\mu)(1+2\mu)}.
\label{B1B2B3B4}
\end{align}
Thus, adding everything and multiplying the necessary factor, $-2/N$,
which we have ignored above, we arrived at
\begin{align}
f^{(a)}(1)\bigr|_{\rm subtracted}
&= -\,\frac{2}{N}\,\frac{C_{\Xi}\,w^{(0)}(1)}{16\mu(1+2\mu)(3+2\mu)}
\frac{\Gamma\left(\frac{\textstyle1}{\textstyle2}-\mu\right)
\Gamma\left(\frac{\textstyle5}{\textstyle2}+2\mu\right)}
{\Gamma(3+\mu)\Gamma(1+\mu)\Gamma(1-\mu)}
\frac{\mu}{2(2+2\mu)}
\nonumber\\
&\times \left[ \frac{2\!\sqrt{\pi}\Gamma(4+4\mu)}{\Gamma(2+2\mu)2^{4+4\mu}}
\frac{2\!\sqrt{\pi}\Gamma(-1-2\mu)}{2^{-1-2\mu}}
\frac{2}{\frac{\textstyle5}{\textstyle2}+\mu}
+ \frac{\pi}{2^{2\mu}\left(-\frac{\textstyle1}{\textstyle2}+\mu\right)(1+2\mu)} \right].
\label{f^a_S}
\end{align}
Then, the self-energy of Figure (b) is evaluated as
\begin{align}
\Sigma^{(b)}_q(y,y') &= -\,f^{(a)}(1)\bigr|_{\rm subtracted}\, \delta(y - y')
\int_0^{\infty}\Pi^{-1}_{q=0}(y,y'')\frac{dy''}{y''^{d-2}}
\nonumber\\
&= -\,f^{(a)}(1)\bigr|_{\rm subtracted}\, \delta(y - y')\,
\frac{D_{3-d,\,\mu}}{y^2}
= \frac{A}{C_f}\, \frac{1}{y^2}\delta(y - y'),
\label{Sigma^b_S}
\end{align}
where we used the Mellin transform of $\Pi^{-1}_{q=0}(y,y')$
with $\sigma=3-d=-\,2-2\mu$ in eqs.~(\ref{Pi^-1=D}) and (\ref{Mellin}).
The constant $D_{3-d,\,\mu}$ is given by
\begin{align}
D_{3-d,\,\mu}
%&= C_{\Pi}^{-1} \frac{\Gamma\left(\frac{\textstyle3-\sigma}{\textstyle2}\right)
%\Gamma\left(\frac{\textstyle3+\sigma}{\textstyle2}+\mu\right)}
%{2^{1+2\mu}\left(\frac{\textstyle\sigma}{\textstyle2}\right)
%\left(\mu+\frac{\textstyle\sigma}{\textstyle2}\right)
%\Gamma\left(1+2\mu+\frac{\textstyle\sigma}{\textstyle2}\right)
%\Gamma\left(1+\mu-\frac{\textstyle\sigma}{\textstyle2}\right)}
%\nonumber\\
%&= C_{\Pi}^{-1} \frac{\Gamma\left(\frac{\textstyle5+2\mu}{\textstyle2}\right)
%\Gamma\left(\frac{\textstyle1-2\mu}{\textstyle2}+\mu\right)}
%{2^{1+2\mu}(-1-\mu)(\mu-1-\mu)\Gamma(1+2\mu-1-\mu)\Gamma(1+\mu+1+\mu)}
%\nonumber\\
%&= -\, C_{\Pi}^{-1} \frac{\Gamma\left(\frac{\textstyle5}{\textstyle2}+\mu\right)
%\Gamma\left(\frac{\textstyle1}{\textstyle2}\right)}
%{2^{1+2\mu}(-1-\mu)\Gamma(\mu)\Gamma(2+2\mu)}
%\nonumber\\
%
&= -\, \frac{\Gamma\left(\frac{\textstyle5}{\textstyle2}+\mu\right)
\Gamma\left(\frac{\textstyle1}{\textstyle2}\right)}
{2^{1+2\mu}(-1-\mu)\Gamma(\mu)\Gamma(2+2\mu)}
\frac{2^{3+2\mu}\Gamma(3+2\mu)}
{\Gamma\left(\frac{\textstyle3}{\textstyle2}+\mu\right)
\Gamma\left(-\frac{\textstyle1}{\textstyle2}-\mu\right)}
\,\frac{1}{C_f}
\nonumber\\
&
= \frac{\Gamma\left(\frac{\textstyle5}{\textstyle2}+\mu\right)
\Gamma(1+\mu)2^{4+2\mu}}
{\Gamma(\mu)
\Gamma(2+2\mu)
\Gamma\left(-\frac{\textstyle1}{\textstyle2}-\mu\right)}
\,\frac{1}{C_f}.
\nonumber
\end{align}
The constant $A$ in eq.~(\ref{Sigma^b_S}) is equal to
$-\,f^{(a)}(1)\bigr|_{\rm subtracted}\,D_{3-d,\,\mu}\,C_f$.
Integrating eq.~(\ref{Dg=h}) is trivial due to the delta function in eq.~(\ref{Sigma^b_S}),
which leads to
\begin{align}
g^{(b)}(\upsilon) = \frac{A}{C_f}C_f g^{(0)}(\upsilon)
 = \frac{A\upsilon}{(\upsilon^2-1)^{d/2-1}}.
\end{align}
The contribution to the correlation function, $f^{(b)}(\upsilon)$,
can be calculated using eq.~(\ref{f=wgg}).
The critical anomaly arises from the second term only.
Using the first expression in eq.~(\ref{w_S}) for $w^{(0)}(\upsilon)$ results in 
\begin{align}
f^{(b)}(\upsilon) &=  
%\frac{\Gamma\left(\frac{\textstyle1}{\textstyle2}-\mu\right)\Gamma(-\mu)}
%{e^{i(1/2+\mu)\pi}2^{1+2\mu}\!\sqrt{\pi}}\,
-\,\frac{1}{2\mu}
g^{(0)}(\upsilon) \int_{c'}^{\upsilon} (\upsilon'^2-1)^{d/2-1}
g^{(b)}(\upsilon') w^{(0)}(\upsilon') d\upsilon'
%\nonumber\\
%&= \frac{\Gamma\left(\frac{\textstyle1}{\textstyle2}-\mu\right)\Gamma(-\mu)}
%{e^{i(1/2+\mu)\pi}2^{1+2\mu}\!\sqrt{\pi}}
%\frac{2^{1/2+\mu}}{\Gamma\left(\frac{\textstyle1}{\textstyle2}-\mu\right)}
%\frac{e^{i(1/2+\mu)\pi}\sqrt{\pi}}{2^{1/2-\mu}\Gamma(1-\mu)}
%\frac{A}{(\upsilon^2-1)^{1/2+\mu}}
%\int_{c'}^{\upsilon}
%\frac{1}{\upsilon'}F\left(1,\,\frac{1}{2}\,1-\mu;\,\frac{1}{\upsilon'^2}\right)d\upsilon'
\nonumber\\
&= -\,\frac{A}{2\mu\,(\upsilon^2-1)^{d/2-1}} \int_{c'}^{\upsilon}
\frac{1}{\upsilon'}F\left(\frac{3}{2},\,1,\,1-\mu;\,\frac{1}{\upsilon'^2}\right)d\upsilon'
\nonumber\\
&\rightarrow  -\,\frac{A\,\upsilon}{2\mu\,(\upsilon^2-1)^{d/2-1}} \log\upsilon,
\;\;\;\; \textrm{for $\upsilon\rightarrow\infty$}.
\end{align}
Thus, we found
\begin{align}
\hat{\gamma}^S &= - \frac{NA}{C_f}\,\frac{1}{2\mu}
\nonumber\\
&= \frac{N}{C_f}\,\frac{1}{2\mu}\, f^{(a)}(1)\bigr|_{\rm subtracted}\,
\,D_{3-d,\,\mu}\,C_f
\nonumber\\
&= \frac{2\Gamma(-1-2\mu)}{\Gamma(2+2\mu)}
\left[ \frac{(1-2\mu)\Gamma(4+4\mu)}{(5+2\mu)\Gamma(2+2\mu)}
+ \frac{1}{\Gamma(-2\mu)} \right],
\end{align}
which is identical to eq.~(\ref{Eq_S}).

\end{document}